\documentclass{article}[12pt]
\usepackage{latexsym}
\begin{document}
\pagenumbering{roman}
\title{Statistical Distribution of Crystallographic
Groups for Inorganic Crystal 
Structure Database 
}
\date{June 29, 2011}
\author{Miyako Fujiwara 
\and Yoshiaki Itoh\thanks{The Institute of Statistical Mathematics and The Graduate University for Advanced Studies, 10-3 Midori-cho, Tachikawa,  Tokyo 190-8562, Japan,    e-mail : itoh@ism.ac.jp} 
\and Takeo Matsumoto\thanks {Kanazawa University, Kakuma-machi, Kanazawa-shi, Ishikawa 920-1164, Japan} 
\and Hiroshi Takeda\thanks{ University of Tokyo,  7-3-1 Hongo, Bunkyo-ku, Tokyo 113-0033, Japan}}
\maketitle
{\footnotesize
\newpage
\tableofcontents
\newpage
\pagenumbering{arabic}

\section{Introduction}
Applying the method in the previous  monograph  (\cite{fuji98}), here  we show that the 
number of formula 
units  in a unit   cell gives a natural classification to understand the 
statistical distribution of Crystallographic groups.

	We introduced a method that defines the species 
(representatives) of inorganic compounds, and studied the statistical 
distribution of the defined species among space groups (distribution of space groups), 
by using ICSD (Inorganic Crystal Structure Database (1990)),
(\cite{fuji98}).

	ICSD is a database of the CRYSTIN (Crystal Structure Information System),
 for complete structural information of non-metal inorganic crystals. 
FIZ Karlsruhe and GMELIN Institute Frankfurt are in charge of making this 
database, ~in cooperation with the Institute for Inorganic Chemistry of the 
University of Bonn.

	CRYSTIN includes two other databases. CSD (Cambridge Structual Database, 
 Organic and Organo-Metallic Compounds) made by Cambridge Crystallographic Data 
Cwntre, and CRYSTMET (The NRCC Metals Crystallographic Data File, Metallic
phases) made by Canadian Scientific Numeric Database Service and National 
Research Council of Canada.

	ICSD includes compound names, chemical formula, unit cell which is a  
basic element of a compound, atomic position of each atom in the unit cell, 
 symmetry operation parameter which derives entire positions of atoms in the cell from those in an asymmetric 
unit of the cell, and bibliographic data, etc.
Some descriptors in the database are as follows (Table 1.):

\vspace{1cm}
\begin{tabbing}
xxxxxxxx \= xxxxxxxx \= xxxxxxxxxxxxxxxxxxxxxxxxxxxxxxxxxxxxxxxxxxxxx \= \kill
\> ANX \> FORMULA TYPE \\
\> AUT \> AUTHOR'S NAME \\
\> CLAS \> CRYSTAL CLASS NAME \\
\> CODN \> CODEN \\
\> COL \> COLLECTION CODE \\
\> CVOL \> CELL VOLUME \\
\> D \> MINIMUM INTERATOMIC DISTANCE \\
\> DATE \> DATE OF RECORDING OR CORRECTION \\
\> ELC \> NUMBER OF DIFFERENT CHEMICAL ELEMENTS PRESENT\\
\> ELE \> KEYWORD \\
\> LAST	\> DESCRIPTOR FOR THE SUBSET SELECTED BY THE \\
\>      \> LAST `FIND' COMMAND \\ 
\> LAUE \> LAUE CLASS SYMBOL \\
\> MINR \> MINERAL NAME \\
\> PRS \> PEARSON SYMBOL \\
\> REL \> RANGE OF RELATIVE ENTRY NUMBERS \\
\> REM \> REMARK SYMBOL \\
\> RFCD \> REFCODE \\
\> RVAL \> R VALUE RANGE \\
\> SGR \> SPACE GROUP \\
\> SYPR \> SYMMETRY PROPERTY \\
\> SYST \> CRYSTAL SYSTEM \\
\> TEST \> FLAG \\
\> USRN \> NAME OF A SAVED SUBSET \\
\> YEAR \> YEAR OF PUBLICATION \\
\end{tabbing}
\begin{center}
		Table 1. List of Descriptor Classes 
		(For more information, See ICSD manual)
\end{center}

	The statistical distribution of the symmetry groups has been drawing 
the attention of crystallographers \cite{bel},~\cite{mac},~\cite{matu66},~\cite{matu68},~
\cite{matu69a},~\cite{matu69b},~
\cite{matu88},~\cite{nowa67a},~\cite{nowa67b},~\cite{take},~\cite{tane},~

\noindent 
\cite{wils88},~\cite{wils90},~\cite{wils93}. 
Stochastic models are introduced to explain the statistical distribution 
\cite{itoh86a},~\cite{itoh91}. 
Here we make use of the database ICSD to study the statistical distribution improving our  
 previous works \cite{fuji93},~\cite{fuji94},~\cite{fuji96},~\cite{fuji98}.
Applying the idea of sequential coding \cite{itoh86b}, we sequentially define 
 species by considering i) space group, ii) chemical component 
 and iii) unit cell volume. 

	This work was performed on the HITAC M-680 computer (OS/VOS3) at The Institute of Statistical Mathematics in the year of 2001-2002.

\newpage

\section{Algorithm}

	Making use of the ICSD database, we classify crystalline 
substances by using some of descriptors, define species (representatives) 
for each class of substances, and obtain the statistical distribution of space groups. 
Here we use five descriptors of ICSD (Table 2.).
We define species sequentially as will be given later.

\begin{tabbing}
xxxxxxxxxx \= xxxxxxxxxx \= xxx \= xxxxxxxxxxxxxxxxxxxxxxxxxxxxxxxxxxxxxxxxxxxxxxxxxx \= \kill
\> MINERAL \> : \> Mineral name \\
\> NAME \> : \> Compound name \\
\> FORM \> : \> Structured chemical formula and empirical formula \\
\> CELL \> : \> Unit cell dimensions, unit cell volume, number of \\
\>  \> 	 \> formula units, measured density \\
\> SGR \> : \> Hermann-Mauguin space group symbol \\
\end{tabbing}
\begin{center}
		Table 2. Five Descriptors used
\end{center}

	We take information on these five descriptors from all the substances in ICSD,
 and copy them into a file.  
Then we transform Hermann-Mauguin symbols into 230 Schoenflies symbols 
and we add the number of lattice point per unit cell to the substance.

	We say two substances are mutually close to each other, when the 
following three conditions i), ii), and iii) are satisfied.\\

i) They have the same space group.\\ 

ii) They have mutually similar chemical formula.\\

	Namely if the ratio for the common elements in both substances, $WDIF1$,\\
\hspace*{0.5cm}  is more than 75\%, these two substances are said to be close to each other.\\ 

$$ WDIF1 = \frac{ICHEM*2}{(NC(S_1)+NC(S_2))} $$ \\

where $NC(S_1)$ is the number of elements included in substance 1, $NC(S_2)$ is the number of elements in substance 2, and $ICHEM$ is the number of the common elements of the two substances.\\

\newpage

  Example.\\

  substance 1 ($S_1$) : \(Na Al Si_3 O_8\) ,        the number of elements = 4 $(NC(S_1))$

  substance 2 ($S_2$) : \(Na_{0.667} K_{0.333} Al Si_3 O_8\) ,  the number of elements = 5 $(NC(S_2))$

  the number of the common elements of the two substances = 4 $(ICHEM)$\\

  Hence we obtain $$ WDIF1 = \frac{8}{9}~~~, $$ 
\hspace*{0.5cm}	in this case.\\

iii) The primitive unit cell volume of the substances are mutually close 
to each other. \\

\vspace{1em}

    \hspace*{0.5cm} Consider the following table (Table 3.).\\

\begin{tabular}{llcc}
space & lattice & lattice points & Primitive lattice volume \\
& & of unit cell & /unit cell volume \\

P & Primitive & 1 & 1 \\
C & C-face centered & 2 & 1/2 \\
A & A-face centered & 2 & 1/2 \\
B & B-face centered & 2 & 1/2 \\
I & Body centered & 2 & 1/2 \\
F & All-face centered & 4 & 1/4 \\
R(R) & Rhombohedral setting & 1 & 1 \\
RH & Hexagonal setting(obverse) & 3 & 1/3 \\
RHR & Hexagonal setting(reverse) & 3 & 1/3 \\
\end{tabular}

\begin{center}
Table 3. Lattice Points of Unit Cell
\end{center}

	If the following value of $WDIF2$ for the primitive unit cell volumes
of the two substances is smaller than the given value of $DIFF2$, we say the two primitive unit cell  
volumes are mutually close to each other with respect to $DIFF2$,

  $$ WDIF2 = \frac{\bigl| (VOL[S_1]-VOL[S_2]) \bigr|}{(VOL[S_1]+VOL[S_2])}~~~, $$ 

\noindent for the primitive unit cell volume of substance 1 ($S_1$), $VOL[S_1]$, and for that of 
substance 2 ($S_2$), $VOL[S_2]$.  (\cite{deza},see page 118.)\\

	The procedure of determining the representatives in this method is as 
follows.
Take the necessary information from the ICSD and copy 
them into a file. When $N$ substances are already defined to be 
representatives, we compare the next substance with these $N$ substances. 
If this substance is not close to the $N$ substances, we define this as the 
$N+1$th representative.  
If this is close to one of the defined $N$ substances, we do not define it as the $N+1$th  
substance.  
We print out tables of the distributions of point groups, for the 
species which we think as representatives of crystalline substances.\\

	The following Hermann-Mauguin symbols which seems to be unclearly stated in ICSD
are not transformed to Schoenflies symbols.\\ 

\begin{tabular}{llcc}
B1N1 &  I121S &  I12/A1S\\
\end{tabular}

\newpage
\section{Examples}

	We studied the statistical distribution of point groups for ICSD by 
the classification of the following categories \cite{fuji96},\cite{fuji98}.
\vspace*{1ex}\\
\hspace*{2em}1.ELEMENTS\\
\hspace*{2em}2.CRYSTAL FAMILY\\
\hspace*{2em}3.ANX SYMBOL\\

We add a new category, which is\
\hspace*{1em}4.ANX SYMBOL and NUMBER of FORMULA UNITS per UNIT CELL(Z)\
We give the result for ICSD. In this paper, we report two examples
such as 1.ELEMENTS and 4.ANX SYMBOL and Z.

\subsection{ELEMENTS}
	We define `Elements' as a species 
whose chemical formula has one chemical element. We use the ICSD command 
`FIND ELC=1'. ELC is defined as `Number of different chemical elements'.  
	Take Elements from ICSD, and count the number of substances.  
We obtain 279 substances by this command.\\

        \subsubsection{Operation}
The procedure is as follows.\

  1. Execute ICSD  to find out the number of substances in ICSD to edit
     ITOH4XC.@DATANO.DATA file.\\
\begin{verbatim}

        >>EX #ITOH4XC.MACRO.CLIST(ICSD)(Enter)

        ***********   Crystal Structure Information System
        * CRYSTIN *   (Release 2.51/Mar 90)
        ***********   Written by Hundt & Sievers (University of Bonn)

        DB2201 - Please give BASE (or HELP) command.
        Command?
        BASE 1(Enter)
        Command?
        FIND ELC=1(Enter)
        Current subset ("LAST") contains    279 entries.
        Command?
        QUIT(Enter)
        >> 
\end{verbatim}

Type `BASE 1' at first. The command to finish ICSD is `QUIT'.\\

For our study we edit following two files by ASPEN.\\

  2. ITOH4XC.@COMMAND.DATA (80A1)
\begin{verbatim}
        >>ASP #ITOH4XC.@COMMAND.DATA(Enter)

               ----+----1----+----2----+----3----+----4----+----5----+----6--
        000100 BASE 1
        000110 FIND ELC=1
        000120 PRINT 40 MINR,NAME,FORM,SGR,CELL
        000130 QUIT

\end{verbatim}

  3. ITOH4XC.@DATANO.DATA (2F5.2)($DIFF1,~DIFF2$)\\
\hspace*{14em}(3(I5,~A18,~1X)) (numbers of data)
\begin{verbatim}
        >>ASP #ITOH4XC.@DATANO.DATA(Enter)

               ----+----1----+----2----+----3----+----4----+----5----+----6--
        000100  0.75 0.025
        000110   279ELC=1(0.025)           0                       0

\end{verbatim}

We have to put 0(zero) at the 5 or 29 or 53th column.\\

  4. Execute ITOH4XC.MACRO.CLIST(ICSDMET).
\begin{verbatim}
        >>EX #ITOH4XC.MACRO.CLIST(ICSDMET)(Enter)

        *** JOB MENU ***
        1. ICSD(INORGANIC CRYSTAL STRUCTURE DATABASE)
        2. CRYSTMET(THE NRCC METALS CRYSTALLOGRAPHIC DATA FILE)
        ENTER NUMBER =1(Enter)

        ***********   Crystal Structure Information System
        * CRYSTIN *   (Release 2.51/Mar 90)
        ***********   Written by Hundt & Sievers (University of Bonn)
        DB2201 - Please give BASE (or HELP) command.
        Command?
        Command?
        Current subset ("LAST") contains    279 entries.
        Command?
        Command?
        JOB ITOH4XCB(J Q26950) SUBMITTED
        >>
\end{verbatim}

        We obtain the statistical distribution of point groups for Elements.
Output files and print out are as follows.
\newpage
\subsubsection{Data}
Output file from ICSD (ITOH1XX.@SPG0.DATA)\\

To see this file on ismmain,
\begin{verbatim}
        >>ASP #ITOH1XX.@SPG0.DATA(Enter)

 000100  0000      870    576  2 3 2 3
 000110   100   Octasulfur
 000120   400   S8
 000130   800    10.92600   2.E-3 10.85500   2.E-3 10.79000   3.E-3 90.000  -1.E
 000140   810    95.920   2.E-2 90.000  -1.E 0  8 2.010  -1.E 00.12729E+04
 000150   900   P121/C1
 000160  9900      870    576
 000170  0000      972    669  2 4 2 4
 000180   100   BORON - \-beta
 000190   400   B314.7
 000200   800    10.92510   2.E-4 10.92510   2.E-4 23.81429   8.E-4 90.000  -1.E
 000210   810    90.000  -1.E 0120.000  -1.E 0  1-1.000  -1.E 00.24617E+04
 000220   900   R3-MH
 000230  9900      972    669
 000240  0000     1402   1070  2 2 4 4
 000250   100   cyclo-Heptasulfur
 000260   400   S7
 000270   800    15.10500   5.E-3  5.99800   7.E-3 15.09600   5.E-3 90.000  -1.E
 000280   810    92.150   5.E-2 90.000  -1.E 0  8-1.000  -1.E 00.13667E+04
 000220   900   R3-MH
 000230  9900      972    669

\end{verbatim}

The small letters of ITOH1XX.@SPG0.DATA are transformed into capital letters,
 and Hermann-Mauguin symbols are transformed into Schoenflies symbols.
(ITOH1XX.@SPGRP1.DATA)

To see this file on ismmain,

\begin{verbatim}
        >>ASP #ITOH1XX.@SPGRP1.DATA(Enter)

 000100  0000      870    576  2 3 2 3
 000110   100   OCTASULFUR
 000120   400   S8
 000130   800    10.92600   2.E-3 10.85500   2.E-3 10.79000   3.E-3 90.000  -1.E
 000140   810    95.920   2.E-2 90.000  -1.E 0  8 2.010  -1.E 00.12729E+04
 000150   900   P121/C1     C2H-5      1
 000160  9900      870    576
 000170  0000      972    669  2 4 2 4
 000180   100   BORON - \-BETA
 000190   400   B314.7
 000200   800    10.92510   2.E-4 10.92510   2.E-4 23.81429   8.E-4 90.000  -1.E
 000210   810    90.000  -1.E 0120.000  -1.E 0  1-1.000  -1.E 00.24617E+04
 000220   900   R3-MH       D3D-5      3
 000230  9900      972    669
 000240  0000     1402   1070  2 2 4 4
 000250   100   CYCLO-HEPTASULFUR
 000260   400   S7
 000270   800    15.10500   5.E-3  5.99800   7.E-3 15.09600   5.E-3 90.000  -1.E
 000280   810    92.150   5.E-2 90.000  -1.E 0  8-1.000  -1.E 00.13667E+04

\end{verbatim}

\newpage

	Now we explain about the records.
\begin{verbatim}
<0000> Column
       1-8  ` 0000   ' Record label
       9-14 Collection code
      15    free
      16-20 Relative entry number
      22-80 free
< 100>,< 400>,< 900>
       1    free
       2-3  ` 1' Compound name
            ` 4' Chemical formula
            ` 9' Space group symbol after Hermann-Mauguin
       4    0
       5    Number of continuation records. 0 if the text contains
            less than 73 characters.
       6-8  free
       9-80 Text. After each 72 chacters a continuation record is
            used.
< 800>
       1-8  `  800   ' Record label
       9-17 a(A)
      18-25 Sigma(a)
      26-34 b(A)
      35-42 Sigma(b)
      43-51 c(A)
      52-59 Sigma(c)
      60-66 Alpha(degrees)
      67-74 Sigma(alpha)
      75-80 free
< 810>
       1-8  `  810   ' Record label
       9-15 Beta(degrees)
      16-23 Sigma(beta)
      24-30 Gamma(degrees)
      31-38 Sigma(gamma)
      39-41 Number of formula units
      42-47 Measured density
      48-55 Sigma(density)
      55-66 Unit cell volume (A**3)
      56-80 free

\end{verbatim}
\newpage

	Take the first substance (OCTASULFUR) as an example.
\begin{verbatim}
*First record
        0000            RECORD LABEL
        870             COLLECTION CODE
        576             RELATIVE ENTRY NUMBER

*Second record
        100             RECORD LABEL
        OCTASULFUR      COMPOUND NAME

*Third record
        400             RECORD LABEL
        S8              CHEMICAL FORMULA

*Fourth record
        800             RECORD LABEL
        10.92600        a(ANGSTROM)
        2.E-3           SIGMA(a)=2.0*(10**(-3))
        10.85500        b(ANGSTROM)
        2.E-3           SIGMA(b)=2.0*(10**(-3))
        10.79000        c(ANGSTROM)
        3.E-3           SIGMA(c)=3.0*(10**(-3))
        90.000          alpha(DEGREES)
        -1.E 0          SIGMA(alpha)=0

*Fifth record
        810             RECORD LABEL
        95.920          beta(DEGREES)
        2.E-2           SIGMA(beta)=2.0*(10**(-2))
        90.000          gamma(DEGREES)
        -1.E 0          SIGMA(gamma)=0
        8               Z
        2.010           density(measured)
        -1.E 0          SIGMA(density)=0
        0.12729E+0.4    unit cell volume(ANGSTROM**3)=0.12729*(10**4)

*Sixth record
        900             RECORD LABEL
        P121/C1         HERMANN-MAUGUIN SYMBOL

*Seventh record
        9900            RECORD LABEL
        870             COLLECTION CODE
        576             RELATIVE ENTRY NUMBER
\end{verbatim}

\newpage
\subsubsection{Results}
	We define the species (representatives) of ELEMENTS (ITOH1XX.@SPGRP2.DATA).
We consider that the number just after a serial number is a collection
code of ICSD as a representative substance. The numbers of its rightside are concluded  
 to be close to the representative. By using the ratio of the 
differences ($DIFF2$) of two primitive unit cell volumes, we obtain the 
statistical distribution of point groups. (cf. 2 Algorithm)\\

$DIFF2$ : 0.025\\
To see this file on ismmain,



Consequently, the 279 crystalline substances have 147 representatives (species).
\newpage
Result1. Distribution of space groups and point groups.\\

$DIFF2:0.025$

To see this file on ismmain,


Result2. Distribution of point groups. (PRINT OUT)

To see this file on ismmain,
\begin{verbatim}
	>>ASP(Enter)

From the ASPEN menu, choose 8(@SLIST) (Enter).
Choose 10 * (jobnumber-1) (FT06F001 GO) (Enter).


    C1     CI     C2     CS    C2H     D2    C2V    D2H     C4     S4    C4H
    D4    C4V    D2D    D4H     C3    C3I     D3    C3V    D3D     C6    C3H
   C6H     D6    C6V    D3H    D6H     T      TH     O      TD     OH  TOTAL
 ELC=1(0.025)
     0      1      3      0     17      1      2     16      2      0      2
     1      1      1      4      1      1      6      0     12      0      0
     0      0      4      0     24      4      8      0      3     24    138
 TOTAL=
     0      1      3      0     17      1      2     16      2      0      2
     1      1      1      4      1      1      6      0     12      0      0
     0      0      4      0     24      4      8      0      3     24    138

To print out these data,
use this command.

	>>OUTPUT * DEST(L) (Enter)
or
	>>CHGD *B   (or  CHGD itoh4xcB) (Enter)
or
To print out a specific job,
use this command.

	>>CHGD *B(Jxxxxxx)  (Enter)  x:number

\end{verbatim}
	The reasons why the total number is less than the number of 
defined species, are as follows. Firstly some substances have Hermann-Mauguin 
symbols which cannot be transformed into Shoenflies symbols. Secondly there  
are some crystalline substances which do not have Hermann-Mauguin symbols, in ICSD. 
We omit the substances without Hermann-Mauguin symbols hereafter.

\newpage
\subsection{ANX SYMBOL and Z}
	ICSD has the descriptor `ANX' (formula type). This descriptor is to 
describe a chemical structure roughly by its oxidation numbers of the chemical 
elements. (e.g. ANX=AX describes \(NaCl\), \(CaCl\), and ANX=ABX3 describes \(KNO_3\), \(CaTiO_3\)
etc.) That is, alphabets A-M describe elements whose oxidation states are
positive, ~X-Z,~W,~V,~U,~T,~S describe elements whose oxidation states are negative, 
and N-R describe elements whose oxidation states are 0. Here, hydrogen and its 
isotopes, ~if their oxidation states are positive, ~are disregarded. If one 
element in the structure has some different oxidation states, ~ICSD uses 
different alphabetical letters respectively. Within a group, the chemical 
elements are arranged according to increasing indices.\\
Examples for the ANX symbol in ICSD are as follows.

\begin{tabbing}
xxxxxxxxxxxxxxxxxxxxxxxx \= xxxxxxxxxxx \= \kill
\(Ca_2 Si O_4\) \> ANX=AB2X4 \\
\(Ca S O_4 (H_2 O)_2\) \> ANX=ABX6 \\
\((Na_{0.2} K_{0.8}) (Al Si_3) O_8\) \> ANX=AB4X8 \\
\(Na Tl\) \> ANX=NO \\
\(H_2 O\) \> ANX=X \\
\(Ca H_2\) \> ANX=AX2 \\
\(Na_2 S_2 O_3\) \> ANX=AB2XY3 \\
\(Fe_3 O_4\) \> ANX=AB2X4 \\
\end{tabbing}

	ICSD 1996 has 5545 ANX symbols. We study the statistical distributions
of point groups for 63 ANX symbols, which have more than 100 substances in ICSD.\\

	We studied the above before \cite{fuji98}.
This time we add another descriptor 'Z' to choose the data, after we use the descriptor 'ANX'.
'Z' is defined as the number of formula units per unit cell. 
\subsubsection{Operation}

  1. To count the number of substances per ANX symbol, we use DISPLAY command of ICSD.

\newpage
  2. We edit two files below, by using these data.
\begin{verbatim}
        >>ASP #ITOH4XC.@COMMAND.DATA(Enter)

        ----+----1----+----2----+----3----+----4----+----5----+----6
        ** top of the data **
        BASE 1
        FIND ANX=AX AND Z=1
        PRINT 40 MINR,NAME,FORM,SGR,CELL
        FIND ANX=AX AND Z=2
        PRINT 40 MINR,NAME,FORM,SGR,CELL
        FIND ANX=AX AND Z=3
        PRINT 40 MINR,NAME,FORM,SGR,CELL
        FIND ANX=AX AND Z=4
        PRINT 40 MINR,NAME,FORM,SGR,CELL
        FIND ANX=AX AND Z=6
        PRINT 40 MINR,NAME,FORM,SGR,CELL
         :
        FIND ANX=AX AND Z=32
        PRINT 40 MINR,NAME,FORM,SGR,CELL
        QUIT
\end{verbatim}
\begin{verbatim}
        >>ASP #ITOH4XC.@DATANO.DATA(Enter)

        ----+----1----+----2----+----3----+----4----+----5----+----6
        ** top of the data **
         0.75 0.025
          100ANX=AX,Z=1(0.025)    225ANX=AX,Z=2            40ANX=AX,Z=3
          636ANX=AX,Z=4            38ANX=AX,Z=6             1ANX=AX,Z=7
           53ANX=AX,Z=8             1ANX=AX,Z=9             2ANX=AX,Z=10
           14ANX=AX,Z=12            1ANX=AX,Z=15           11ANX=AX,Z=16
            1ANX=AX,Z=20            1ANX=AX,Z=24            1ANX=AX,Z=27
            2ANX=AX,Z=30            5ANX=AX,Z=32            0
            0                       0
          451ANX=ABX3,Z=1(0.025)  256ANX=ABX3,Z=2          27ANX=ABX3,Z=3
          584ANX=ABX3,Z=4           5ANX=ABX3,Z=5         185ANX=ABX3,Z=6
          154ANX=ABX3,Z=8          19ANX=ABX3,Z=9           5ANX=ABX3,Z=10
           16ANX=ABX3,Z=12          2ANX=ABX3,Z=14          1ANX=ABX3,Z=15
           51ANX=ABX3,Z=16          4ANX=ABX3,Z=18          2ANX=ABX3,Z=20
            4ANX=ABX3,Z=24          1ANX=ABX3,Z=27          5ANX=ABX3,Z=28
            3ANX=ABX3,Z=32          0                       0
              :
              :
\end{verbatim}

We have to put 0(zero) at the 5 or 29 or 53th column.\\

  3. Execute ITOH4XC.MACRO.CLIST(ICSDMET).\\
\begin{verbatim}
        >>EX #ITOH4XC.MACRO.CLIST(ICSDMET)(Enter)

        *** JOB MENU ***
        1. ICSD(INORGANIC CRYSTAL STRUCTURE DATABASE)
        2. CRYSTMET(THE NRCC METALS CRYSTALLOGRAPHIC DATA FILE)
        ENTER NUMBER =1(Enter)
         :
         :
\end{verbatim}
\newpage
\subsubsection{Data} 

       For ANX symbols and Z, 
the numbers of the original species of ICSD are as follows.\\
 
{\small
        \begin{tabular}{*{9}{|r}|}
        \hline
ANX=\hspace*{0.6em} & AX\hspace*{1.9em} & ABX3\hspace*{1.0em} & AX2\hspace*{1.6em} & AX3\hspace*{1.6em} & AB2X4\hspace*{0.4em} & ABX4\hspace*{1.0em} & AB2X6\hspace*{0.4em} & ABX2\hspace*{1.0em}\\
 \hline
	\hline
TOTAL & 1137 & 1776 & 1520 & 604 & 2059 & 1159 & 855 & 692\\
        \hline
Z=         1 & 100 & 451 & 150 & 51 & 110 & 42 & 88 & 72\\
        \hline
2 & 225 & 256 & 366 & 182 & 463 & 148 & 164 & 107\\
        \hline
3 & 40 & 27 & 137 & 20 & 31 & 70 & 32 & 155\\
        \hline
4 & 636 & 584 & 651 & 173 & 771 & 777 & 349 & 273\\
        \hline
5 & 0 & 5 & 1 & 0 & 0 & 0 & 0 & 0\\
        \hline
6 & 38 & 185 & 15 & 75 & 7 & 14 & 3 & 13\\
        \hline
7 & 1 & 0 & 4 & 0 & 0 & 0 & 0 & 0\\
        \hline
8 & 53 & 154 & 71 & 76 & 587 & 52 & 200 & 31\\
        \hline
9 & 1 & 19 & 1 & 0 & 1 & 1 & 6 & 0\\
        \hline
10 & 2 & 5 & 17 & 1 & 1 & 1 & 1 & 0\\
        \hline
11 & 0 & 0 & 0 & 0 & 0 & 0 & 0 & 0\\
        \hline
12 & 14 & 16 & 14 & 9 & 32 & 9 & 1 & 1\\
        \hline
13 & 0 & 0 & 4 & 0 & 0 & 0 & 0 & 0\\
        \hline
14 & 0 & 2 & 0 & 0 & 0 & 0 & 0 & 0\\
        \hline
15 & 1 & 1 & 0 & 0 & 0 & 0 & 0 & 0\\
        \hline
16 & 11 & 51 & 36 & 10 & 18 & 21 & 6 & 28\\
        \hline
17 & 0 & 0 & 0 & 0 & 0 & 0 & 0 & 0\\
        \hline
18 & 0 & 4 & 3 & 1 & 24 & 16 & 0 & 6\\
        \hline
19 & 0 & 0 & 0 & 0 & 0 & 0 & 0 & 0\\
        \hline
20 & 1 & 2 & 1 & 0 & 0 & 3 & 0 & 0\\
        \hline
21 & 0 & 0 & 0 & 0 & 0 & 0 & 0 & 0\\
        \hline
22 & 0 & 0 & 0 & 0 & 0 & 0 & 0 & 0\\
        \hline
23 & 0 & 0 & 0 & 0 & 0 & 0 & 0 & 0\\
        \hline
24 & 1 & 4 & 9 & 1 & 0 & 4 & 4 & 0\\
        \hline
25 & 0 & 0 & 0 & 0 & 0 & 0 & 0 & 0\\
        \hline
26 & 0 & 0 & 0 & 0 & 0 & 0 & 0 & 0\\
        \hline
27 & 1 & 1 & 0 & 0 & 0 & 0 & 0 & 0\\
        \hline
28 & 0 & 5 & 1 & 0 & 0 & 0 & 0 & 0\\
        \hline
29 & 0 & 0 & 0 & 0 & 0 & 0 & 0 & 0\\
        \hline
30 & 2 & 0 & 0 & 0 & 0 & 0 & 0 & 0\\
        \hline
31 & 0 & 0 & 0 & 0 & 0 & 0 & 0 & 0\\
        \hline
32 & 5 & 3 & 22 & 3 & 11 & 0 & 0 & 4\\
        \hline
        \end{tabular}
}
\newpage
{\small
        \begin{tabular}{*{9}{|r}|}
        \hline
ANX=\hspace*{0.6em} & ABC2X6 & NO2\hspace*{1.6em} & A2X3\hspace*{1.0em} & NO\hspace*{1.9em} & AXY\hspace*{1.6em} & AB2X5\hspace*{0.4em} & A3X4\hspace*{1.0em} & A2B2X7\\
        \hline
        \hline
TOTAL & 641 & 461 & 421 & 327 & 304 & 319 & 329 & 344\\
        \hline
Z=         1 & 58 & 85 & 26 & 21 & 12 & 12 & 47 & 4\\
        \hline
2 & 87 & 118 & 54 & 35 & 173 & 43 & 52 & 77\\
        \hline
3 & 50 & 26 & 16 & 16 & 2 & 0 & 4 & 0\\
        \hline
4 & 373 & 172 & 78 & 214 & 90 & 211 & 37 & 75\\
        \hline
5 & 0 & 0 & 0 & 0 & 0 & 0 & 0 & 0\\
        \hline
6 & 11 & 9 & 128 & 10 & 8 & 15 & 3 & 3\\
        \hline
7 & 0 & 0 & 0 & 0 & 0 & 0 & 0 & 0\\
        \hline
8 & 39 & 34 & 14 & 16 & 11 & 27 & 185 & 176\\
        \hline
9 & 0 & 0 & 0 & 0 & 0 & 0 & 0 & 0\\
        \hline
10 & 1 & 0 & 0 & 1 & 0 & 0 & 0 & 2\\
        \hline
11 & 0 & 0 & 0 & 0 & 0 & 0 & 0 & 0\\
        \hline
12 & 4 & 6 & 4 & 6 & 2 & 2 & 1 & 2\\
        \hline
13 & 0 & 0 & 0 & 0 & 0 & 0 & 0 & 0\\
        \hline
14 & 0 & 0 & 0 & 0 & 0 & 0 & 0 & 0\\
        \hline
15 & 0 & 0 & 0 & 0 & 0 & 0 & 0 & 0\\
        \hline
16 & 16 & 7 & 101 & 2 & 4 & 6 & 0 & 5\\
        \hline
17 & 0 & 0 & 0 & 0 & 0 & 0 & 0 & 0\\
        \hline
18 & 1 & 1 & 0 & 0 & 0 & 0 & 0 & 0\\
        \hline
19 & 0 & 0 & 0 & 0 & 0 & 0 & 0 & 0\\
        \hline
20 & 0 & 1 & 0 & 1 & 1 & 0 & 0 & 0\\
        \hline
21 & 0 & 0 & 0 & 0 & 0 & 0 & 0 & 0\\
        \hline
22 & 0 & 0 & 0 & 0 & 0 & 1 & 0 & 0\\
        \hline
23 & 0 & 0 & 0 & 0 & 0 & 0 & 0 & 0\\
        \hline
24 & 0 & 0 & 0 & 1 & 0 & 1 & 0 & 0\\
        \hline
25 & 0 & 0 & 0 & 0 & 0 & 0 & 0 & 0\\
        \hline
26 & 0 & 0 & 0 & 0 & 0 & 0 & 0 & 0\\
        \hline
27 & 0 & 0 & 0 & 0 & 0 & 0 & 0 & 0\\
        \hline
28 & 0 & 0 & 0 & 0 & 0 & 0 & 0 & 0\\
        \hline
29 & 0 & 0 & 0 & 0 & 0 & 0 & 0 & 0\\
        \hline
30 & 0 & 0 & 0 & 0 & 0 & 0 & 0 & 0\\
        \hline
31 & 0 & 0 & 0 & 0 & 0 & 0 & 0 & 0\\
        \hline
32 & 1 & 1 & 0 & 2 & 1 & 0 & 0 & 0\\
        \hline
        \end{tabular}
}
\newpage
\subsubsection{Results}

       For ANX symbols and Z,
the numbers of the defined species are as follows.\\

$DIFF2$ : 0.025\\
{\small
        \begin{tabular}{*{9}{|r}|}
        \hline
ANX=\hspace*{0.6em} & AX\hspace*{1.9em} & ABX3\hspace*{1.0em} & AX2\hspace*{1.6em} & AX3\hspace*{1.6em} & AB2X4\hspace*{0.4em} & ABX4\hspace*{1.0em} & AB2X6\hspace*{0.4em} & ABX2\hspace*{1.0em}\\
        \hline
        \hline
TOTAL & 595 & 902 & 653 & 347 & 763 & 623 & 486 & 456\\
        \hline
Z=         1 & 66 & 193 & 89 & 28 & 42 & 23 & 72 & 53\\
        \hline
2 & 119 & 160 & 131 & 96 & 193 & 87 & 111 & 71\\
        \hline
3 & 23 & 22 & 47 & 18 & 18 & 24 & 26 & 106\\
        \hline
4 & 301 & 298 & 275 & 102 & 288 & 405 & 200 & 162\\
        \hline
5 & 0 & 2 & 1 & 0 & 0 & 0 & 0 & 0\\
        \hline
6 & 23 & 86 & 9 & 47 & 7 & 11 & 3 & 12\\
        \hline
7 & 1 & 0 & 2 & 0 & 0 & 0 & 0 & 0\\
        \hline
8 & 38 & 84 & 42 & 39 & 157 & 33 & 62 & 27\\
        \hline
9 & 1 & 8 & 1 & 0 & 1 & 1 & 5 & 0\\
        \hline
10 & 2 & 3 & 7 & 1 & 1 & 1 & 1 & 0\\
        \hline
11 & 0 & 0 & 0 & 0 & 0 & 0 & 0 & 0\\
        \hline
12 & 4 & 11 & 12 & 6 & 22 & 7 & 1 & 1\\
        \hline
13 & 0 & 0 & 1 & 0 & 0 & 0 & 0 & 0\\
        \hline
14 & 0 & 2 & 0 & 0 & 0 & 0 & 0 & 0\\
        \hline
15 & 1 & 1 & 0 & 0 & 0 & 0 & 0 & 0\\
        \hline
16 & 8 & 20 & 16 & 7 & 12 & 16 & 4 & 15\\
        \hline
17 & 0 & 0 & 0 & 0 & 0 & 0 & 0 & 0\\
        \hline
18 & 0 & 4 & 3 & 1 & 11 & 9 & 0 & 6\\
        \hline
19 & 0 & 0 & 0 & 0 & 0 & 0 & 0 & 0\\
        \hline
20 & 1 & 2 & 1 & 0 & 0 & 3 & 0 & 0\\
        \hline
21 & 0 & 0 & 0 & 0 & 0 & 0 & 0 & 0\\
        \hline
22 & 0 & 0 & 0 & 0 & 0 & 0 & 0 & 0\\
        \hline
23 & 0 & 0 & 0 & 0 & 0 & 0 & 0 & 0\\
        \hline
24 & 1 & 2 & 6 & 1 & 0 & 3 & 1 & 0\\
        \hline
25 & 0 & 0 & 0 & 0 & 0 & 0 & 0 & 0\\
        \hline
26 & 0 & 0 & 0 & 0 & 0 & 0 & 0 & 0\\
        \hline
27 & 1 & 1 & 0 & 0 & 0 & 0 & 0 & 0\\
        \hline
28 & 0 & 1 & 1 & 0 & 0 & 0 & 0 & 0\\
        \hline
29 & 0 & 0 & 0 & 0 & 0 & 0 & 0 & 0\\
        \hline
30 & 2 & 0 & 0 & 0 & 0 & 0 & 0 & 0\\
        \hline
31 & 0 & 0 & 0 & 0 & 0 & 0 & 0 & 0\\
        \hline
32 & 3 & 2 & 9 & 1 & 11 & 0 & 0 & 3\\
        \hline
        \end{tabular}
}
\newpage
{\small
        \begin{tabular}{*{9}{|r}|}
        \hline
ANX=\hspace*{0.6em} & ABC2X6 & NO2\hspace*{1.6em} & A2X3\hspace*{1.0em} & NO\hspace*{1.9em} & AXY\hspace*{1.6em} & AB2X5\hspace*{0.4em} & A3X4\hspace*{1.0em} & A2B2X7\\
        \hline
        \hline
TOTAL & 251 & 340 & 185 & 202 & 200 & 192 & 127 & 187\\
        \hline
Z=         1 & 18 & 73 & 21 & 19 & 11 & 8 & 29 & 4\\
        \hline
2 & 59 & 84 & 26 & 25 & 112 & 24 & 17 & 52\\
        \hline
3 & 15 & 23 & 10 & 15 & 2 & 0 & 4 & 0\\
	\hline
4 & 124 & 113 & 52 & 107 & 53 & 124 & 18 & 52\\
        \hline
5 & 0 & 0 & 0 & 0 & 0 & 0 & 0 & 0\\
        \hline
6 & 5 & 8 & 24 & 8 & 7 & 6 & 3 & 2\\
        \hline
7 & 0 & 0 & 0 & 0 & 0 & 0 & 0 & 0\\
        \hline
8 & 18 & 25 & 12 & 16 & 7 & 21 & 55 & 68\\
        \hline
9 & 0 & 0 & 0 & 0 & 0 & 0 & 0 & 0\\
        \hline
10 & 1 & 0 & 0 & 1 & 0 & 0 & 0 & 2\\
        \hline
11 & 0 & 0 & 0 & 0 & 0 & 0 & 0 & 0\\
        \hline
12 & 2 & 6 & 3 & 5 & 2 & 2 & 1 & 2\\
        \hline
13 & 0 & 0 & 0 & 0 & 0 & 0 & 0 & 0\\
        \hline
14 & 0 & 0 & 0 & 0 & 0 & 0 & 0 & 0\\
        \hline
15 & 0 & 0 & 0 & 0 & 0 & 0 & 0 & 0\\
        \hline
16 & 7 & 5 & 37 & 2 & 4 & 5 & 0 & 5\\
        \hline
17 & 0 & 0 & 0 & 0 & 0 & 0 & 0 & 0\\
        \hline
18 & 1 & 1 & 0 & 0 & 0 & 0 & 0 & 0\\
        \hline
19 & 0 & 0 & 0 & 0 & 0 & 0 & 0 & 0\\
        \hline
20 & 0 & 1 & 0 & 1 & 1 & 0 & 0 & 0\\
        \hline
21 & 0 & 0 & 0 & 0 & 0 & 0 & 0 & 0\\
        \hline
22 & 0 & 0 & 0 & 0 & 0 & 1 & 0 & 0\\
        \hline
23 & 0 & 0 & 0 & 0 & 0 & 0 & 0 & 0\\
        \hline
24 & 0 & 0 & 0 & 1 & 0 & 1 & 0 & 0\\
        \hline
25 & 0 & 0 & 0 & 0 & 0 & 0 & 0 & 0\\
        \hline
26 & 0 & 0 & 0 & 0 & 0 & 0 & 0 & 0\\
        \hline
27 & 0 & 0 & 0 & 0 & 0 & 0 & 0 & 0\\
        \hline
28 & 0 & 0 & 0 & 0 & 0 & 0 & 0 & 0\\
        \hline
29 & 0 & 0 & 0 & 0 & 0 & 0 & 0 & 0\\
        \hline
30 & 0 & 0 & 0 & 0 & 0 & 0 & 0 & 0\\
        \hline
31 & 0 & 0 & 0 & 0 & 0 & 0 & 0 & 0\\
        \hline
32 & 1 & 1 & 0 & 2 & 1 & 0 & 0 & 0\\
        \hline
        \end{tabular}
}
\newpage
\setlength{\oddsidemargin}{-0.1cm}
These are the distributions of point groups and space groups.\\

ANX=AX\\
{\small
        \begin{tabular}{*{18}{|r}|}
        \hline
AX & Z=1 & 2 & \hspace*{0.9em}3 & 4 & \hspace*{0.9em}5 & \hspace*{0.9em}6 & \hspace*{0.9em}7 & \hspace*{0.9em}8 & \hspace*{0.9em}9 & \hspace*{0.7em}10 & \hspace*{0.7em}11 & \hspace*{0.7em}12 & \hspace*{0.7em}13 & \hspace*{0.7em}14 & \hspace*{0.7em}15 & \hspace*{0.7em}16 & \hspace*{0.7em}17\\
        \hline
        \hline
C1 & 0 & 0 & 0 & 0 & 0 & 0 & 0 & 0 & 0 & 0 & 0 & 0 & 0 & 0 & 0 & 0 & 0\\
        \hline
CI & 1 & 1 & 0 & 1 & 0 & 0 & 0 & 2 & 0 & 0 & 0 & 0 & 0 & 0 & 0 & 0 & 0\\
        \hline
C2 & 1 & 0 & 0 & 3 & 0 & 0 & 0 & 0 & 0 & 0 & 0 & 0 & 0 & 0 & 0 & 0 & 0\\
        \hline
CS & 0 & 0 & 0 & 1 & 0 & 0 & 0 & 0 & 0 & 0 & 0 & 0 & 0 & 0 & 0 & 0 & 0\\
        \hline
C2H & 0 & 19 & 0 & 14 & 0 & 0 & 0 & 5 & 0 & 1 & 0 & 0 & 0 & 0 & 0 & 3 & 0\\
        \hline
D2 & 0 & 0 & 0 & 2 & 0 & 0 & 0 & 2 & 0 & 0 & 0 & 0 & 0 & 0 & 0 & 0 & 0\\
        \hline
C2V & 0 & 2 & 0 & 16 & 0 & 0 & 0 & 6 & 0 & 0 & 0 & 0 & 0 & 0 & 0 & 1 & 0\\
        \hline
D2H & 2 & 6 & 0 & 46 & 0 & 0 & 0 & 2 & 0 & 0 & 0 & 2 & 0 & 0 & 0 & 2 & 0\\
        \hline
C4 & 1 & 0 & 0 & 1 & 0 & 0 & 0 & 0 & 0 & 0 & 0 & 0 & 0 & 0 & 0 & 0 & 0\\
        \hline
S4 & 0 & 1 & 0 & 0 & 0 & 0 & 0 & 3 & 0 & 0 & 0 & 0 & 0 & 0 & 0 & 0 & 0\\
        \hline
C4H & 1 & 1 & 0 & 0 & 0 & 0 & 0 & 1 & 0 & 0 & 0 & 0 & 0 & 0 & 0 & 1 & 0\\
        \hline
D4 & 0 & 0 & 0 & 0 & 0 & 0 & 0 & 0 & 0 & 0 & 0 & 0 & 0 & 0 & 0 & 0 & 0\\
        \hline
C4V & 0 & 0 & 0 & 1 & 0 & 0 & 0 & 0 & 0 & 0 & 0 & 0 & 0 & 0 & 0 & 0 & 0\\
        \hline
D2D & 0 & 4 & 0 & 0 & 0 & 0 & 0 & 0 & 0 & 0 & 0 & 0 & 0 & 0 & 0 & 0 & 0\\
        \hline
D4H & 1 & 26 & 0 & 3 & 0 & 0 & 0 & 6 & 0 & 0 & 0 & 0 & 0 & 0 & 0 & 0 & 0\\
        \hline
C3 & 0 & 0 & 0 & 0 & 0 & 0 & 0 & 0 & 0 & 0 & 0 & 0 & 0 & 0 & 0 & 0 & 0\\
        \hline
C3I & 0 & 0 & 0 & 0 & 0 & 0 & 0 & 0 & 0 & 0 & 0 & 0 & 0 & 0 & 0 & 0 & 0\\
        \hline
D3 & 0 & 0 & 6 & 0 & 0 & 0 & 0 & 0 & 0 & 0 & 0 & 0 & 0 & 0 & 0 & 0 & 0\\
        \hline
C3V & 0 & 0 & 3 & 0 & 0 & 3 & 1 & 0 & 1 & 0 & 0 & 0 & 0 & 0 & 1 & 0 & 0\\
        \hline
D3D & 8 & 2 & 3 & 0 & 0 & 12 & 0 & 1 & 0 & 0 & 0 & 1 & 0 & 0 & 0 & 0 & 0\\
        \hline
C6 & 0 & 0 & 0 & 0 & 0 & 2 & 0 & 0 & 0 & 0 & 0 & 0 & 0 & 0 & 0 & 0 & 0\\
        \hline
C3H & 0 & 2 & 0 & 1 & 0 & 0 & 0 & 1 & 0 & 0 & 0 & 0 & 0 & 0 & 0 & 0 & 0\\
        \hline
C6H & 0 & 0 & 0 & 0 & 0 & 0 & 0 & 0 & 0 & 0 & 0 & 0 & 0 & 0 & 0 & 0 & 0\\
        \hline
D6 & 0 & 0 & 0 & 0 & 0 & 0 & 0 & 0 & 0 & 0 & 0 & 0 & 0 & 0 & 0 & 0 & 0\\
        \hline
C6V & 5 & 20 & 0 & 2 & 0 & 1 & 0 & 4 & 0 & 1 & 0 & 0 & 0 & 0 & 0 & 0 & 0\\
        \hline
D3H & 1 & 1 & 8 & 1 & 0 & 3 & 0 & 0 & 0 & 0 & 0 & 1 & 0 & 0 & 0 & 0 & 0\\
        \hline
D6H & 4 & 14 & 1 & 2 & 0 & 2 & 0 & 0 & 0 & 0 & 0 & 0 & 0 & 0 & 0 & 0 & 0\\
        \hline
T & 2 & 0 & 0 & 30 & 0 & 0 & 0 & 0 & 0 & 0 & 0 & 0 & 0 & 0 & 0 & 0 & 0\\
        \hline
TH & 0 & 0 & 0 & 0 & 0 & 0 & 0 & 2 & 0 & 0 & 0 & 0 & 0 & 0 & 0 & 0 & 0\\
        \hline
O & 0 & 0 & 0 & 0 & 0 & 0 & 0 & 0 & 0 & 0 & 0 & 0 & 0 & 0 & 0 & 0 & 0\\
        \hline
TD & 0 & 2 & 0 & 34 & 0 & 0 & 0 & 0 & 0 & 0 & 0 & 0 & 0 & 0 & 0 & 0 & 0\\
        \hline
OH & 39 & 18 & 2 & 139 & 0 & 0 & 0 & 3 & 0 & 0 & 0 & 0 & 0 & 0 & 0 & 1 & 0\\
        \hline
TOTAL & 66 & 119 & 23 & 297 & 0 & 23 & 1 & 38 & 1 & 2 & 0 & 4 & 0 & 0 & 1 & 8 & 0\\
        \hline
{\footnotesize Original } & 100 & 225 & 40 & 636 & 0 & 38 & 1 & 53 & 1 & 2 & 0 & 14 & 0 & 0 & 1 & 11 & 0\\
        \hline
        \end{tabular}
}
\newpage
\setlength{\oddsidemargin}{-0.3cm}
{\small
        \begin{tabular}{*{17}{|r}|}
        \hline
AX & Z=18 & \hspace*{0.7em}19 & \hspace*{0.7em}20 & \hspace*{0.7em}21 & \hspace*{0.7em}22 & \hspace*{0.7em}23 & \hspace*{0.7em}24 & \hspace*{0.7em}25 & \hspace*{0.7em}26 & \hspace*{0.7em}27 & \hspace*{0.7em}28 & \hspace*{0.7em}29 & \hspace*{0.7em}30 & \hspace*{0.7em}31 & \hspace*{0.7em}32 & {\scriptsize TOTAL }\\
        \hline
        \hline
C1 & 0 & 0 & 0 & 0 & 0 & 0 & 0 & 0 & 0 & 0 & 0 & 0 & 0 & 0 & 0 & 0\\
        \hline
CI & 0 & 0 & 0 & 0 & 0 & 0 & 0 & 0 & 0 & 0 & 0 & 0 & 0 & 0 & 0 & 5\\
        \hline
C2 & 0 & 0 & 0 & 0 & 0 & 0 & 0 & 0 & 0 & 0 & 0 & 0 & 0 & 0 & 0 & 4\\
        \hline
CS & 0 & 0 & 0 & 0 & 0 & 0 & 0 & 0 & 0 & 0 & 0 & 0 & 0 & 0 & 0 & 1\\
        \hline
C2H & 0 & 0 & 0 & 0 & 0 & 0 & 0 & 0 & 0 & 0 & 0 & 0 & 0 & 0 & 0 & 42\\
        \hline
D2 & 0 & 0 & 0 & 0 & 0 & 0 & 0 & 0 & 0 & 0 & 0 & 0 & 0 & 0 & 0 & 4\\
        \hline
C2V & 0 & 0 & 1 & 0 & 0 & 0 & 0 & 0 & 0 & 0 & 0 & 0 & 0 & 0 & 0 & 26\\
        \hline
D2H & 0 & 0 & 0 & 0 & 0 & 0 & 0 & 0 & 0 & 0 & 0 & 0 & 0 & 0 & 0 & 60\\
        \hline
C4 & 0 & 0 & 0 & 0 & 0 & 0 & 0 & 0 & 0 & 0 & 0 & 0 & 0 & 0 & 0 & 2\\
        \hline
S4 & 0 & 0 & 0 & 0 & 0 & 0 & 0 & 0 & 0 & 0 & 0 & 0 & 0 & 0 & 0 & 4\\
        \hline
C4H & 0 & 0 & 0 & 0 & 0 & 0 & 0 & 0 & 0 & 0 & 0 & 0 & 0 & 0 & 0 & 4\\
        \hline
D4 & 0 & 0 & 0 & 0 & 0 & 0 & 0 & 0 & 0 & 0 & 0 & 0 & 0 & 0 & 0 & 0\\
        \hline
C4V & 0 & 0 & 0 & 0 & 0 & 0 & 0 & 0 & 0 & 0 & 0 & 0 & 0 & 0 & 0 & 1\\
        \hline
D2D & 0 & 0 & 0 & 0 & 0 & 0 & 0 & 0 & 0 & 0 & 0 & 0 & 0 & 0 & 0 & 4\\
        \hline
D4H & 0 & 0 & 0 & 0 & 0 & 0 & 0 & 0 & 0 & 0 & 0 & 0 & 0 & 0 & 0 & 36\\
        \hline
C3 & 0 & 0 & 0 & 0 & 0 & 0 & 0 & 0 & 0 & 0 & 0 & 0 & 0 & 0 & 0 & 0\\
        \hline
C3I & 0 & 0 & 0 & 0 & 0 & 0 & 0 & 0 & 0 & 0 & 0 & 0 & 0 & 0 & 0 & 0\\
        \hline
D3 & 0 & 0 & 0 & 0 & 0 & 0 & 0 & 0 & 0 & 0 & 0 & 0 & 0 & 0 & 0 & 6\\
        \hline
C3V & 0 & 0 & 0 & 0 & 0 & 0 & 1 & 0 & 0 & 1 & 0 & 0 & 0 & 0 & 0 & 11\\
        \hline
D3D & 0 & 0 & 0 & 0 & 0 & 0 & 0 & 0 & 0 & 0 & 0 & 0 & 0 & 0 & 0 & 27\\
        \hline
C6 & 0 & 0 & 0 & 0 & 0 & 0 & 0 & 0 & 0 & 0 & 0 & 0 & 0 & 0 & 0 & 2\\
        \hline
C3H & 0 & 0 & 0 & 0 & 0 & 0 & 0 & 0 & 0 & 0 & 0 & 0 & 0 & 0 & 0 & 4\\
        \hline
C6H & 0 & 0 & 0 & 0 & 0 & 0 & 0 & 0 & 0 & 0 & 0 & 0 & 0 & 0 & 0 & 0\\
        \hline
D6 & 0 & 0 & 0 & 0 & 0 & 0 & 0 & 0 & 0 & 0 & 0 & 0 & 0 & 0 & 0 & 0\\
        \hline
C6V & 0 & 0 & 0 & 0 & 0 & 0 & 0 & 0 & 0 & 0 & 0 & 0 & 0 & 0 & 0 & 33\\
        \hline
D3H & 0 & 0 & 0 & 0 & 0 & 0 & 0 & 0 & 0 & 0 & 0 & 0 & 0 & 0 & 0 & 15\\
        \hline
D6H & 0 & 0 & 0 & 0 & 0 & 0 & 0 & 0 & 0 & 0 & 0 & 0 & 0 & 0 & 0 & 23\\
        \hline
T & 0 & 0 & 0 & 0 & 0 & 0 & 0 & 0 & 0 & 0 & 0 & 0 & 0 & 0 & 1 & 33\\
        \hline
TH & 0 & 0 & 0 & 0 & 0 & 0 & 0 & 0 & 0 & 0 & 0 & 0 & 0 & 0 & 0 & 2\\
        \hline
O & 0 & 0 & 0 & 0 & 0 & 0 & 0 & 0 & 0 & 0 & 0 & 0 & 0 & 0 & 0 & 0\\
        \hline
TD & 0 & 0 & 0 & 0 & 0 & 0 & 0 & 0 & 0 & 0 & 0 & 0 & 0 & 0 & 1 & 37\\
        \hline
OH & 0 & 0 & 0 & 0 & 0 & 0 & 0 & 0 & 0 & 0 & 0 & 0 & 2 & 0 & 1 & 205\\
        \hline
TOTAL & 0 & 0 & 1 & 0 & 0 & 0 & 1 & 0 & 0 & 1 & 0 & 0 & 2 & 0 & 3 & 591\\
        \hline
{\footnotesize Original } & 0 & 0 & 1 & 0 & 0 & 0 & 1 & 0 & 0 & 1 & 0 & 0 & 2 & 0 & 5 & 1132\\
        \hline
        \end{tabular}
}
\newpage
\setlength{\oddsidemargin}{2.54cm}

\newpage

\newpage
 
\newpage

\newpage

\newpage

\newpage

\newpage

\newpage

\newpage

\newpage

\newpage

\newpage

\newpage

\newpage

\newpage

\newpage
                            
\setlength{\oddsidemargin}{-0.1cm}
ANX=ABX3\\
{\small
        \begin{tabular}{*{18}{|r}|}
        \hline
ABX3 & Z=1 & 2 & \hspace*{0.9em}3 & 4 & \hspace*{0.9em}5 & \hspace*{0.9em}6 & \hspace*{0.9em}7 & \hspace*{0.9em}8 & \hspace*{0.9em}9 & \hspace*{0.7em}10 & \hspace*{0.7em}11 & \hspace*{0.7em}12 & \hspace*{0.7em}13 & \hspace*{0.7em}14 & \hspace*{0.7em}15 & \hspace*{0.7em}16 & \hspace*{0.7em}17\\
 \hline
        \hline
C1 & 1 & 0 & 0 & 3 & 0 & 1 & 0 & 0 & 0 & 0 & 0 & 0 & 0 & 0 & 0 & 0 & 0\\
        \hline
CI & 1 & 6 & 0 & 4 & 0 & 3 & 0 & 2 & 0 & 1 & 0 & 2 & 0 & 1 & 0 & 0 & 0\\
        \hline
C2 & 0 & 0 & 0 & 2 & 0 & 0 & 0 & 3 & 0 & 0 & 0 & 0 & 0 & 0 & 0 & 0 & 0\\
        \hline
CS & 0 & 4 & 0 & 3 & 0 & 0 & 0 & 3 & 0 & 0 & 0 & 0 & 0 & 0 & 0 & 0 & 0\\
        \hline
C2H & 3 & 19 & 0 & 44 & 0 & 0 & 0 & 27 & 0 & 0 & 0 & 6 & 0 & 0 & 0 & 1 & 0\\
        \hline
D2 & 0 & 0 & 0 & 7 & 0 & 0 & 0 & 4 & 0 & 0 & 0 & 0 & 0 & 0 & 0 & 1 & 0\\
        \hline
C2V & 0 & 5 & 0 & 19 & 0 & 0 & 0 & 7 & 0 & 1 & 0 & 0 & 0 & 0 & 0 & 0 & 0\\
        \hline
D2H & 3 & 16 & 0 & 187 & 0 & 0 & 0 & 21 & 0 & 0 & 0 & 1 & 0 & 0 & 0 & 14 & 0\\
        \hline
C4 & 0 & 0 & 0 & 1 & 0 & 0 & 0 & 0 & 0 & 0 & 0 & 0 & 0 & 0 & 0 & 0 & 0\\
        \hline
S4 & 1 & 0 & 0 & 1 & 0 & 0 & 0 & 0 & 0 & 0 & 0 & 0 & 0 & 0 & 0 & 0 & 0\\
        \hline
C4H & 1 & 0 & 0 & 1 & 0 & 0 & 0 & 3 & 0 & 0 & 0 & 0 & 0 & 0 & 0 & 1 & 0\\
        \hline
D4 & 0 & 0 & 0 & 0 & 0 & 0 & 0 & 0 & 0 & 0 & 0 & 0 & 0 & 0 & 0 & 0 & 0\\
        \hline
C4V & 11 & 0 & 0 & 0 & 0 & 0 & 0 & 1 & 0 & 0 & 0 & 0 & 0 & 0 & 0 & 0 & 0\\
        \hline
D2D & 2 & 1 & 0 & 1 & 0 & 0 & 0 & 0 & 0 & 0 & 0 & 0 & 0 & 0 & 0 & 0 & 0\\
        \hline
D4H & 8 & 8 & 0 & 9 & 0 & 0 & 0 & 3 & 0 & 0 & 0 & 0 & 0 & 0 & 0 & 0 & 0\\
        \hline
C3 & 4 & 0 & 3 & 0 & 0 & 0 & 0 & 0 & 3 & 0 & 0 & 0 & 0 & 0 & 0 & 0 & 0\\
        \hline
C3I & 1 & 11 & 3 & 0 & 0 & 16 & 0 & 0 & 0 & 0 & 0 & 0 & 0 & 0 & 0 & 0 & 0\\
        \hline
D3 & 0 & 0 & 2 & 0 & 0 & 0 & 0 & 0 & 0 & 0 & 0 & 0 & 0 & 0 & 0 & 0 & 0\\
        \hline
C3V & 18 & 3 & 8 & 0 & 1 & 11 & 0 & 0 & 2 & 0 & 0 & 0 & 0 & 0 & 0 & 0 & 0\\
        \hline
D3D & 5 & 19 & 6 & 1 & 1 & 25 & 0 & 0 & 3 & 0 & 0 & 1 & 0 & 0 & 1 & 0 & 0\\
        \hline
C6 & 0 & 3 & 0 & 0 & 0 & 3 & 0 & 0 & 0 & 0 & 0 & 0 & 0 & 0 & 0 & 0 & 0\\
        \hline
C3H & 0 & 0 & 0 & 0 & 0 & 0 & 0 & 0 & 0 & 0 & 0 & 0 & 0 & 0 & 0 & 0 & 0\\
        \hline
C6H & 2 & 1 & 0 & 0 & 0 & 0 & 0 & 0 & 0 & 0 & 0 & 0 & 0 & 0 & 0 & 0 & 0\\
        \hline
D6 & 0 & 2 & 0 & 1 & 0 & 2 & 0 & 0 & 0 & 0 & 0 & 0 & 0 & 0 & 0 & 0 & 0\\
        \hline
C6V & 0 & 8 & 0 & 0 & 0 & 10 & 0 & 0 & 0 & 0 & 0 & 0 & 0 & 0 & 0 & 0 & 0\\
        \hline
D3H & 0 & 2 & 0 & 0 & 0 & 3 & 0 & 0 & 0 & 0 & 0 & 0 & 0 & 0 & 0 & 0 & 0\\
        \hline
D6H & 0 & 50 & 0 & 7 & 0 & 12 & 0 & 1 & 0 & 1 & 0 & 0 & 0 & 1 & 0 & 0 & 0\\
        \hline
T & 5 & 0 & 0 & 2 & 0 & 0 & 0 & 0 & 0 & 0 & 0 & 0 & 0 & 0 & 0 & 0 & 0\\
        \hline
TH & 1 & 0 & 0 & 0 & 0 & 0 & 0 & 2 & 0 & 0 & 0 & 1 & 0 & 0 & 0 & 0 & 0\\
        \hline
O & 1 & 0 & 0 & 0 & 0 & 0 & 0 & 0 & 0 & 0 & 0 & 0 & 0 & 0 & 0 & 0 & 0\\
        \hline
TD & 2 & 0 & 0 & 0 & 0 & 0 & 0 & 1 & 0 & 0 & 0 & 0 & 0 & 0 & 0 & 0 & 0\\
        \hline
OH & 123 & 2 & 0 & 1 & 0 & 0 & 0 & 6 & 0 & 0 & 0 & 0 & 0 & 0 & 0 & 3 & 0\\
        \hline
TOTAL & 193 & 160 & 22 & 294 & 2 & 86 & 0 & 84 & 8 & 3 & 0 & 11 & 0 & 2 & 1 & 20
& 0\\
        \hline
{\footnotesize Original }& 451 & 256 & 27 & 584 & 5 & 185 & 0 & 154 & 19 & 5 & 0 & 16 & 0 & 2 & 1 & 51 & 0\\
        \hline
        \end{tabular}
}
\newpage
\setlength{\oddsidemargin}{-0.3cm}
{\small
        \begin{tabular}{*{17}{|r}|}
        \hline
ABX3 & Z=18 & \hspace*{0.7em}19 & \hspace*{0.7em}20 & \hspace*{0.7em}21 & \hspace*{0.7em}22 & \hspace*{0.7em}23 & \hspace*{0.7em}24 & \hspace*{0.7em}25 & \hspace*{0.7em}26 & \hspace*{0.7em}27 & \hspace*{0.7em}28 & \hspace*{0.7em}29 & \hspace*{0.7em}30 & \hspace*{0.7em}31 & \hspace*{0.7em}32 & {\scriptsize TOTAL }\\
        \hline
        \hline
C1 & 0 & 0 & 0 & 0 & 0 & 0 & 0 & 0 & 0 & 0 & 0 & 0 & 0 & 0 & 0 & 5\\
        \hline
CI & 1 & 0 & 1 & 0 & 0 & 0 & 0 & 0 & 0 & 0 & 1 & 0 & 0 & 0 & 0 & 23\\
        \hline
C2 & 0 & 0 & 0 & 0 & 0 & 0 & 0 & 0 & 0 & 0 & 0 & 0 & 0 & 0 & 0 & 5\\
        \hline
CS & 0 & 0 & 1 & 0 & 0 & 0 & 0 & 0 & 0 & 0 & 0 & 0 & 0 & 0 & 0 & 11\\
        \hline
C2H & 0 & 0 & 0 & 0 & 0 & 0 & 1 & 0 & 0 & 0 & 0 & 0 & 0 & 0 & 0 & 101\\
        \hline
D2 & 0 & 0 & 0 & 0 & 0 & 0 & 0 & 0 & 0 & 0 & 0 & 0 & 0 & 0 & 0 & 12\\
        \hline
C2V & 0 & 0 & 0 & 0 & 0 & 0 & 0 & 0 & 0 & 0 & 0 & 0 & 0 & 0 & 0 & 32\\
        \hline
D2H & 0 & 0 & 0 & 0 & 0 & 0 & 0 & 0 & 0 & 0 & 0 & 0 & 0 & 0 & 0 & 242\\
        \hline
C4 & 0 & 0 & 0 & 0 & 0 & 0 & 0 & 0 & 0 & 0 & 0 & 0 & 0 & 0 & 0 & 1\\
        \hline
S4 & 0 & 0 & 0 & 0 & 0 & 0 & 0 & 0 & 0 & 0 & 0 & 0 & 0 & 0 & 0 & 2\\
        \hline
C4H & 0 & 0 & 0 & 0 & 0 & 0 & 0 & 0 & 0 & 0 & 0 & 0 & 0 & 0 & 2 & 8\\
        \hline
D4 & 0 & 0 & 0 & 0 & 0 & 0 & 0 & 0 & 0 & 0 & 0 & 0 & 0 & 0 & 0 & 0\\
        \hline
C4V & 0 & 0 & 0 & 0 & 0 & 0 & 0 & 0 & 0 & 0 & 0 & 0 & 0 & 0 & 0 & 12\\
        \hline
D2D & 0 & 0 & 0 & 0 & 0 & 0 & 0 & 0 & 0 & 0 & 0 & 0 & 0 & 0 & 0 & 4\\
        \hline
D4H & 0 & 0 & 0 & 0 & 0 & 0 & 0 & 0 & 0 & 0 & 0 & 0 & 0 & 0 & 0 & 28\\
        \hline
C3 & 0 & 0 & 0 & 0 & 0 & 0 & 0 & 0 & 0 & 0 & 0 & 0 & 0 & 0 & 0 & 10\\
        \hline
C3I & 2 & 0 & 0 & 0 & 0 & 0 & 0 & 0 & 0 & 0 & 0 & 0 & 0 & 0 & 0 & 33\\
        \hline
D3 & 0 & 0 & 0 & 0 & 0 & 0 & 0 & 0 & 0 & 0 & 0 & 0 & 0 & 0 & 0 & 2\\
        \hline
C3V & 0 & 0 & 0 & 0 & 0 & 0 & 1 & 0 & 0 & 0 & 0 & 0 & 0 & 0 & 0 & 44\\
        \hline
D3D & 0 & 0 & 0 & 0 & 0 & 0 & 0 & 0 & 0 & 1 & 0 & 0 & 0 & 0 & 0 & 63\\
        \hline
C6 & 0 & 0 & 0 & 0 & 0 & 0 & 0 & 0 & 0 & 0 & 0 & 0 & 0 & 0 & 0 & 6\\
        \hline
C3H & 0 & 0 & 0 & 0 & 0 & 0 & 0 & 0 & 0 & 0 & 0 & 0 & 0 & 0 & 0 & 0\\
        \hline
C6H & 0 & 0 & 0 & 0 & 0 & 0 & 0 & 0 & 0 & 0 & 0 & 0 & 0 & 0 & 0 & 3\\
        \hline
D6 & 0 & 0 & 0 & 0 & 0 & 0 & 0 & 0 & 0 & 0 & 0 & 0 & 0 & 0 & 0 & 5\\
        \hline
C6V & 0 & 0 & 0 & 0 & 0 & 0 & 0 & 0 & 0 & 0 & 0 & 0 & 0 & 0 & 0 & 18\\
        \hline
D3H & 1 & 0 & 0 & 0 & 0 & 0 & 0 & 0 & 0 & 0 & 0 & 0 & 0 & 0 & 0 & 6\\
        \hline
D6H & 0 & 0 & 0 & 0 & 0 & 0 & 0 & 0 & 0 & 0 & 0 & 0 & 0 & 0 & 0 & 72\\
        \hline
T & 0 & 0 & 0 & 0 & 0 & 0 & 0 & 0 & 0 & 0 & 0 & 0 & 0 & 0 & 0 & 7\\
        \hline
TH & 0 & 0 & 0 & 0 & 0 & 0 & 0 & 0 & 0 & 0 & 0 & 0 & 0 & 0 & 0 & 4\\
        \hline
O & 0 & 0 & 0 & 0 & 0 & 0 & 0 & 0 & 0 & 0 & 0 & 0 & 0 & 0 & 0 & 1\\
        \hline
TD & 0 & 0 & 0 & 0 & 0 & 0 & 0 & 0 & 0 & 0 & 0 & 0 & 0 & 0 & 0 & 3\\
        \hline
OH & 0 & 0 & 0 & 0 & 0 & 0 & 0 & 0 & 0 & 0 & 0 & 0 & 0 & 0 & 0 & 135\\
        \hline
TOTAL & 4 & 0 & 2 & 0 & 0 & 0 & 2 & 0 & 0 & 1 & 1 & 0 & 0 & 0 & 2 & 898\\
        \hline
{\footnotesize Original } & 4 & 0 & 2 & 0 & 0 & 0 & 4 & 0 & 0 & 1 & 5 & 0 & 0 & 0 & 3 & 1775\\
        \hline
        \end{tabular}
}
\newpage
\setlength{\oddsidemargin}{2.54cm}

\newpage

\newpage

\newpage

\newpage

\newpage

\newpage

\newpage

\newpage

\newpage

\newpage

\newpage

\newpage

\newpage

\newpage

\newpage

\newpage

\newpage

\newpage

\newpage
\setlength{\oddsidemargin}{-0.1cm}
ANX=AX2\\
{\small
        \begin{tabular}{*{18}{|r}|}
        \hline
AX2 & Z=1 & 2 & \hspace*{0.9em}3 & 4 & \hspace*{0.9em}5 & \hspace*{0.9em}6 & \hspace*{0.9em}7 & \hspace*{0.9em}8 & \hspace*{0.9em}9 & \hspace*{0.7em}10 & \hspace*{0.7em}11 & \hspace*{0.7em}12 & \hspace*{0.7em}13 & \hspace*{0.7em}14 & \hspace*{0.7em}15 & \hspace*{0.7em}16 & \hspace*{0.7em}17\\
	\hline
        \hline
C1 & 0 & 0 & 1 & 6 & 0 & 0 & 0 & 0 & 0 & 0 & 0 & 0 & 0 & 0 & 0 & 0 & 0\\
        \hline
CI & 1 & 1 & 0 & 3 & 0 & 0 & 0 & 1 & 0 & 0 & 0 & 0 & 0 & 0 & 0 & 0 & 0\\
        \hline
C2 & 1 & 4 & 0 & 3 & 0 & 0 & 0 & 2 & 0 & 0 & 0 & 1 & 0 & 0 & 0 & 0 & 0\\
        \hline
CS & 0 & 0 & 0 & 3 & 0 & 0 & 0 & 2 & 0 & 0 & 0 & 1 & 0 & 0 & 0 & 0 & 0\\
        \hline
C2H & 2 & 15 & 0 & 27 & 0 & 1 & 0 & 8 & 0 & 0 & 0 & 3 & 0 & 0 & 0 & 6 & 0\\
        \hline
D2 & 1 & 1 & 0 & 7 & 0 & 0 & 0 & 4 & 0 & 0 & 0 & 0 & 0 & 0 & 0 & 1 & 0\\
        \hline
C2V & 0 & 4 & 0 & 24 & 0 & 0 & 0 & 3 & 0 & 0 & 0 & 1 & 0 & 0 & 0 & 0 & 0\\
        \hline
D2H & 0 & 23 & 0 & 58 & 0 & 0 & 0 & 9 & 0 & 0 & 0 & 3 & 0 & 0 & 0 & 1 & 0\\
        \hline
C4 & 0 & 0 & 0 & 0 & 0 & 0 & 0 & 0 & 0 & 0 & 0 & 0 & 0 & 0 & 0 & 1 & 0\\
        \hline
S4 & 0 & 0 & 0 & 1 & 0 & 0 & 0 & 1 & 0 & 0 & 0 & 0 & 0 & 0 & 0 & 0 & 0\\
        \hline
C4H & 0 & 3 & 0 & 0 & 0 & 0 & 0 & 2 & 0 & 4 & 0 & 0 & 0 & 0 & 0 & 0 & 0\\
        \hline
D4 & 0 & 1 & 0 & 8 & 0 & 0 & 0 & 2 & 0 & 0 & 0 & 1 & 0 & 0 & 0 & 0 & 0\\
        \hline
C4V & 0 & 0 & 0 & 0 & 0 & 0 & 0 & 0 & 0 & 0 & 0 & 0 & 0 & 0 & 0 & 0 & 0\\
        \hline
D2D & 0 & 1 & 0 & 7 & 0 & 0 & 0 & 1 & 0 & 0 & 0 & 0 & 0 & 0 & 0 & 0 & 0\\
        \hline
D4H & 16 & 56 & 0 & 11 & 0 & 0 & 0 & 1 & 0 & 1 & 0 & 0 & 0 & 0 & 0 & 2 & 0\\
        \hline
C3 & 0 & 0 & 0 & 0 & 0 & 0 & 0 & 0 & 0 & 0 & 0 & 0 & 0 & 0 & 0 & 0 & 0\\
        \hline
C3I & 1 & 0 & 1 & 0 & 0 & 0 & 0 & 0 & 0 & 0 & 0 & 0 & 0 & 0 & 0 & 0 & 0\\
        \hline
D3 & 4 & 0 & 16 & 0 & 0 & 0 & 0 & 0 & 0 & 0 & 0 & 0 & 0 & 0 & 0 & 0 & 0\\
        \hline
C3V & 0 & 0 & 10 & 1 & 1 & 2 & 2 & 1 & 1 & 2 & 0 & 0 & 1 & 0 & 0 & 1 & 0\\
        \hline
D3D & 40 & 1 & 14 & 0 & 0 & 4 & 0 & 0 & 0 & 0 & 0 & 0 & 0 & 0 & 0 & 0 & 0\\
        \hline
C6 & 0 & 0 & 0 & 0 & 0 & 0 & 0 & 0 & 0 & 0 & 0 & 0 & 0 & 0 & 0 & 0 & 0\\
        \hline
C3H & 1 & 0 & 0 & 0 & 0 & 0 & 0 & 0 & 0 & 0 & 0 & 0 & 0 & 0 & 0 & 0 & 0\\
        \hline
C6H & 1 & 0 & 0 & 0 & 0 & 0 & 0 & 0 & 0 & 0 & 0 & 0 & 0 & 0 & 0 & 0 & 0\\
        \hline
D6 & 1 & 0 & 3 & 1 & 0 & 0 & 0 & 0 & 0 & 0 & 0 & 0 & 0 & 0 & 0 & 0 & 0\\
        \hline
C6V & 0 & 3 & 0 & 2 & 0 & 0 & 0 & 0 & 0 & 0 & 0 & 0 & 0 & 0 & 0 & 0 & 0\\
        \hline
D3H & 0 & 0 & 2 & 1 & 0 & 0 & 0 & 0 & 0 & 0 & 0 & 0 & 0 & 0 & 0 & 0 & 0\\
        \hline
D6H & 1 & 8 & 0 & 3 & 0 & 1 & 0 & 0 & 0 & 0 & 0 & 1 & 0 & 0 & 0 & 0 & 0\\
        \hline
T & 0 & 0 & 0 & 1 & 0 & 0 & 0 & 1 & 0 & 0 & 0 & 0 & 0 & 0 & 0 & 0 & 0\\
        \hline
TH & 1 & 1 & 0 & 28 & 0 & 1 & 0 & 0 & 0 & 0 & 0 & 0 & 0 & 0 & 0 & 0 & 0\\
        \hline
O & 0 & 0 & 0 & 0 & 0 & 0 & 0 & 1 & 0 & 0 & 0 & 0 & 0 & 0 & 0 & 0 & 0\\
        \hline
TD & 1 & 0 & 0 & 0 & 0 & 0 & 0 & 0 & 0 & 0 & 0 & 0 & 0 & 0 & 0 & 1 & 0\\
        \hline
OH & 17 & 8 & 0 & 79 & 0 & 0 & 0 & 1 & 0 & 0 & 0 & 1 & 0 & 0 & 0 & 3 & 0\\
        \hline
TOTAL & 89 & 130 & 47 & 274 & 1 & 9 & 2 & 40 & 1 & 7 & 0 & 12 & 1 & 0 & 0 & 16 & 0\\
        \hline
{\footnotesize Original } & 150 & 366 & 137 & 651 & 1 & 15 & 4 & 71 & 1 & 17 & 0 & 14 & 4 & 0 & 0 & 36 & 0\\
        \hline
        \end{tabular}
}
\newpage
\setlength{\oddsidemargin}{-0.3cm}
{\small
        \begin{tabular}{*{17}{|r}|}
        \hline
AX2 & Z=18 & \hspace*{0.7em}19 & \hspace*{0.7em}20 & \hspace*{0.7em}21 & \hspace*{0.7em}22 & \hspace*{0.7em}23 & \hspace*{0.7em}24 & \hspace*{0.7em}25 & \hspace*{0.7em}26 & \hspace*{0.7em}27 & \hspace*{0.7em}28 & \hspace*{0.7em}29 & \hspace*{0.7em}30 & \hspace*{0.7em}31 & \hspace*{0.7em}32 & {\scriptsize TOTAL }\\
        \hline
        \hline
C1 & 0 & 0 & 0 & 0 & 0 & 0 & 0 & 0 & 0 & 0 & 0 & 0 & 0 & 0 & 0 & 7\\
        \hline
CI & 0 & 0 & 0 & 0 & 0 & 0 & 0 & 0 & 0 & 0 & 0 & 0 & 0 & 0 & 0 & 6\\
        \hline
C2 & 0 & 0 & 0 & 0 & 0 & 0 & 0 & 0 & 0 & 0 & 0 & 0 & 0 & 0 & 0 & 11\\
        \hline
CS & 0 & 0 & 0 & 0 & 0 & 0 & 0 & 0 & 0 & 0 & 0 & 0 & 0 & 0 & 0 & 6\\
        \hline
C2H & 0 & 0 & 0 & 0 & 0 & 0 & 0 & 0 & 0 & 0 & 1 & 0 & 0 & 0 & 0 & 63\\
        \hline
D2 & 0 & 0 & 0 & 0 & 0 & 0 & 1 & 0 & 0 & 0 & 0 & 0 & 0 & 0 & 1 & 16\\
        \hline
C2V & 0 & 0 & 0 & 0 & 0 & 0 & 2 & 0 & 0 & 0 & 0 & 0 & 0 & 0 & 0 & 34\\
        \hline
D2H & 0 & 0 & 1 & 0 & 0 & 0 & 1 & 0 & 0 & 0 & 0 & 0 & 0 & 0 & 0 & 96\\
        \hline
C4 & 0 & 0 & 0 & 0 & 0 & 0 & 0 & 0 & 0 & 0 & 0 & 0 & 0 & 0 & 0 & 1\\
        \hline
S4 & 0 & 0 & 0 & 0 & 0 & 0 & 0 & 0 & 0 & 0 & 0 & 0 & 0 & 0 & 0 & 2\\
        \hline
C4H & 0 & 0 & 0 & 0 & 0 & 0 & 0 & 0 & 0 & 0 & 0 & 0 & 0 & 0 & 2 & 11\\
        \hline
D4 & 0 & 0 & 0 & 0 & 0 & 0 & 0 & 0 & 0 & 0 & 0 & 0 & 0 & 0 & 0 & 12\\
        \hline
C4V & 0 & 0 & 0 & 0 & 0 & 0 & 0 & 0 & 0 & 0 & 0 & 0 & 0 & 0 & 0 & 0\\
        \hline
D2D & 0 & 0 & 0 & 0 & 0 & 0 & 0 & 0 & 0 & 0 & 0 & 0 & 0 & 0 & 0 & 9\\
        \hline
D4H & 0 & 0 & 0 & 0 & 0 & 0 & 0 & 0 & 0 & 0 & 0 & 0 & 0 & 0 & 4 & 91\\
        \hline
C3 & 0 & 0 & 0 & 0 & 0 & 0 & 0 & 0 & 0 & 0 & 0 & 0 & 0 & 0 & 0 & 0\\
        \hline
C3I & 0 & 0 & 0 & 0 & 0 & 0 & 0 & 0 & 0 & 0 & 0 & 0 & 0 & 0 & 0 & 2\\
        \hline
D3 & 0 & 0 & 0 & 0 & 0 & 0 & 0 & 0 & 0 & 0 & 0 & 0 & 0 & 0 & 0 & 20\\
        \hline
C3V & 1 & 0 & 0 & 0 & 0 & 0 & 0 & 0 & 0 & 0 & 0 & 0 & 0 & 0 & 0 & 23\\
        \hline
D3D & 0 & 0 & 0 & 0 & 0 & 0 & 0 & 0 & 0 & 0 & 0 & 0 & 0 & 0 & 0 & 59\\
        \hline
C6 & 0 & 0 & 0 & 0 & 0 & 0 & 0 & 0 & 0 & 0 & 0 & 0 & 0 & 0 & 0 & 0\\
        \hline
C3H & 0 & 0 & 0 & 0 & 0 & 0 & 0 & 0 & 0 & 0 & 0 & 0 & 0 & 0 & 0 & 1\\
        \hline
C6H & 0 & 0 & 0 & 0 & 0 & 0 & 0 & 0 & 0 & 0 & 0 & 0 & 0 & 0 & 0 & 1\\
        \hline
D6 & 0 & 0 & 0 & 0 & 0 & 0 & 0 & 0 & 0 & 0 & 0 & 0 & 0 & 0 & 0 & 5\\
        \hline
C6V & 1 & 0 & 0 & 0 & 0 & 0 & 0 & 0 & 0 & 0 & 0 & 0 & 0 & 0 & 0 & 6\\
        \hline
D3H & 0 & 0 & 0 & 0 & 0 & 0 & 0 & 0 & 0 & 0 & 0 & 0 & 0 & 0 & 0 & 3\\
        \hline
D6H & 0 & 0 & 0 & 0 & 0 & 0 & 0 & 0 & 0 & 0 & 0 & 0 & 0 & 0 & 0 & 14\\
        \hline
T & 0 & 0 & 0 & 0 & 0 & 0 & 0 & 0 & 0 & 0 & 0 & 0 & 0 & 0 & 0 & 2\\
        \hline
TH &  1 & 0 & 0 & 0 & 0 & 0 & 0 & 0 & 0 & 0 & 0 & 0 & 0 & 0 & 1 & 33\\
        \hline
O & 0 & 0 & 0 & 0 & 0 & 0 & 0 & 0 & 0 & 0 & 0 & 0 & 0 & 0 & 0 & 1\\
        \hline
TD & 0 & 0 & 0 & 0 & 0 & 0 & 1 & 0 & 0 & 0 & 0 & 0 & 0 & 0 & 1 & 4\\
        \hline
OH & 0 & 0 & 0 & 0 & 0 & 0 & 0 & 0 & 0 & 0 & 0 & 0 & 0 & 0 & 0 & 109\\
        \hline
TOTAL & 3 & 0 & 1 & 0 & 0 & 0 & 5 & 0 & 0 & 0 & 1 & 0 & 0 & 0 & 9 & 648\\
        \hline
{\footnotesize Original } & 3 & 0 & 1 & 0 & 0 & 0 & 9 & 0 & 0 & 0 & 1 & 0 & 0 & 0 & 22 & 1503\\
        \hline
        \end{tabular}
}
\newpage
\setlength{\oddsidemargin}{2.54cm}

\newpage

\newpage

\newpage

\newpage

\newpage

\newpage

\newpage

\newpage

\newpage

\newpage

\newpage

\newpage

\newpage

\newpage

\newpage

\newpage

\newpage

\newpage
\setlength{\oddsidemargin}{-0.1cm}
ANX=AX3\\
{\small
        \begin{tabular}{*{18}{|r}|}
        \hline
AX3 & Z=1 & 2 & \hspace*{0.9em}3 & 4 & \hspace*{0.9em}5 & \hspace*{0.9em}6 & \hspace*{0.9em}7 & \hspace*{0.9em}8 & \hspace*{0.9em}9 & \hspace*{0.7em}10 & \hspace*{0.7em}11 & \hspace*{0.7em}12 & \hspace*{0.7em}13 & \hspace*{0.7em}14 & \hspace*{0.7em}15 & \hspace*{0.7em}16 & \hspace*{0.7em}17\\
        \hline
        \hline
C1 & 0 & 0 & 0 & 0 & 0 & 0 & 0 & 0 & 0 & 0 & 0 & 0 & 0 & 0 & 0 & 0 & 0\\
        \hline
CI & 2 & 4 & 0 & 4 & 0 & 0 & 0 & 3 & 0 & 0 & 0 & 0 & 0 & 0 & 0 & 0 & 0\\
        \hline
C2 & 0 & 0 & 0 & 0 & 0 & 0 & 0 & 0 & 0 & 1 & 0 & 0 & 0 & 0 & 0 & 0 & 0\\
        \hline
CS & 0 & 0 & 0 & 0 & 0 & 0 & 0 & 0 & 0 & 0 & 0 & 0 & 0 & 0 & 0 & 0 & 0\\
        \hline
C2H & 1 & 9 & 0 & 27 & 0 & 2 & 0 & 10 & 0 & 0 & 0 & 3 & 0 & 0 & 0 & 2 & 0\\
        \hline
D2 & 0 & 0 & 0 & 10 & 0 & 0 & 0 & 1 & 0 & 0 & 0 & 0 & 0 & 0 & 0 & 0 & 0\\
        \hline
C2V & 0 & 3 & 0 & 8 & 0 & 0 & 0 & 5 & 0 & 0 & 0 & 1 & 0 & 0 & 0 & 0 & 0\\
        \hline
D2H & 1 & 2 & 0 & 39 & 0 & 0 & 0 & 1 & 0 & 0 & 0 & 2 & 0 & 0 & 0 & 2 & 0\\
        \hline
C4 & 0 & 0 & 0 & 0 & 0 & 0 & 0 & 0 & 0 & 0 & 0 & 0 & 0 & 0 & 0 & 0 & 0\\
        \hline
S4 & 0 & 0 & 0 & 0 & 0 & 0 & 0 & 0 & 0 & 0 & 0 & 0 & 0 & 0 & 0 & 0 & 0\\
        \hline
C4H & 0 & 0 & 0 & 0 & 0 & 0 & 0 & 1 & 0 & 0 & 0 & 0 & 0 & 0 & 0 & 0 & 0\\
        \hline
D4 & 0 & 0 & 0 & 0 & 0 & 0 & 0 & 0 & 0 & 0 & 0 & 0 & 0 & 0 & 0 & 0 & 0\\
        \hline
C4V & 0 & 0 & 0 & 0 & 0 & 0 & 0 & 0 & 0 & 0 & 0 & 0 & 0 & 0 & 0 & 0 & 0\\
        \hline
D2D & 0 & 2 & 0 & 0 & 0 & 0 & 0 & 0 & 0 & 0 & 0 & 0 & 0 & 0 & 0 & 0 & 0\\
        \hline
D4H & 1 & 1 & 0 & 7 & 0 & 0 & 0 & 3 & 0 & 0 & 0 & 0 & 0 & 0 & 0 & 2 & 0\\
        \hline
C3 & 0 & 0 & 0 & 0 & 0 & 3 & 0 & 0 & 0 & 0 & 0 & 0 & 0 & 0 & 0 & 0 & 0\\
        \hline
C3I & 0 & 0 & 1 & 0 & 0 & 10 & 0 & 0 & 0 & 0 & 0 & 0 & 0 & 0 & 0 & 0 & 0\\
        \hline
D3 & 1 & 1 & 5 & 0 & 0 & 5 & 0 & 0 & 0 & 0 & 0 & 0 & 0 & 0 & 0 & 0 & 0\\
        \hline
C3V & 1 & 1 & 0 & 0 & 0 & 0 & 0 & 0 & 0 & 0 & 0 & 0 & 0 & 0 & 0 & 0 & 0\\
        \hline
D3D & 4 & 17 & 10 & 1 & 0 & 17 & 0 & 0 & 0 & 0 & 0 & 0 & 0 & 0 & 0 & 0 & 0\\
        \hline
C6 & 0 & 2 & 0 & 0 & 0 & 1 & 0 & 0 & 0 & 0 & 0 & 0 & 0 & 0 & 0 & 0 & 0\\
        \hline
C3H & 0 & 0 & 0 & 0 & 0 & 0 & 0 & 0 & 0 & 0 & 0 & 0 & 0 & 0 & 0 & 0 & 0\\
        \hline
C6H & 0 & 39 & 0 & 0 & 0 & 0 & 0 & 0 & 0 & 0 & 0 & 0 & 0 & 0 & 0 & 0 & 0\\
        \hline
D6 & 0 & 1 & 0 & 0 & 0 & 6 & 0 & 0 & 0 & 0 & 0 & 0 & 0 & 0 & 0 & 0 & 0\\
        \hline
C6V & 0 & 2 & 0 & 0 & 0 & 3 & 0 & 0 & 0 & 0 & 0 & 0 & 0 & 0 & 0 & 0 & 0\\
        \hline
D3H & 0 & 1 & 0 & 0 & 0 & 0 & 0 & 0 & 0 & 0 & 0 & 0 & 0 & 0 & 0 & 0 & 0\\
        \hline
D6H & 1 & 11 & 1 & 0 & 0 & 0 & 0 & 0 & 0 & 0 & 0 & 0 & 0 & 0 & 0 & 0 & 0\\
        \hline
T & 0 & 0 & 1 & 0 & 0 & 0 & 0 & 0 & 0 & 0 & 0 & 0 & 0 & 0 & 0 & 0 & 0\\
        \hline
TH & 0 & 0 & 0 & 0 & 0 & 0 & 0 & 8 & 0 & 0 & 0 & 0 & 0 & 0 & 0 & 0 & 0\\
        \hline
O & 0 & 0 & 0 & 0 & 0 & 0 & 0 & 0 & 0 & 0 & 0 & 0 & 0 & 0 & 0 & 0 & 0\\
        \hline
TD & 0 & 0 & 0 & 1 & 0 & 0 & 0 & 0 & 0 & 0 & 0 & 0 & 0 & 0 & 0 & 0 & 0\\
        \hline
OH & 16 & 0 & 0 & 5 & 0 & 0 & 0 & 7 & 0 & 0 & 0 & 0 & 0 & 0 & 0 & 1 & 0\\
        \hline
TOTAL & 28 & 96 & 18 & 102 & 0 & 47 & 0 & 39 & 0 & 1 & 0 & 6 & 0 & 0 & 0 & 7 & 0\\
        \hline
{\footnotesize Original } & 51 & 182 & 20 & 173 & 0 & 75 & 0 & 76 & 0 & 1 & 0 & 9 & 0 & 0 & 0 & 10 & 0\\
        \hline
        \end{tabular}
}
\newpage
\setlength{\oddsidemargin}{-0.3cm}
{\small
        \begin{tabular}{*{17}{|r}|}
        \hline
AX3 & Z=18 & \hspace*{0.7em}19 & \hspace*{0.7em}20 & \hspace*{0.7em}21 & \hspace*{0.7em}22 & \hspace*{0.7em}23 & \hspace*{0.7em}24 & \hspace*{0.7em}25 & \hspace*{0.7em}26 & \hspace*{0.7em}27 & \hspace*{0.7em}28 & \hspace*{0.7em}29 & \hspace*{0.7em}30 & \hspace*{0.7em}31 & \hspace*{0.7em}32 & {\scriptsize TOTAL }\\
        \hline
        \hline
C1 & 0 & 0 & 0 & 0 & 0 & 0 & 0 & 0 & 0 & 0 & 0 & 0 & 0 & 0 & 0 & 0\\
        \hline
CI & 0 & 0 & 0 & 0 & 0 & 0 & 0 & 0 & 0 & 0 & 0 & 0 & 0 & 0 & 0 & 13\\
        \hline
C2 & 0 & 0 & 0 & 0 & 0 & 0 & 0 & 0 & 0 & 0 & 0 & 0 & 0 & 0 & 0 & 1\\
        \hline
CS & 0 & 0 & 0 & 0 & 0 & 0 & 0 & 0 & 0 & 0 & 0 & 0 & 0 & 0 & 0 & 0\\
        \hline
C2H & 0 & 0 & 0 & 0 & 0 & 0 & 0 & 0 & 0 & 0 & 0 & 0 & 0 & 0 & 0 & 54\\
        \hline
D2 & 0 & 0 & 0 & 0 & 0 & 0 & 0 & 0 & 0 & 0 & 0 & 0 & 0 & 0 & 0 & 11\\
        \hline
C2V & 0 & 0 & 0 & 0 & 0 & 0 & 0 & 0 & 0 & 0 & 0 & 0 & 0 & 0 & 0 & 17\\
        \hline
D2H & 0 & 0 & 0 & 0 & 0 & 0 & 0 & 0 & 0 & 0 & 0 & 0 & 0 & 0 & 1 & 48\\
        \hline
C4 & 0 & 0 & 0 & 0 & 0 & 0 & 0 & 0 & 0 & 0 & 0 & 0 & 0 & 0 & 0 & 0\\
        \hline
S4 & 0 & 0 & 0 & 0 & 0 & 0 & 0 & 0 & 0 & 0 & 0 & 0 & 0 & 0 & 0 & 0\\
        \hline
C4H & 0 & 0 & 0 & 0 & 0 & 0 & 0 & 0 & 0 & 0 & 0 & 0 & 0 & 0 & 0 & 1\\
        \hline
D4 & 0 & 0 & 0 & 0 & 0 & 0 & 0 & 0 & 0 & 0 & 0 & 0 & 0 & 0 & 0 & 0\\
        \hline
C4V & 0 & 0 & 0 & 0 & 0 & 0 & 0 & 0 & 0 & 0 & 0 & 0 & 0 & 0 & 0 & 0\\
        \hline
D2D & 0 & 0 & 0 & 0 & 0 & 0 & 0 & 0 & 0 & 0 & 0 & 0 & 0 & 0 & 0 & 2\\
        \hline
D4H & 0 & 0 & 0 & 0 & 0 & 0 & 0 & 0 & 0 & 0 & 0 & 0 & 0 & 0 & 0 & 14\\
        \hline
C3 & 0 & 0 & 0 & 0 & 0 & 0 & 0 & 0 & 0 & 0 & 0 & 0 & 0 & 0 & 0 & 3\\
        \hline
C3I &  0 & 0 & 0 & 0 & 0 & 0 & 1 & 0 & 0 & 0 & 0 & 0 & 0 & 0 & 0 & 12\\
        \hline
D3 & 0 & 0 & 0 & 0 & 0 & 0 & 0 & 0 & 0 & 0 & 0 & 0 & 0 & 0 & 0 & 12\\
        \hline
C3V & 0 & 0 & 0 & 0 & 0 & 0 & 0 & 0 & 0 & 0 & 0 & 0 & 0 & 0 & 0 & 2\\
        \hline
D3D & 1 & 0 & 0 & 0 & 0 & 0 & 0 & 0 & 0 & 0 & 0 & 0 & 0 & 0 & 0 & 50\\
        \hline
C6 & 0 & 0 & 0 & 0 & 0 & 0 & 0 & 0 & 0 & 0 & 0 & 0 & 0 & 0 & 0 & 3\\
        \hline
C3H & 0 & 0 & 0 & 0 & 0 & 0 & 0 & 0 & 0 & 0 & 0 & 0 & 0 & 0 & 0 & 0\\
        \hline
C6H & 0 & 0 & 0 & 0 & 0 & 0 & 0 & 0 & 0 & 0 & 0 & 0 & 0 & 0 & 0 & 39\\
        \hline
D6 & 0 & 0 & 0 & 0 & 0 & 0 & 0 & 0 & 0 & 0 & 0 & 0 & 0 & 0 & 0 & 7\\
        \hline
C6V & 0 & 0 & 0 & 0 & 0 & 0 & 0 & 0 & 0 & 0 & 0 & 0 & 0 & 0 & 0 & 5\\
        \hline
D3H & 0 & 0 & 0 & 0 & 0 & 0 & 0 & 0 & 0 & 0 & 0 & 0 & 0 & 0 & 0 & 1\\
        \hline
D6H & 0 & 0 & 0 & 0 & 0 & 0 & 0 & 0 & 0 & 0 & 0 & 0 & 0 & 0 & 0 & 13\\
        \hline
T & 0 & 0 & 0 & 0 & 0 & 0 & 0 & 0 & 0 & 0 & 0 & 0 & 0 & 0 & 0 & 1\\
        \hline
TH & 0 & 0 & 0 & 0 & 0 & 0 & 0 & 0 & 0 & 0 & 0 & 0 & 0 & 0 & 0 & 8\\
        \hline
O & 0 & 0 & 0 & 0 & 0 & 0 & 0 & 0 & 0 & 0 & 0 & 0 & 0 & 0 & 0 & 0\\
        \hline
TD & 0 & 0 & 0 & 0 & 0 & 0 & 0 & 0 & 0 & 0 & 0 & 0 & 0 & 0 & 0 & 1\\
        \hline
OH & 0 & 0 & 0 & 0 & 0 & 0 & 0 & 0 & 0 & 0 & 0 & 0 & 0 & 0 & 0 & 29\\
        \hline
TOTAL & 1 & 0 & 0 & 0 & 0 & 0 & 1 & 0 & 0 & 0 & 0 & 0 & 0 & 0 & 1 & 347\\
        \hline
{\footnotesize Original } & 1 & 0 & 0 & 0 & 0 & 0 & 1 & 0 & 0 & 0 & 0 & 0 & 0 & 0 & 3 & 602\\
        \hline
        \end{tabular}
}
\newpage
\setlength{\oddsidemargin}{2.54cm}

\newpage

\newpage

\newpage

\newpage

\newpage

\newpage

\newpage

\newpage

\newpage

\newpage

\newpage

\newpage

\setlength{\oddsidemargin}{-0.1cm}
ANX=AB2X4\\
{\small
        \begin{tabular}{*{18}{|r}|}
        \hline
{\footnotesize AB2X4 } & Z=1 & 2 & \hspace*{0.9em}3 & 4 & \hspace*{0.9em}5 & \hspace*{0.9em}6 & \hspace*{0.9em}7 & \hspace*{0.9em}8 & \hspace*{0.9em}9 & \hspace*{0.7em}10 & \hspace*{0.7em}11 & \hspace*{0.7em}12 & \hspace*{0.7em}13 & \hspace*{0.7em}14 & \hspace*{0.7em}15 & \hspace*{0.7em}16 & \hspace*{0.7em}17\\
        \hline
        \hline
C1 & 0 & 2 & 0 & 1 & 0 & 0 & 0 & 1 & 0 & 0 & 0 & 0 & 0 & 0 & 0 & 0 & 0\\
        \hline
CI & 0 & 4 & 0 & 3 & 0 & 0 & 0 & 1 & 0 & 0 & 0 & 0 & 0 & 0 & 0 & 1 & 0\\
        \hline
C2 & 1 & 7 & 0 & 1 & 0 & 0 & 0 & 0 & 0 & 0 & 0 & 0 & 0 & 0 & 0 & 1 & 0\\
        \hline
CS & 0 & 1 & 0 & 1 & 0 & 0 & 0 & 2 & 0 & 0 & 0 & 0 & 0 & 0 & 0 & 0 & 0\\
        \hline
C2H & 6 & 27 & 0 & 37 & 0 & 0 & 0 & 7 & 0 & 0 & 0 & 4 & 0 & 0 & 0 & 2 & 0\\
        \hline
D2 & 0 & 1 & 0 & 2 & 0 & 0 & 0 & 0 & 0 & 0 & 0 & 0 & 0 & 0 & 0 & 0 & 0\\
        \hline
C2V & 0 & 3 & 0 & 15 & 0 & 0 & 0 & 7 & 0 & 0 & 0 & 9 & 0 & 0 & 0 & 0 & 0\\
        \hline
D2H & 3 & 22 & 2 & 176 & 0 & 0 & 0 & 16 & 0 & 1 & 0 & 1 & 0 & 0 & 0 & 1 & 0\\
        \hline
C4 & 1 & 0 & 0 & 1 & 0 & 0 & 0 & 0 & 0 & 0 & 0 & 0 & 0 & 0 & 0 & 0 & 0\\
        \hline
S4 & 0 & 23 & 0 & 0 & 0 & 0 & 0 & 0 & 0 & 0 & 0 & 0 & 0 & 0 & 0 & 0 & 0\\
        \hline
C4H & 6 & 2 & 0 & 2 & 0 & 0 & 0 & 0 & 0 & 0 & 0 & 0 & 0 & 0 & 0 & 0 & 0\\
        \hline
D4 & 0 & 0 & 0 & 4 & 0 & 0 & 0 & 0 & 0 & 0 & 0 & 0 & 0 & 0 & 0 & 0 & 0\\
        \hline
C4V & 0 & 0 & 0 & 1 & 0 & 0 & 0 & 0 & 0 & 0 & 0 & 0 & 0 & 0 & 0 & 0 & 0\\
        \hline
D2D & 4 & 5 & 0 & 4 & 0 & 0 & 0 & 2 & 0 & 0 & 0 & 2 & 0 & 0 & 0 & 0 & 0\\
        \hline
D4H & 7 & 75 & 0 & 26 & 0 & 0 & 0 & 3 & 0 & 0 & 0 & 0 & 0 & 0 & 0 & 2 & 0\\
        \hline
C3 & 0 & 0 & 0 & 0 & 0 & 0 & 0 & 0 & 0 & 0 & 0 & 0 & 0 & 0 & 0 & 0 & 0\\
        \hline
C3I & 0 & 0 & 0 & 0 & 0 & 3 & 0 & 0 & 0 & 0 & 0 & 0 & 0 & 0 & 0 & 0 & 0\\
        \hline
D3 & 0 & 0 & 1 & 0 & 0 & 0 & 0 & 0 & 0 & 0 & 0 & 0 & 0 & 0 & 0 & 0 & 0\\
        \hline
C3V & 0 & 0 & 1 & 0 & 0 & 2 & 0 & 0 & 0 & 0 & 0 & 0 & 0 & 0 & 0 & 0 & 0\\
        \hline
D3D & 9 & 6 & 12 & 1 & 0 & 0 & 0 & 1 & 0 & 0 & 0 & 0 & 0 & 0 & 0 & 0 & 0\\
        \hline
C6 & 0 & 0 & 0 & 0 & 0 & 1 & 0 & 2 & 0 & 0 & 0 & 0 & 0 & 0 & 0 & 0 & 0\\
        \hline
C3H & 0 & 0 & 0 & 0 & 0 & 0 & 0 & 0 & 0 & 0 & 0 & 0 & 0 & 0 & 0 & 0 & 0\\
        \hline
C6H & 0 & 0 & 0 & 0 & 0 & 0 & 0 & 0 & 1 & 0 & 0 & 1 & 0 & 0 & 0 & 0 & 0\\
        \hline
D6 & 0 & 3 & 0 & 0 & 0 & 1 & 0 & 1 & 0 & 0 & 0 & 0 & 0 & 0 & 0 & 0 & 0\\
        \hline
C6V & 0 & 3 & 0 & 1 & 0 & 0 & 0 & 0 & 0 & 0 & 0 & 0 & 0 & 0 & 0 & 0 & 0\\
        \hline
D3H & 0 & 0 & 2 & 0 & 0 & 0 & 0 & 0 & 0 & 0 & 0 & 0 & 0 & 0 & 0 & 0 & 0\\
        \hline
D6H & 0 & 7 & 0 & 0 & 0 & 0 & 0 & 0 & 0 & 0 & 0 & 0 & 0 & 0 & 0 & 0 & 0\\
        \hline
T & 0 & 0 & 0 & 0 & 0 & 0 & 0 & 1 & 0 & 0 & 0 & 0 & 0 & 0 & 0 & 0 & 0\\
        \hline
TH & 0 & 0 & 0 & 0 & 0 & 0 & 0 & 0 & 0 & 0 & 0 & 4 & 0 & 0 & 0 & 1 & 0\\
        \hline
O & 0 & 0 & 0 & 0 & 0 & 0 & 0 & 0 & 0 & 0 & 0 & 0 & 0 & 0 & 0 & 0 & 0\\
        \hline
TD & 0 & 1 & 0 & 1 & 0 & 0 & 0 & 4 & 0 & 0 & 0 & 0 & 0 & 0 & 0 & 0 & 0\\
        \hline
OH & 5 & 1 & 0 & 8 & 0 & 0 & 0 & 109 & 0 & 0 & 0 & 1 & 0 & 0 & 0 & 4 & 0\\
        \hline
TOTAL & 42 & 193 & 18 & 285 & 0 & 7 & 0 & 157 & 1 & 1 & 0 & 22 & 0 & 0 & 0 & 12 & 0\\
        \hline
{\footnotesize Original } & 110 & 463 & 31 & 771 & 0 & 7 & 0 & 587 & 1 & 1 & 0 & 32 & 0 & 0 & 0 & 18 & 0\\
        \hline
        \end{tabular}
}
\newpage
\setlength{\oddsidemargin}{-0.3cm}
{\small
        \begin{tabular}{*{17}{|r}|}
        \hline
{\footnotesize AB2X4 } & Z=18 & \hspace*{0.7em}19 & \hspace*{0.7em}20 & \hspace*{0.7em}21 & \hspace*{0.7em}22 & \hspace*{0.7em}23 & \hspace*{0.7em}24 & \hspace*{0.7em}25 & \hspace*{0.7em}26 & \hspace*{0.7em}27 & \hspace*{0.7em}28 & \hspace*{0.7em}29 & \hspace*{0.7em}30 & \hspace*{0.7em}31 & \hspace*{0.7em}32 & {\scriptsize TOTAL }\\
        \hline
        \hline
C1 & 0 & 0 & 0 & 0 & 0 & 0 & 0 & 0 & 0 & 0 & 0 & 0 & 0 & 0 & 0 & 4\\
        \hline
CI & 0 & 0 & 0 & 0 & 0 & 0 & 0 & 0 & 0 & 0 & 0 & 0 & 0 & 0 & 0 & 9\\
        \hline
C2 & 0 & 0 & 0 & 0 & 0 & 0 & 0 & 0 & 0 & 0 & 0 & 0 & 0 & 0 & 0 & 10\\
        \hline
CS & 0 & 0 & 0 & 0 & 0 & 0 & 0 & 0 & 0 & 0 & 0 & 0 & 0 & 0 & 0 & 4\\
        \hline
C2H & 0 & 0 & 0 & 0 & 0 & 0 & 0 & 0 & 0 & 0 & 0 & 0 & 0 & 0 & 0 & 83\\
        \hline
D2 & 0 & 0 & 0 & 0 & 0 & 0 & 0 & 0 & 0 & 0 & 0 & 0 & 0 & 0 & 0 & 3\\
        \hline
C2V & 0 & 0 & 0 & 0 & 0 & 0 & 0 & 0 & 0 & 0 & 0 & 0 & 0 & 0 & 0 & 34\\
        \hline
D2H & 0 & 0 & 0 & 0 & 0 & 0 & 0 & 0 & 0 & 0 & 0 & 0 & 0 & 0 & 10 & 232\\
        \hline
C4 & 0 & 0 & 0 & 0 & 0 & 0 & 0 & 0 & 0 & 0 & 0 & 0 & 0 & 0 & 0 & 2\\
        \hline
S4 & 0 & 0 & 0 & 0 & 0 & 0 & 0 & 0 & 0 & 0 & 0 & 0 & 0 & 0 & 0 & 23\\
        \hline
C4H & 0 & 0 & 0 & 0 & 0 & 0 & 0 & 0 & 0 & 0 & 0 & 0 & 0 & 0 & 0 & 10\\
        \hline
D4 & 0 & 0 & 0 & 0 & 0 & 0 & 0 & 0 & 0 & 0 & 0 & 0 & 0 & 0 & 0 & 4\\
        \hline
C4V & 0 & 0 & 0 & 0 & 0 & 0 & 0 & 0 & 0 & 0 & 0 & 0 & 0 & 0 & 0 & 1\\
        \hline
D2D & 0 & 0 & 0 & 0 & 0 & 0 & 0 & 0 & 0 & 0 & 0 & 0 & 0 & 0 & 0 & 17\\
        \hline
D4H & 0 & 0 & 0 & 0 & 0 & 0 & 0 & 0 & 0 & 0 & 0 & 0 & 0 & 0 & 0 & 113\\
        \hline
C3 & 0 & 0 & 0 & 0 & 0 & 0 & 0 & 0 & 0 & 0 & 0 & 0 & 0 & 0 & 0 & 0\\
        \hline
C3I & 9 & 0 & 0 & 0 & 0 & 0 & 0 & 0 & 0 & 0 & 0 & 0 & 0 & 0 & 0 & 12\\
        \hline
D3 & 0 & 0 & 0 & 0 & 0 & 0 & 0 & 0 & 0 & 0 & 0 & 0 & 0 & 0 & 0 & 1\\
        \hline
C3V & 1 & 0 & 0 & 0 & 0 & 0 & 0 & 0 & 0 & 0 & 0 & 0 & 0 & 0 & 0 & 4\\
        \hline
D3D & 1 & 0 & 0 & 0 & 0 & 0 & 0 & 0 & 0 & 0 & 0 & 0 & 0 & 0 & 0 & 30\\
        \hline
C6 & 0 & 0 & 0 & 0 & 0 & 0 & 0 & 0 & 0 & 0 & 0 & 0 & 0 & 0 & 0 & 3\\
        \hline
C3H & 0 & 0 & 0 & 0 & 0 & 0 & 0 & 0 & 0 & 0 & 0 & 0 & 0 & 0 & 0 & 0\\
        \hline
C6H & 0 & 0 & 0 & 0 & 0 & 0 & 0 & 0 & 0 & 0 & 0 & 0 & 0 & 0 & 0 & 2\\
        \hline
D6 & 0 & 0 & 0 & 0 & 0 & 0 & 0 & 0 & 0 & 0 & 0 & 0 & 0 & 0 & 0 & 5\\
        \hline
C6V & 0 & 0 & 0 & 0 & 0 & 0 & 0 & 0 & 0 & 0 & 0 & 0 & 0 & 0 & 0 & 4\\
        \hline
D3H & 0 & 0 & 0 & 0 & 0 & 0 & 0 & 0 & 0 & 0 & 0 & 0 & 0 & 0 & 0 & 2\\
        \hline
D6H & 0 & 0 & 0 & 0 & 0 & 0 & 0 & 0 & 0 & 0 & 0 & 0 & 0 & 0 & 0 & 7\\
        \hline
T & 0 & 0 & 0 & 0 & 0 & 0 & 0 & 0 & 0 & 0 & 0 & 0 & 0 & 0 & 0 & 1\\
        \hline
TH & 0 & 0 & 0 & 0 & 0 & 0 & 0 & 0 & 0 & 0 & 0 & 0 & 0 & 0 & 1 & 6\\
        \hline
O & 0 & 0 & 0 & 0 & 0 & 0 & 0 & 0 & 0 & 0 & 0 & 0 & 0 & 0 & 0 & 0\\
        \hline
TD & 0 & 0 & 0 & 0 & 0 & 0 & 0 & 0 & 0 & 0 & 0 & 0 & 0 & 0 & 0 & 6\\
        \hline
OH & 0 & 0 & 0 & 0 & 0 & 0 & 0 & 0 & 0 & 0 & 0 & 0 & 0 & 0 & 0 & 128\\
        \hline
TOTAL & 11 & 0 & 0 & 0 & 0 & 0 & 0 & 0 & 0 & 0 & 0 & 0 & 0 & 0 & 11 & 760\\
        \hline
{\footnotesize Original } & 24 & 0 & 0 & 0 & 0 & 0 & 0 & 0 & 0 & 0 & 0 & 0 & 0 & 0 & 11 & 2056\\
        \hline
        \end{tabular}
}
\newpage
\setlength{\oddsidemargin}{2.54cm}

\newpage

\newpage

\newpage

\newpage

\newpage

\newpage

\newpage

\newpage

\newpage

\newpage

\newpage

\newpage
\setlength{\oddsidemargin}{-0.1cm}
ANX=ABX4\\
{\small
        \begin{tabular}{*{18}{|r}|}
        \hline
ABX4 & Z=1 & 2 & \hspace*{0.9em}3 & 4 & \hspace*{0.9em}5 & \hspace*{0.9em}6 & \hspace*{0.9em}7 & \hspace*{0.9em}8 & \hspace*{0.9em}9 & \hspace*{0.7em}10 & \hspace*{0.7em}11 & \hspace*{0.7em}12 & \hspace*{0.7em}13 & \hspace*{0.7em}14 & \hspace*{0.7em}15 & \hspace*{0.7em}16 & \hspace*{0.7em}17\\
        \hline
        \hline
C1 & 1 & 0 & 1 & 1 & 0 & 0 & 0 & 0 & 0 & 0 & 0 & 0 & 0 & 0 & 0 & 0 & 0\\
        \hline
CI & 6 & 6 & 0 & 6 & 0 & 6 & 0 & 0 & 0 & 0 & 0 & 1 & 0 & 0 & 0 & 0 & 0\\
        \hline
C2 & 0 & 2 & 0 & 4 & 0 & 1 & 0 & 2 & 0 & 1 & 0 & 0 & 0 & 0 & 0 & 0 & 0\\
        \hline
CS & 0 & 2 & 0 & 1 & 0 & 0 & 0 & 0 & 0 & 0 & 0 & 0 & 0 & 0 & 0 & 0 & 0\\
        \hline
C2H & 0 & 38 & 0 & 95 & 0 & 0 & 0 & 13 & 0 & 0 & 0 & 1 & 0 & 0 & 0 & 1 & 0\\
        \hline
D2 & 0 & 3 & 0 & 15 & 0 & 0 & 0 & 0 & 0 & 0 & 0 & 0 & 0 & 0 & 0 & 1 & 0\\
        \hline
C2V & 0 & 3 & 0 & 23 & 0 & 0 & 0 & 6 & 0 & 0 & 0 & 2 & 0 & 0 & 0 & 2 & 0\\
        \hline
D2H & 0 & 7 & 0 & 109 & 0 & 0 & 0 & 9 & 0 & 0 & 0 & 3 & 0 & 0 & 0 & 4 & 0\\
        \hline
C4 & 0 & 0 & 0 & 1 & 0 & 0 & 0 & 0 & 0 & 0 & 0 & 0 & 0 & 0 & 0 & 0 & 0\\
        \hline
S4 & 0 & 11 & 0 & 0 & 0 & 0 & 0 & 0 & 0 & 0 & 0 & 0 & 0 & 0 & 0 & 0 & 0\\
        \hline
C4H & 2 & 3 & 0 & 43 & 0 & 0 & 0 & 0 & 0 & 0 & 0 & 0 & 0 & 0 & 0 & 0 & 0\\
        \hline
D4 & 0 & 0 & 0 & 0 & 0 & 0 & 0 & 0 & 0 & 0 & 0 & 0 & 0 & 0 & 0 & 0 & 0\\
        \hline
C4V & 0 & 0 & 0 & 0 & 0 & 0 & 0 & 0 & 0 & 0 & 0 & 0 & 0 & 0 & 0 & 0 & 0\\
        \hline
D2D & 0 & 3 & 0 & 8 & 0 & 0 & 0 & 1 & 0 & 0 & 0 & 0 & 0 & 0 & 0 & 0 & 0\\
        \hline
D4H & 4 & 4 & 0 & 73 & 0 & 0 & 0 & 0 & 0 & 0 & 0 & 0 & 0 & 0 & 0 & 4 & 0\\
        \hline
C3 & 0 & 0 & 1 & 0 & 0 & 1 & 0 & 0 & 1 & 0 & 0 & 0 & 0 & 0 & 0 & 0 & 0\\
        \hline
C3I & 1 & 0 & 1 & 0 & 0 & 0 & 0 & 0 & 0 & 0 & 0 & 0 & 0 & 0 & 0 & 0 & 0\\
        \hline
D3 & 0 & 0 & 12 & 0 & 0 & 0 & 0 & 0 & 0 & 0 & 0 & 0 & 0 & 0 & 0 & 0 & 0\\
        \hline
C3V & 1 & 0 & 0 & 0 & 0 & 0 & 0 & 0 & 0 & 0 & 0 & 0 & 0 & 0 & 0 & 0 & 0\\
        \hline
D3D & 7 & 0 & 2 & 0 & 0 & 0 & 0 & 0 & 0 & 0 & 0 & 0 & 0 & 0 & 0 & 0 & 0\\
        \hline
C6 & 0 & 0 & 0 & 0 & 0 & 1 & 0 & 0 & 0 & 0 & 0 & 0 & 0 & 0 & 0 & 0 & 0\\
        \hline
C3H & 0 & 0 & 0 & 0 & 0 & 0 & 0 & 0 & 0 & 0 & 0 & 0 & 0 & 0 & 0 & 0 & 0\\
        \hline
C6H & 0 & 0 & 0 & 0 & 0 & 0 & 0 & 0 & 0 & 0 & 0 & 0 & 0 & 0 & 0 & 0 & 0\\
        \hline
D6 & 0 & 0 & 5 & 0 & 0 & 0 & 0 & 0 & 0 & 0 & 0 & 0 & 0 & 0 & 0 & 0 & 0\\
        \hline
C6V & 0 & 0 & 0 & 0 & 0 & 0 & 0 & 0 & 0 & 0 & 0 & 0 & 0 & 0 & 0 & 0 & 0\\
        \hline
D3H & 0 & 0 & 2 & 0 & 0 & 0 & 0 & 0 & 0 & 0 & 0 & 0 & 0 & 0 & 0 & 0 & 0\\
        \hline
D6H & 0 & 0 & 0 & 0 & 0 & 0 & 0 & 0 & 0 & 0 & 0 & 0 & 0 & 0 & 0 & 0 & 0\\
        \hline
T & 0 & 0 & 0 & 3 & 0 & 0 & 0 & 0 & 0 & 0 & 0 & 0 & 0 & 0 & 0 & 0 & 0\\
        \hline
TH & 0 & 0 & 0 & 0 & 0 & 0 & 0 & 0 & 0 & 0 & 0 & 0 & 0 & 0 & 0 & 0 & 0\\
        \hline
O & 0 & 0 & 0 & 0 & 0 & 0 & 0 & 0 & 0 & 0 & 0 & 0 & 0 & 0 & 0 & 0 & 0\\
        \hline
TD & 0 & 0 & 0 & 10 & 0 & 0 & 0 & 0 & 0 & 0 & 0 & 0 & 0 & 0 & 0 & 0 & 0\\
        \hline
OH & 1 & 0 & 0 & 9 & 0 & 0 & 0 & 2 & 0 & 0 & 0 & 0 & 0 & 0 & 0 & 4 & 0\\
        \hline
TOTAL & 23 & 82 & 24 & 401 & 0 & 9 & 0 & 33 & 1 & 1 & 0 & 7 & 0 & 0 & 0 & 16 & 0\\
        \hline
{\footnotesize Original } & 42 & 148 & 70 & 777 & 0 & 14 & 0 & 52 & 1 & 1 & 0 & 9 & 0 & 0 & 0 & 21 & 0\\
        \hline
        \end{tabular}
}
\newpage
\setlength{\oddsidemargin}{-0.3cm}
{\small
        \begin{tabular}{*{17}{|r}|}
        \hline
ABX4 & Z=18 & \hspace*{0.7em}19 & \hspace*{0.7em}20 & \hspace*{0.7em}21 & \hspace*{0.7em}22 & \hspace*{0.7em}23 & \hspace*{0.7em}24 & \hspace*{0.7em}25 & \hspace*{0.7em}26 & \hspace*{0.7em}27 & \hspace*{0.7em}28 & \hspace*{0.7em}29 & \hspace*{0.7em}30 & \hspace*{0.7em}31 & \hspace*{0.7em}32 & {\scriptsize TOTAL }\\
        \hline
        \hline
C1 & 0 & 0 & 0 & 0 & 0 & 0 & 0 & 0 & 0 & 0 & 0 & 0 & 0 & 0 & 0 & 3\\
        \hline
CI & 0 & 0 & 0 & 0 & 0 & 0 & 0 & 0 & 0 & 0 & 0 & 0 & 0 & 0 & 0 & 25\\
        \hline
C2 & 0 & 0 & 0 & 0 & 0 & 0 & 0 & 0 & 0 & 0 & 0 & 0 & 0 & 0 & 0 & 10\\
        \hline
CS & 0 & 0 & 0 & 0 & 0 & 0 & 0 & 0 & 0 & 0 & 0 & 0 & 0 & 0 & 0 & 3\\
        \hline
C2H & 0 & 0 & 0 & 0 & 0 & 0 & 1 & 0 & 0 & 0 & 0 & 0 & 0 & 0 & 0 & 149\\
        \hline
D2 & 0 & 0 & 0 & 0 & 0 & 0 & 0 & 0 & 0 & 0 & 0 & 0 & 0 & 0 & 0 & 19\\
        \hline
C2V & 0 & 0 & 0 & 0 & 0 & 0 & 0 & 0 & 0 & 0 & 0 & 0 & 0 & 0 & 0 & 36\\
        \hline
D2H & 0 & 0 & 0 & 0 & 0 & 0 & 1 & 0 & 0 & 0 & 0 & 0 & 0 & 0 & 0 & 133\\
        \hline
C4 & 0 & 0 & 0 & 0 & 0 & 0 & 0 & 0 & 0 & 0 & 0 & 0 & 0 & 0 & 0 & 1\\
        \hline
S4 & 0 & 0 & 0 & 0 & 0 & 0 & 0 & 0 & 0 & 0 & 0 & 0 & 0 & 0 & 0 & 11\\
        \hline
C4H & 0 & 0 & 0 & 0 & 0 & 0 & 0 & 0 & 0 & 0 & 0 & 0 & 0 & 0 & 0 & 48\\
        \hline
D4 & 0 & 0 & 0 & 0 & 0 & 0 & 0 & 0 & 0 & 0 & 0 & 0 & 0 & 0 & 0 & 0\\
        \hline
C4V & 0 & 0 & 0 & 0 & 0 & 0 & 0 & 0 & 0 & 0 & 0 & 0 & 0 & 0 & 0 & 0\\
        \hline
D2D & 0 & 0 & 2 & 0 & 0 & 0 & 0 & 0 & 0 & 0 & 0 & 0 & 0 & 0 & 0 & 14\\
        \hline
D4H & 0 & 0 & 0 & 0 & 0 & 0 & 0 & 0 & 0 & 0 & 0 & 0 & 0 & 0 & 0 & 85\\
        \hline
C3 & 1 & 0 & 0 & 0 & 0 & 0 & 0 & 0 & 0 & 0 & 0 & 0 & 0 & 0 & 0 & 4\\
        \hline
C3I & 2 & 0 & 0 & 0 & 0 & 0 & 0 & 0 & 0 & 0 & 0 & 0 & 0 & 0 & 0 & 4\\
        \hline
D3 & 0 & 0 & 0 & 0 & 0 & 0 & 0 & 0 & 0 & 0 & 0 & 0 & 0 & 0 & 0 & 12\\
        \hline
C3V & 4 & 0 & 0 & 0 & 0 & 0 & 0 & 0 & 0 & 0 & 0 & 0 & 0 & 0 & 0 & 5\\
        \hline
D3D & 0 & 0 & 0 & 0 & 0 & 0 & 0 & 0 & 0 & 0 & 0 & 0 & 0 & 0 & 0 & 9\\
        \hline
C6 & 0 & 0 & 0 & 0 & 0 & 0 & 0 & 0 & 0 & 0 & 0 & 0 & 0 & 0 & 0 & 1\\
        \hline
C3H & 1 & 0 & 0 & 0 & 0 & 0 & 0 & 0 & 0 & 0 & 0 & 0 & 0 & 0 & 0 & 1\\
        \hline
C6H & 0 & 0 & 0 & 0 & 0 & 0 & 0 & 0 & 0 & 0 & 0 & 0 & 0 & 0 & 0 & 0\\
        \hline
D6 & 0 & 0 & 0 & 0 & 0 & 0 & 0 & 0 & 0 & 0 & 0 & 0 & 0 & 0 & 0 & 5\\
        \hline
C6V & 1 & 0 & 0 & 0 & 0 & 0 & 0 & 0 & 0 & 0 & 0 & 0 & 0 & 0 & 0 & 1\\
        \hline
D3H & 0 & 0 & 0 & 0 & 0 & 0 & 0 & 0 & 0 & 0 & 0 & 0 & 0 & 0 & 0 & 2\\
        \hline
D6H & 0 & 0 & 0 & 0 & 0 & 0 & 0 & 0 & 0 & 0 & 0 & 0 & 0 & 0 & 0 & 0\\
        \hline
T & 0 & 0 & 0 & 0 & 0 & 0 & 0 & 0 & 0 & 0 & 0 & 0 & 0 & 0 & 0 & 3\\
        \hline
TH & 0 & 0 & 0 & 0 & 0 & 0 & 0 & 0 & 0 & 0 & 0 & 0 & 0 & 0 & 0 & 0\\
        \hline
O & 0 & 0 & 0 & 0 & 0 & 0 & 0 & 0 & 0 & 0 & 0 & 0 & 0 & 0 & 0 & 0\\
        \hline
TD & 0 & 0 & 0 & 0 & 0 & 0 & 1 & 0 & 0 & 0 & 0 & 0 & 0 & 0 & 0 & 11\\
        \hline
OH & 0 & 0 & 0 & 0 & 0 & 0 & 0 & 0 & 0 & 0 & 0 & 0 & 0 & 0 & 0 & 16\\
        \hline
TOTAL & 9 & 0 & 2 & 0 & 0 & 0 & 3 & 0 & 0 & 0 & 0 & 0 & 0 & 0 & 0 & 611\\
        \hline
{\footnotesize Original } & 16 & 0 & 3 & 0 & 0 & 0 & 4 & 0 & 0 & 0 & 0 & 0 & 0 & 0 & 0 & 1158\\
        \hline
        \end{tabular}
}
\newpage
\setlength{\oddsidemargin}{2.54cm}

\newpage

\newpage

\newpage

\newpage

\newpage

\newpage

\newpage

\newpage

\newpage

\newpage

\newpage

\newpage

\newpage
\setlength{\oddsidemargin}{-0.1cm}
ANX=AB2X6\\
{\small
        \begin{tabular}{*{18}{|r}|}
        \hline
{\footnotesize AB2X6 } & Z=1 & 2 & \hspace*{0.9em}3 & 4 & \hspace*{0.9em}5 & \hspace*{0.9em}6 & \hspace*{0.9em}7 & \hspace*{0.9em}8 & \hspace*{0.9em}9 & \hspace*{0.7em}10 & \hspace*{0.7em}11 & \hspace*{0.7em}12 & \hspace*{0.7em}13 & \hspace*{0.7em}14 & \hspace*{0.7em}15 & \hspace*{0.7em}16 & \hspace*{0.7em}17\\
        \hline
        \hline
C1 & 0 & 1 & 0 & 0 & 0 & 0 & 0 & 0 & 0 & 0 & 0 & 0 & 0 & 0 & 0 & 0 & 0\\
        \hline
CI & 5 & 3 & 1 & 2 & 0 & 0 & 0 & 0 & 0 & 0 & 0 & 0 & 0 & 0 & 0 & 1 & 0\\
        \hline
C2 & 0 & 10 & 0 & 0 & 0 & 0 & 0 & 0 & 0 & 0 & 0 & 0 & 0 & 0 & 0 & 0 & 0\\
        \hline
CS & 0 & 1 & 0 & 1 & 0 & 0 & 0 & 0 & 0 & 0 & 0 & 0 & 0 & 0 & 0 & 0 & 0\\
        \hline
C2H & 1 & 35 & 0 & 42 & 0 & 0 & 0 & 11 & 0 & 0 & 0 & 0 & 0 & 0 & 0 & 1 & 0\\
        \hline
D2 & 0 & 2 & 0 & 12 & 0 & 2 & 0 & 4 & 0 & 0 & 0 & 0 & 0 & 0 & 0 & 0 & 0\\
        \hline
C2V & 0 & 0 & 0 & 6 & 0 & 0 & 0 & 8 & 0 & 1 & 0 & 0 & 0 & 0 & 0 & 1 & 0\\
        \hline
D2H & 1 & 8 & 0 & 38 & 0 & 0 & 0 & 5 & 0 & 0 & 0 & 0 & 0 & 0 & 0 & 0 & 0\\
        \hline
C4 & 0 & 0 & 0 & 0 & 0 & 0 & 0 & 0 & 0 & 0 & 0 & 0 & 0 & 0 & 0 & 0 & 0\\
        \hline
S4 & 0 & 0 & 0 & 0 & 0 & 0 & 0 & 0 & 0 & 0 & 0 & 0 & 0 & 0 & 0 & 0 & 0\\
        \hline
C4H & 0 & 5 & 0 & 1 & 0 & 0 & 0 & 0 & 0 & 0 & 0 & 0 & 0 & 0 & 0 & 0 & 0\\
        \hline
D4 & 0 & 4 & 0 & 1 & 0 & 0 & 0 & 0 & 0 & 0 & 0 & 0 & 0 & 0 & 0 & 0 & 0\\
        \hline
C4V & 0 & 0 & 0 & 0 & 0 & 0 & 0 & 1 & 0 & 0 & 0 & 0 & 0 & 0 & 0 & 0 & 0\\
        \hline
D2D & 0 & 2 & 0 & 1 & 0 & 0 & 0 & 1 & 0 & 0 & 0 & 0 & 0 & 0 & 0 & 0 & 0\\
        \hline
D4H & 4 & 32 & 0 & 0 & 0 & 0 & 0 & 3 & 0 & 0 & 0 & 0 & 0 & 0 & 0 & 1 & 0\\
        \hline
C3 & 0 & 1 & 1 & 0 & 0 & 0 & 0 & 0 & 0 & 0 & 0 & 0 & 0 & 0 & 0 & 0 & 0\\
        \hline
C3I & 7 & 0 & 3 & 0 & 0 & 0 & 0 & 0 & 0 & 0 & 0 & 0 & 0 & 0 & 0 & 0 & 0\\
        \hline
D3 & 9 & 0 & 9 & 0 & 0 & 0 & 0 & 0 & 1 & 0 & 0 & 0 & 0 & 0 & 0 & 0 & 0\\
        \hline
C3V & 2 & 0 & 3 & 0 & 0 & 0 & 0 & 0 & 1 & 0 & 0 & 0 & 0 & 0 & 0 & 0 & 0\\
        \hline
D3D & 33 & 0 & 8 & 1 & 0 & 0 & 0 & 0 & 3 & 0 & 0 & 0 & 0 & 0 & 0 & 0 & 0\\
        \hline
C6 & 1 & 0 & 0 & 0 & 0 & 1 & 0 & 0 & 0 & 0 & 0 & 0 & 0 & 0 & 0 & 0 & 0\\
        \hline
C3H & 1 & 0 & 0 & 0 & 0 & 0 & 0 & 0 & 0 & 0 & 0 & 0 & 0 & 0 & 0 & 0 & 0\\
        \hline
C6H & 1 & 0 & 0 & 0 & 0 & 0 & 0 & 0 & 0 & 0 & 0 & 0 & 0 & 0 & 0 & 0 & 0\\
        \hline
D6 & 0 & 0 & 0 & 0 & 0 & 0 & 0 & 0 & 0 & 0 & 0 & 0 & 0 & 0 & 0 & 0 & 0\\
        \hline
C6V & 0 & 6 & 0 & 0 & 0 & 0 & 0 & 0 & 0 & 0 & 0 & 0 & 0 & 0 & 0 & 0 & 0\\
        \hline
D3H & 7 & 0 & 0 & 0 & 0 & 0 & 0 & 0 & 0 & 0 & 0 & 0 & 0 & 0 & 0 & 0 & 0\\
        \hline
D6H & 0 & 0 & 1 & 0 & 0 & 0 & 0 & 0 & 0 & 0 & 0 & 1 & 0 & 0 & 0 & 0 & 0\\
        \hline
T & 0 & 0 & 0 & 1 & 0 & 0 & 0 & 0 & 0 & 0 & 0 & 0 & 0 & 0 & 0 & 0 & 0\\
        \hline
TH & 0 & 0 & 0 & 8 & 0 & 0 & 0 & 1 & 0 & 0 & 0 & 0 & 0 & 0 & 0 & 0 & 0\\
        \hline
O & 0 & 0 & 0 & 2 & 0 & 0 & 0 & 0 & 0 & 0 & 0 & 0 & 0 & 0 & 0 & 0 & 0\\
        \hline
TD & 0 & 0 & 0 & 0 & 0 & 0 & 0 & 0 & 0 & 0 & 0 & 0 & 0 & 0 & 0 & 0 & 0\\
        \hline
OH & 0 & 0 & 0 & 84 & 0 & 0 & 0 & 28 & 0 & 0 & 0 & 0 & 0 & 0 & 0 & 0 & 0\\
        \hline
TOTAL & 72 & 110 & 26 & 200 & 0 & 3 & 0 & 62 & 5 & 1 & 0 & 1 & 0 & 0 & 0 & 4 & 0\\
        \hline
{\footnotesize Original } & 88 & 164 & 32 & 349 & 0 & 3 & 0 & 200 & 6 & 1 & 0 & 1 & 0 & 0 & 0 & 6 & 0\\
        \hline
        \end{tabular}
}
\newpage
\setlength{\oddsidemargin}{-0.3cm}
{\small
        \begin{tabular}{*{17}{|r}|}
        \hline
{\footnotesize AB2X6 } & Z=18 & \hspace*{0.7em}19 & \hspace*{0.7em}20 & \hspace*{0.7em}21 & \hspace*{0.7em}22 & \hspace*{0.7em}23 & \hspace*{0.7em}24 & \hspace*{0.7em}25 & \hspace*{0.7em}26 & \hspace*{0.7em}27 & \hspace*{0.7em}28 & \hspace*{0.7em}29 & \hspace*{0.7em}30 & \hspace*{0.7em}31 & \hspace*{0.7em}32 & {\scriptsize TOTAL }\\
        \hline
        \hline
C1 & 0 & 0 & 0 & 0 & 0 & 0 & 0 & 0 & 0 & 0 & 0 & 0 & 0 & 0 & 0 & 1\\
        \hline
CI & 0 & 0 & 0 & 0 & 0 & 0 & 0 & 0 & 0 & 0 & 0 & 0 & 0 & 0 & 0 & 12\\
        \hline
C2 & 0 & 0 & 0 & 0 & 0 & 0 & 0 & 0 & 0 & 0 & 0 & 0 & 0 & 0 & 0 & 10\\
        \hline
CS & 0 & 0 & 0 & 0 & 0 & 0 & 0 & 0 & 0 & 0 & 0 & 0 & 0 & 0 & 0 & 2\\
        \hline
C2H & 0 & 0 & 0 & 0 & 0 & 0 & 0 & 0 & 0 & 0 & 0 & 0 & 0 & 0 & 0 & 90\\
        \hline
D2 & 0 & 0 & 0 & 0 & 0 & 0 & 0 & 0 & 0 & 0 & 0 & 0 & 0 & 0 & 0 & 20\\
        \hline
C2V & 0 & 0 & 0 & 0 & 0 & 0 & 1 & 0 & 0 & 0 & 0 & 0 & 0 & 0 & 0 & 17\\
        \hline
D2H & 0 & 0 & 0 & 0 & 0 & 0 & 0 & 0 & 0 & 0 & 0 & 0 & 0 & 0 & 0 & 52\\
        \hline
C4 & 0 & 0 & 0 & 0 & 0 & 0 & 0 & 0 & 0 & 0 & 0 & 0 & 0 & 0 & 0 & 0\\
        \hline
S4 & 0 & 0 & 0 & 0 & 0 & 0 & 0 & 0 & 0 & 0 & 0 & 0 & 0 & 0 & 0 & 0\\
        \hline
C4H & 0 & 0 & 0 & 0 & 0 & 0 & 0 & 0 & 0 & 0 & 0 & 0 & 0 & 0 & 0 & 6\\
        \hline
D4 & 0 & 0 & 0 & 0 & 0 & 0 & 0 & 0 & 0 & 0 & 0 & 0 & 0 & 0 & 0 & 5\\
        \hline
C4V & 0 & 0 & 0 & 0 & 0 & 0 & 0 & 0 & 0 & 0 & 0 & 0 & 0 & 0 & 0 & 1\\
        \hline
D2D & 0 & 0 & 0 & 0 & 0 & 0 & 0 & 0 & 0 & 0 & 0 & 0 & 0 & 0 & 0 & 4\\
        \hline
D4H & 0 & 0 & 0 & 0 & 0 & 0 & 0 & 0 & 0 & 0 & 0 & 0 & 0 & 0 & 0 & 40\\
        \hline
C3 & 0 & 0 & 0 & 0 & 0 & 0 & 0 & 0 & 0 & 0 & 0 & 0 & 0 & 0 & 0 & 2\\
        \hline
C3I & 0 & 0 & 0 & 0 & 0 & 0 & 0 & 0 & 0 & 0 & 0 & 0 & 0 & 0 & 0 & 10\\
        \hline
D3 & 0 & 0 & 0 & 0 & 0 & 0 & 0 & 0 & 0 & 0 & 0 & 0 & 0 & 0 & 0 & 19\\
        \hline
C3V & 0 & 0 & 0 & 0 & 0 & 0 & 0 & 0 & 0 & 0 & 0 & 0 & 0 & 0 & 0 & 6\\
        \hline
D3D & 0 & 0 & 0 & 0 & 0 & 0 & 0 & 0 & 0 & 0 & 0 & 0 & 0 & 0 & 0 & 45\\
        \hline
C6 & 0 & 0 & 0 & 0 & 0 & 0 & 0 & 0 & 0 & 0 & 0 & 0 & 0 & 0 & 0 & 2\\
        \hline
C3H & 0 & 0 & 0 & 0 & 0 & 0 & 0 & 0 & 0 & 0 & 0 & 0 & 0 & 0 & 0 & 1\\
        \hline
C6H & 0 & 0 & 0 & 0 & 0 & 0 & 0 & 0 & 0 & 0 & 0 & 0 & 0 & 0 & 0 & 1\\
        \hline
D6 & 0 & 0 & 0 & 0 & 0 & 0 & 0 & 0 & 0 & 0 & 0 & 0 & 0 & 0 & 0 & 0\\
        \hline
C6V & 0 & 0 & 0 & 0 & 0 & 0 & 0 & 0 & 0 & 0 & 0 & 0 & 0 & 0 & 0 & 6\\
        \hline
D3H & 0 & 0 & 0 & 0 & 0 & 0 & 0 & 0 & 0 & 0 & 0 & 0 & 0 & 0 & 0 & 7\\
        \hline
D6H & 0 & 0 & 0 & 0 & 0 & 0 & 0 & 0 & 0 & 0 & 0 & 0 & 0 & 0 & 0 & 2\\
        \hline
T & 0 & 0 & 0 & 0 & 0 & 0 & 0 & 0 & 0 & 0 & 0 & 0 & 0 & 0 & 0 & 1\\
        \hline
TH & 0 & 0 & 0 & 0 & 0 & 0 & 0 & 0 & 0 & 0 & 0 & 0 & 0 & 0 & 0 & 9\\
        \hline
O & 0 & 0 & 0 & 0 & 0 & 0 & 0 & 0 & 0 & 0 & 0 & 0 & 0 & 0 & 0 & 2\\
        \hline
TD & 0 & 0 & 0 & 0 & 0 & 0 & 0 & 0 & 0 & 0 & 0 & 0 & 0 & 0 & 0 & 0\\
        \hline
OH & 0 & 0 & 0 & 0 & 0 & 0 & 0 & 0 & 0 & 0 & 0 & 0 & 0 & 0 & 0 & 112\\
        \hline
TOTAL & 0 & 0 & 0 & 0 & 0 & 0 & 1 & 0 & 0 & 0 & 0 & 0 & 0 & 0 & 0 & 485\\
        \hline
{\footnotesize Original } & 0 & 0 & 0 & 0 & 0 & 0 & 4 & 0 & 0 & 0 & 0 & 0 & 0 & 0 & 0 & 854\\
        \hline
        \end{tabular}
}
\newpage
\setlength{\oddsidemargin}{2.54cm}

\newpage

\newpage

\newpage

\newpage

\newpage

\newpage

\newpage

\newpage

\newpage

\newpage

\newpage
\setlength{\oddsidemargin}{-0.1cm}
ANX=ABX2\\
{\small
        \begin{tabular}{*{18}{|r}|}
        \hline
ABX2 & Z=1 & 2 & \hspace*{0.9em}3 & 4 & \hspace*{0.9em}5 & \hspace*{0.9em}6 & \hspace*{0.9em}7 & \hspace*{0.9em}8 & \hspace*{0.9em}9 & \hspace*{0.7em}10 & \hspace*{0.7em}11 & \hspace*{0.7em}12 & \hspace*{0.7em}13 & \hspace*{0.7em}14 & \hspace*{0.7em}15 & \hspace*{0.7em}16 & \hspace*{0.7em}17\\
        \hline
        \hline
C1 & 0 & 0 & 0 & 2 & 0 & 0 & 0 & 0 & 0 & 0 & 0 & 0 & 0 & 0 & 0 & 0 & 0\\
        \hline
CI & 1 & 1 & 0 & 3 & 0 & 1 & 0 & 0 & 0 & 0 & 0 & 0 & 0 & 0 & 0 & 0 & 0\\
        \hline
C2 & 0 & 0 & 0 & 1 & 0 & 0 & 0 & 0 & 0 & 0 & 0 & 0 & 0 & 0 & 0 & 0 & 0\\
        \hline
CS & 0 & 1 & 0 & 0 & 0 & 0 & 0 & 2 & 0 & 0 & 0 & 0 & 0 & 0 & 0 & 1 & 0\\
        \hline
C2H & 1 & 10 & 0 & 28 & 0 & 0 & 0 & 8 & 0 & 0 & 0 & 1 & 0 & 0 & 0 & 4 & 0\\
        \hline
D2 & 0 & 0 & 0 & 2 & 0 & 0 & 0 & 1 & 0 & 0 & 0 & 0 & 0 & 0 & 0 & 0 & 0\\
        \hline
C2V & 0 & 7 & 0 & 15 & 0 & 0 & 0 & 0 & 0 & 0 & 0 & 0 & 0 & 0 & 0 & 0 & 0\\
        \hline
D2H & 3 & 7 & 0 & 37 & 0 & 1 & 0 & 5 & 0 & 0 & 0 & 0 & 0 & 0 & 0 & 4 & 0\\
        \hline
C4 & 0 & 0 & 0 & 0 & 0 & 0 & 0 & 1 & 0 & 0 & 0 & 0 & 0 & 0 & 0 & 0 & 0\\
        \hline
S4 & 0 & 2 & 0 & 0 & 0 & 0 & 0 & 0 & 0 & 0 & 0 & 0 & 0 & 0 & 0 & 0 & 0\\
        \hline
C4H & 2 & 0 & 0 & 0 & 0 & 0 & 0 & 0 & 0 & 0 & 0 & 0 & 0 & 0 & 0 & 1 & 0\\
        \hline
D4 & 0 & 0 & 0 & 2 & 0 & 0 & 0 & 0 & 0 & 0 & 0 & 0 & 0 & 0 & 0 & 1 & 0\\
        \hline
C4V & 0 & 0 & 0 & 2 & 0 & 0 & 0 & 0 & 0 & 0 & 0 & 0 & 0 & 0 & 0 & 0 & 0\\
        \hline
D2D & 0 & 2 & 0 & 42 & 0 & 0 & 0 & 0 & 0 & 0 & 0 & 0 & 0 & 0 & 0 & 0 & 0\\
        \hline
D4H & 3 & 13 & 0 & 25 & 0 & 0 & 0 & 1 & 0 & 0 & 0 & 0 & 0 & 0 & 0 & 0 & 0\\
        \hline
C3 & 0 & 0 & 2 & 0 & 0 & 0 & 0 & 0 & 0 & 0 & 0 & 0 & 0 & 0 & 0 & 0 & 0\\
        \hline
C3I & 3 & 0 & 1 & 0 & 0 & 0 & 0 & 0 & 0 & 0 & 0 & 0 & 0 & 0 & 0 & 0 & 0\\
        \hline
D3 & 3 & 0 & 3 & 0 & 0 & 0 & 0 & 0 & 0 & 0 & 0 & 0 & 0 & 0 & 0 & 0 & 0\\
        \hline
C3V & 3 & 0 & 6 & 0 & 0 & 0 & 0 & 0 & 0 & 0 & 0 & 0 & 0 & 0 & 0 & 0 & 0\\
        \hline
D3D & 22 & 5 & 93 & 2 & 0 & 8 & 0 & 0 & 0 & 0 & 0 & 0 & 0 & 0 & 0 & 0 & 0\\
        \hline
C6 & 0 & 0 & 0 & 0 & 0 & 0 & 0 & 0 & 0 & 0 & 0 & 0 & 0 & 0 & 0 & 0 & 0\\
        \hline
C3H & 0 & 0 & 0 & 0 & 0 & 0 & 0 & 0 & 0 & 0 & 0 & 0 & 0 & 0 & 0 & 0 & 0\\
        \hline
C6H & 0 & 0 & 0 & 0 & 0 & 0 & 0 & 0 & 0 & 0 & 0 & 0 & 0 & 0 & 0 & 0 & 0\\
        \hline
D6 & 2 & 0 & 1 & 0 & 0 & 1 & 0 & 0 & 0 & 0 & 0 & 0 & 0 & 0 & 0 & 0 & 0\\
        \hline
C6V & 0 & 0 & 0 & 0 & 0 & 0 & 0 & 0 & 0 & 0 & 0 & 0 & 0 & 0 & 0 & 0 & 0\\
        \hline
D3H & 4 & 4 & 0 & 0 & 0 & 0 & 0 & 0 & 0 & 0 & 0 & 0 & 0 & 0 & 0 & 0 & 0\\
        \hline
D6H & 1 & 19 & 0 & 0 & 0 & 1 & 0 & 1 & 0 & 0 & 0 & 0 & 0 & 0 & 0 & 0 & 0\\
        \hline
T & 0 & 0 & 0 & 0 & 0 & 0 & 0 & 0 & 0 & 0 & 0 & 0 & 0 & 0 & 0 & 0 & 0\\
        \hline
TH & 0 & 0 & 0 & 0 & 0 & 0 & 0 & 0 & 0 & 0 & 0 & 0 & 0 & 0 & 0 & 0 & 0\\
        \hline
OH & 0 & 0 & 0 & 0 & 0 & 0 & 0 & 0 & 0 & 0 & 0 & 0 & 0 & 0 & 0 & 1 & 0\\
        \hline
TD & 0 & 0 & 0 & 0 & 0 & 0 & 0 & 1 & 0 & 0 & 0 & 0 & 0 & 0 & 0 & 0 & 0\\
        \hline
OH & 5 & 0 & 0 & 1 & 0 & 0 & 0 & 7 & 0 & 0 & 0 & 0 & 0 & 0 & 0 & 3 & 0\\
        \hline
TOTAL & 53 & 71 & 106 & 162 & 0 & 12 & 0 & 27 & 0 & 0 & 0 & 1 & 0 & 0 & 0 & 15 & 0\\
        \hline
{\footnotesize Original } & 72 & 107 & 155 & 273 & 0 & 13 & 0 & 31 & 0 & 0 & 0 & 1 & 0 & 0 & 0 & 28 & 0\\
        \hline
        \end{tabular}
}
\newpage
\setlength{\oddsidemargin}{-0.3cm}
{\small
        \begin{tabular}{*{17}{|r}|}
        \hline
ABX2 & Z=18 & \hspace*{0.7em}19 & \hspace*{0.7em}20 & \hspace*{0.7em}21 & \hspace*{0.7em}22 & \hspace*{0.7em}23 & \hspace*{0.7em}24 & \hspace*{0.7em}25 & \hspace*{0.7em}26 & \hspace*{0.7em}27 & \hspace*{0.7em}28 & \hspace*{0.7em}29 & \hspace*{0.7em}30 & \hspace*{0.7em}31 & \hspace*{0.7em}32 & {\scriptsize TOTAL }\\
        \hline
        \hline
C1 & 0 & 0 & 0 & 0 & 0 & 0 & 0 & 0 & 0 & 0 & 0 & 0 & 0 & 0 & 0 & 2\\
        \hline
CI & 0 & 0 & 0 & 0 & 0 & 0 & 0 & 0 & 0 & 0 & 0 & 0 & 0 & 0 & 0 & 6\\
        \hline
C2 & 0 & 0 & 0 & 0 & 0 & 0 & 0 & 0 & 0 & 0 & 0 & 0 & 0 & 0 & 0 & 1\\
        \hline
CS & 0 & 0 & 0 & 0 & 0 & 0 & 0 & 0 & 0 & 0 & 0 & 0 & 0 & 0 & 0 & 4\\
        \hline
C2H & 0 & 0 & 0 & 0 & 0 & 0 & 0 & 0 & 0 & 0 & 0 & 0 & 0 & 0 & 1 & 53\\
        \hline
D2 & 0 & 0 & 0 & 0 & 0 & 0 & 0 & 0 & 0 & 0 & 0 & 0 & 0 & 0 & 0 & 3\\
        \hline
C2V & 0 & 0 & 0 & 0 & 0 & 0 & 0 & 0 & 0 & 0 & 0 & 0 & 0 & 0 & 0 & 22\\
        \hline
D2H & 0 & 0 & 0 & 0 & 0 & 0 & 0 & 0 & 0 & 0 & 0 & 0 & 0 & 0 & 1 & 58\\
        \hline
C4 & 0 & 0 & 0 & 0 & 0 & 0 & 0 & 0 & 0 & 0 & 0 & 0 & 0 & 0 & 0 & 1\\
        \hline
S4 & 0 & 0 & 0 & 0 & 0 & 0 & 0 & 0 & 0 & 0 & 0 & 0 & 0 & 0 & 0 & 2\\
        \hline
C4H & 0 & 0 & 0 & 0 & 0 & 0 & 0 & 0 & 0 & 0 & 0 & 0 & 0 & 0 & 0 & 3\\
        \hline
D4 & 0 & 0 & 0 & 0 & 0 & 0 & 0 & 0 & 0 & 0 & 0 & 0 & 0 & 0 & 0 & 3\\
        \hline
C4V & 0 & 0 & 0 & 0 & 0 & 0 & 0 & 0 & 0 & 0 & 0 & 0 & 0 & 0 & 0 & 2\\
        \hline
D2D & 0 & 0 & 0 & 0 & 0 & 0 & 0 & 0 & 0 & 0 & 0 & 0 & 0 & 0 & 0 & 44\\
        \hline
D4H & 0 & 0 & 0 & 0 & 0 & 0 & 0 & 0 & 0 & 0 & 0 & 0 & 0 & 0 & 0 & 42\\
        \hline
C3 & 0 & 0 & 0 & 0 & 0 & 0 & 0 & 0 & 0 & 0 & 0 & 0 & 0 & 0 & 0 & 2\\
        \hline
C3I & 1 & 0 & 0 & 0 & 0 & 0 & 0 & 0 & 0 & 0 & 0 & 0 & 0 & 0 & 0 & 5\\
        \hline
D3 & 0 & 0 & 0 & 0 & 0 & 0 & 0 & 0 & 0 & 0 & 0 & 0 & 0 & 0 & 0 & 6\\
        \hline
C3V & 1 & 0 & 0 & 0 & 0 & 0 & 0 & 0 & 0 & 0 & 0 & 0 & 0 & 0 & 0 & 10\\
        \hline
D3D & 4 & 0 & 0 & 0 & 0 & 0 & 0 & 0 & 0 & 0 & 0 & 0 & 0 & 0 & 0 & 134\\
        \hline
C6 & 0 & 0 & 0 & 0 & 0 & 0 & 0 & 0 & 0 & 0 & 0 & 0 & 0 & 0 & 0 & 0\\
        \hline
C3H & 0 & 0 & 0 & 0 & 0 & 0 & 0 & 0 & 0 & 0 & 0 & 0 & 0 & 0 & 0 & 0\\
        \hline
C6H & 0 & 0 & 0 & 0 & 0 & 0 & 0 & 0 & 0 & 0 & 0 & 0 & 0 & 0 & 0 & 0\\
        \hline
D6 & 0 & 0 & 0 & 0 & 0 & 0 & 0 & 0 & 0 & 0 & 0 & 0 & 0 & 0 & 0 & 4\\
        \hline
C6V & 0 & 0 & 0 & 0 & 0 & 0 & 0 & 0 & 0 & 0 & 0 & 0 & 0 & 0 & 0 & 0\\
        \hline
D3H & 0 & 0 & 0 & 0 & 0 & 0 & 0 & 0 & 0 & 0 & 0 & 0 & 0 & 0 & 0 & 8\\
        \hline
D6H & 0 & 0 & 0 & 0 & 0 & 0 & 0 & 0 & 0 & 0 & 0 & 0 & 0 & 0 & 0 & 22\\
        \hline
T & 0 & 0 & 0 & 0 & 0 & 0 & 0 & 0 & 0 & 0 & 0 & 0 & 0 & 0 & 0 & 0\\
        \hline
TH & 0 & 0 & 0 & 0 & 0 & 0 & 0 & 0 & 0 & 0 & 0 & 0 & 0 & 0 & 0 & 0\\
        \hline
OH & 0 & 0 & 0 & 0 & 0 & 0 & 0 & 0 & 0 & 0 & 0 & 0 & 0 & 0 & 0 & 1\\
        \hline
TD & 0 & 0 & 0 & 0 & 0 & 0 & 0 & 0 & 0 & 0 & 0 & 0 & 0 & 0 & 0 & 1\\
        \hline
OH & 0 & 0 & 0 & 0 & 0 & 0 & 0 & 0 & 0 & 0 & 0 & 0 & 0 & 0 & 1 & 17\\
        \hline
TOTAL & 6 & 0 & 0 & 0 & 0 & 0 & 0 & 0 & 0 & 0 & 0 & 0 & 0 & 0 & 3 & 456\\
        \hline
{\footnotesize Original } & 6 & 0 & 0 & 0 & 0 & 0 & 0 & 0 & 0 & 0 & 0 & 0 & 0 & 0 & 4 & 690\\
        \hline
        \end{tabular}
}
\newpage
\setlength{\oddsidemargin}{2.54cm}

\newpage

\newpage

\newpage

\newpage

\newpage

\newpage

\newpage

\newpage

\newpage

\newpage
\setlength{\oddsidemargin}{-0.1cm}
ANX=ABC2X6\\
{\small
        \begin{tabular}{*{18}{|r}|}
        \hline
{\footnotesize ABC2X6 } & Z=1 & 2 & \hspace*{0.9em}3 & 4 & \hspace*{0.9em}5 & \hspace*{0.9em}6 & \hspace*{0.9em}7 & \hspace*{0.9em}8 & \hspace*{0.9em}9 & \hspace*{0.7em}10 & \hspace*{0.7em}11 & \hspace*{0.7em}12 & \hspace*{0.7em}13 & \hspace*{0.7em}14 & \hspace*{0.7em}15 & \hspace*{0.7em}16 & \hspace*{0.7em}17\\
        \hline
        \hline
C1 & 0 & 0 & 0 & 0 & 0 & 0 & 0 & 0 & 0 & 0 & 0 & 0 & 0 & 0 & 0 & 0 & 0\\
        \hline
CI & 0 & 3 & 0 & 1 & 0 & 0 & 0 & 0 & 0 & 0 & 0 & 1 & 0 & 0 & 0 & 0 & 0\\
        \hline
C2 & 0 & 3 & 0 & 0 & 0 & 0 & 0 & 0 & 0 & 0 & 0 & 0 & 0 & 0 & 0 & 0 & 0\\
        \hline
CS & 0 & 0 & 0 & 1 & 0 & 0 & 0 & 0 & 0 & 0 & 0 & 0 & 0 & 0 & 0 & 0 & 0\\
        \hline
C2H & 3 & 22 & 0 & 34 & 0 & 0 & 0 & 6 & 0 & 0 & 0 & 0 & 0 & 0 & 0 & 1 & 0\\
        \hline
D2 & 0 & 0 & 0 & 2 & 0 & 0 & 0 & 0 & 0 & 0 & 0 & 0 & 0 & 0 & 0 & 0 & 0\\
        \hline
C2V & 0 & 5 & 0 & 3 & 0 & 0 & 0 & 1 & 0 & 0 & 0 & 0 & 0 & 0 & 0 & 0 & 0\\
        \hline
D2H & 1 & 6 & 0 & 12 & 0 & 0 & 0 & 6 & 0 & 0 & 0 & 0 & 0 & 0 & 0 & 2 & 0\\
        \hline
C4 & 0 & 0 & 0 & 0 & 0 & 0 & 0 & 0 & 0 & 0 & 0 & 0 & 0 & 0 & 0 & 0 & 0\\
        \hline
S4 & 0 & 0 & 0 & 0 & 0 & 0 & 0 & 1 & 0 & 0 & 0 & 0 & 0 & 0 & 0 & 0 & 0\\
        \hline
C4H & 0 & 4 & 0 & 0 & 0 & 0 & 0 & 0 & 0 & 0 & 0 & 0 & 0 & 0 & 0 & 2 & 0\\
        \hline
D4 & 0 & 0 & 0 & 0 & 0 & 0 & 0 & 0 & 0 & 0 & 0 & 0 & 0 & 0 & 0 & 0 & 0\\
        \hline
C4V & 0 & 1 & 0 & 0 & 0 & 0 & 0 & 0 & 0 & 0 & 0 & 0 & 0 & 0 & 0 & 0 & 0\\
        \hline
D2D & 0 & 0 & 0 & 1 & 0 & 0 & 0 & 1 & 0 & 0 & 0 & 0 & 0 & 0 & 0 & 0 & 0\\
        \hline
D4H & 5 & 7 & 0 & 4 & 0 & 0 & 0 & 0 & 0 & 0 & 0 & 0 & 0 & 0 & 0 & 2 & 0\\
        \hline
C3 & 0 & 0 & 1 & 0 & 0 & 0 & 0 & 0 & 0 & 0 & 0 & 0 & 0 & 0 & 0 & 0 & 0\\
        \hline
C3I & 1 & 1 & 5 & 1 & 0 & 1 & 0 & 0 & 0 & 0 & 0 & 0 & 0 & 0 & 0 & 0 & 0\\
        \hline
D3 & 0 & 0 & 2 & 0 & 0 & 0 & 0 & 0 & 0 & 0 & 0 & 0 & 0 & 0 & 0 & 0 & 0\\
        \hline
C3V & 0 & 0 & 2 & 0 & 0 & 0 & 0 & 0 & 0 & 0 & 0 & 0 & 0 & 0 & 0 & 0 & 0\\
        \hline
D3D & 2 & 3 & 5 & 1 & 0 & 4 & 0 & 0 & 0 & 0 & 0 & 1 & 0 & 0 & 0 & 0 & 0\\
        \hline
C6 & 0 & 0 & 0 & 0 & 0 & 0 & 0 & 0 & 0 & 0 & 0 & 0 & 0 & 0 & 0 & 0 & 0\\
        \hline
C3H & 0 & 0 & 0 & 0 & 0 & 0 & 0 & 0 & 0 & 0 & 0 & 0 & 0 & 0 & 0 & 0 & 0\\
        \hline
C6H & 0 & 0 & 0 & 0 & 0 & 0 & 0 & 0 & 0 & 0 & 0 & 0 & 0 & 0 & 0 & 0 & 0\\
        \hline
D6 & 0 & 0 & 0 & 0 & 0 & 0 & 0 & 0 & 0 & 0 & 0 & 0 & 0 & 0 & 0 & 0 & 0\\
        \hline
C6V & 0 & 1 & 0 & 0 & 0 & 0 & 0 & 0 & 0 & 0 & 0 & 0 & 0 & 0 & 0 & 0 & 0\\
        \hline
D3H & 0 & 0 & 0 & 0 & 0 & 0 & 0 & 0 & 0 & 0 & 0 & 0 & 0 & 0 & 0 & 0 & 0\\
        \hline
D6H & 0 & 1 & 0 & 1 & 0 & 0 & 0 & 1 & 0 & 0 & 0 & 0 & 0 & 0 & 0 & 0 & 0\\
        \hline
T & 0 & 0 & 0 & 0 & 0 & 0 & 0 & 0 & 0 & 0 & 0 & 0 & 0 & 0 & 0 & 0 & 0\\
        \hline
TH & 1 & 0 & 0 & 5 & 0 & 0 & 0 & 1 & 0 & 0 & 0 & 0 & 0 & 0 & 0 & 0 & 0\\
        \hline
O & 0 & 0 & 0 & 0 & 0 & 0 & 0 & 0 & 0 & 0 & 0 & 0 & 0 & 0 & 0 & 0 & 0\\
        \hline
TD & 0 & 0 & 0 & 0 & 0 & 0 & 0 & 0 & 0 & 0 & 0 & 0 & 0 & 0 & 0 & 0 & 0\\
        \hline
OH & 5 & 2 & 0 & 57 & 0 & 0 & 0 & 1 & 0 & 0 & 0 & 0 & 0 & 0 & 0 & 0 & 0\\
        \hline
TOTAL & 18 & 59 & 15 & 123 & 0 & 5 & 0 & 18 & 0 & 0 & 0 & 2 & 0 & 0 & 0 & 7 & 0\\
        \hline
{\footnotesize Original } & 58 & 87 & 50 & 373 & 0 & 11 & 0 & 39 & 0 & 1 & 0 & 4 & 0 & 0 & 0 & 16 & 0\\
        \hline
        \end{tabular}
}
\newpage
\setlength{\oddsidemargin}{-0.3cm}
{\small
        \begin{tabular}{*{17}{|r}|}
        \hline
{\footnotesize ABC2X6 } & Z=18 & \hspace*{0.7em}19 & \hspace*{0.7em}20 & \hspace*{0.7em}21 & \hspace*{0.7em}22 & \hspace*{0.7em}23 & \hspace*{0.7em}24 & \hspace*{0.7em}25 & \hspace*{0.7em}26 & \hspace*{0.7em}27 & \hspace*{0.7em}28 & \hspace*{0.7em}29 & \hspace*{0.7em}30 & \hspace*{0.7em}31 & \hspace*{0.7em}32 & {\scriptsize TOTAL }\\
        \hline
        \hline
C1 & 0 & 0 & 0 & 0 & 0 & 0 & 0 & 0 & 0 & 0 & 0 & 0 & 0 & 0 & 0 & 0\\
        \hline
CI & 0 & 0 & 0 & 0 & 0 & 0 & 0 & 0 & 0 & 0 & 0 & 0 & 0 & 0 & 0 & 5\\
        \hline
C2 & 0 & 0 & 0 & 0 & 0 & 0 & 0 & 0 & 0 & 0 & 0 & 0 & 0 & 0 & 0 & 3\\
        \hline
CS & 0 & 0 & 0 & 0 & 0 & 0 & 0 & 0 & 0 & 0 & 0 & 0 & 0 & 0 & 0 & 1\\
        \hline
C2H & 0 & 0 & 0 & 0 & 0 & 0 & 0 & 0 & 0 & 0 & 0 & 0 & 0 & 0 & 0 & 66\\
        \hline
D2 & 0 & 0 & 0 & 0 & 0 & 0 & 0 & 0 & 0 & 0 & 0 & 0 & 0 & 0 & 0 & 2\\
        \hline
C2V & 0 & 0 & 0 & 0 & 0 & 0 & 0 & 0 & 0 & 0 & 0 & 0 & 0 & 0 & 1 & 10\\
        \hline
D2H & 0 & 0 & 0 & 0 & 0 & 0 & 0 & 0 & 0 & 0 & 0 & 0 & 0 & 0 & 0 & 27\\
        \hline
C4 & 0 & 0 & 0 & 0 & 0 & 0 & 0 & 0 & 0 & 0 & 0 & 0 & 0 & 0 & 0 & 0\\
        \hline
S4 & 0 & 0 & 0 & 0 & 0 & 0 & 0 & 0 & 0 & 0 & 0 & 0 & 0 & 0 & 0 & 1\\
        \hline
C4H & 0 & 0 & 0 & 0 & 0 & 0 & 0 & 0 & 0 & 0 & 0 & 0 & 0 & 0 & 0 & 6\\
        \hline
D4 & 0 & 0 & 0 & 0 & 0 & 0 & 0 & 0 & 0 & 0 & 0 & 0 & 0 & 0 & 0 & 0\\
        \hline
C4V & 0 & 0 & 0 & 0 & 0 & 0 & 0 & 0 & 0 & 0 & 0 & 0 & 0 & 0 & 0 & 1\\
        \hline
D2D & 0 & 0 & 0 & 0 & 0 & 0 & 0 & 0 & 0 & 0 & 0 & 0 & 0 & 0 & 0 & 2\\
        \hline
D4H & 0 & 0 & 0 & 0 & 0 & 0 & 0 & 0 & 0 & 0 & 0 & 0 & 0 & 0 & 0 & 18\\
        \hline
C3 & 0 & 0 & 0 & 0 & 0 & 0 & 0 & 0 & 0 & 0 & 0 & 0 & 0 & 0 & 0 & 1\\
        \hline
C3I & 1 & 0 & 0 & 0 & 0 & 0 & 0 & 0 & 0 & 0 & 0 & 0 & 0 & 0 & 0 & 10\\
        \hline
D3 & 0 & 0 & 0 & 0 & 0 & 0 & 0 & 0 & 0 & 0 & 0 & 0 & 0 & 0 & 0 & 2\\
        \hline
C3V & 0 & 0 & 0 & 0 & 0 & 0 & 0 & 0 & 0 & 0 & 0 & 0 & 0 & 0 & 0 & 2\\
        \hline
D3D & 0 & 0 & 0 & 0 & 0 & 0 & 0 & 0 & 0 & 0 & 0 & 0 & 0 & 0 & 0 & 16\\
        \hline
C6 & 0 & 0 & 0 & 0 & 0 & 0 & 0 & 0 & 0 & 0 & 0 & 0 & 0 & 0 & 0 & 0\\
        \hline
C3H & 0 & 0 & 0 & 0 & 0 & 0 & 0 & 0 & 0 & 0 & 0 & 0 & 0 & 0 & 0 & 0\\
        \hline
C6H & 0 & 0 & 0 & 0 & 0 & 0 & 0 & 0 & 0 & 0 & 0 & 0 & 0 & 0 & 0 & 0\\
        \hline
D6 & 0 & 0 & 0 & 0 & 0 & 0 & 0 & 0 & 0 & 0 & 0 & 0 & 0 & 0 & 0 & 0\\
        \hline
C6V & 0 & 0 & 0 & 0 & 0 & 0 & 0 & 0 & 0 & 0 & 0 & 0 & 0 & 0 & 0 & 1\\
        \hline
D3H & 0 & 0 & 0 & 0 & 0 & 0 & 0 & 0 & 0 & 0 & 0 & 0 & 0 & 0 & 0 & 0\\
        \hline
D6H & 0 & 0 & 0 & 0 & 0 & 0 & 0 & 0 & 0 & 0 & 0 & 0 & 0 & 0 & 0 & 3\\
        \hline
T & 0 & 0 & 0 & 0 & 0 & 0 & 0 & 0 & 0 & 0 & 0 & 0 & 0 & 0 & 0 & 0\\
        \hline
TH & 0 & 0 & 0 & 0 & 0 & 0 & 0 & 0 & 0 & 0 & 0 & 0 & 0 & 0 & 0 & 7\\
        \hline
O & 0 & 0 & 0 & 0 & 0 & 0 & 0 & 0 & 0 & 0 & 0 & 0 & 0 & 0 & 0 & 0\\
        \hline
TD & 0 & 0 & 0 & 0 & 0 & 0 & 0 & 0 & 0 & 0 & 0 & 0 & 0 & 0 & 0 & 0\\
        \hline
OH & 0 & 0 & 0 & 0 & 0 & 0 & 0 & 0 & 0 & 0 & 0 & 0 & 0 & 0 & 0 & 65\\
        \hline
TOTAL & 1 & 0 & 0 & 0 & 0 & 0 & 0 & 0 & 0 & 0 & 0 & 0 & 0 & 0 & 1 & 249\\
        \hline
{\footnotesize Original } & 1 & 0 & 0 & 0 & 0 & 0 & 0 & 0 & 0 & 0 & 0 & 0 & 0 & 0 & 1 & 641\\
        \hline
        \end{tabular}
}
\newpage
\setlength{\oddsidemargin}{2.54cm}

\newpage

\newpage

\newpage

\newpage

\newpage

\newpage

\newpage

\newpage

\newpage

\newpage

\newpage
\setlength{\oddsidemargin}{-0.1cm}
ANX=NO2\\
{\small
        \begin{tabular}{*{18}{|r}|}
        \hline
NO2 & Z=1 & 2 & \hspace*{0.9em}3 & 4 & \hspace*{0.9em}5 & \hspace*{0.9em}6 & \hspace*{0.9em}7 & \hspace*{0.9em}8 & \hspace*{0.9em}9 & \hspace*{0.7em}10 & \hspace*{0.7em}11 & \hspace*{0.7em}12 & \hspace*{0.7em}13 & \hspace*{0.7em}14 & \hspace*{0.7em}15 & \hspace*{0.7em}16 & \hspace*{0.7em}17\\
        \hline
        \hline
C1 & 0 & 0 & 0 & 0 & 0 & 0 & 0 & 1 & 0 & 0 & 0 & 0 & 0 & 0 & 0 & 0 & 0\\
        \hline
CI & 0 & 0 & 0 & 0 & 0 & 0 & 0 & 0 & 0 & 0 & 0 & 0 & 0 & 0 & 0 & 0 & 0\\
        \hline
C2 & 1 & 1 & 0 & 3 & 0 & 0 & 0 & 0 & 0 & 0 & 0 & 0 & 0 & 0 & 0 & 0 & 0\\
        \hline
CS & 0 & 1 & 0 & 1 & 0 & 0 & 0 & 0 & 0 & 0 & 0 & 0 & 0 & 0 & 0 & 0 & 0\\
        \hline
C2H & 0 & 5 & 0 & 12 & 0 & 3 & 0 & 2 & 0 & 0 & 0 & 0 & 0 & 0 & 0 & 0 & 0\\
        \hline
D2 & 0 & 1 & 0 & 0 & 0 & 0 & 0 & 2 & 0 & 0 & 0 & 0 & 0 & 0 & 0 & 0 & 0\\
        \hline
C2V & 0 & 1 & 0 & 3 & 0 & 0 & 0 & 3 & 0 & 0 & 0 & 1 & 0 & 0 & 0 & 0 & 0\\
        \hline
D2H & 1 & 17 & 1 & 23 & 0 & 0 & 0 & 10 & 0 & 0 & 0 & 3 & 0 & 0 & 0 & 3 & 0\\
        \hline
C4 & 0 & 2 & 0 & 0 & 0 & 0 & 0 & 0 & 0 & 0 & 0 & 0 & 0 & 0 & 0 & 0 & 0\\
        \hline
S4 & 0 & 0 & 0 & 0 & 0 & 0 & 0 & 0 & 0 & 0 & 0 & 0 & 0 & 0 & 0 & 0 & 0\\
        \hline
C4H & 0 & 9 & 0 & 0 & 0 & 0 & 0 & 0 & 0 & 0 & 0 & 0 & 0 & 0 & 0 & 0 & 0\\
        \hline
D4 & 0 & 0 & 0 & 1 & 0 & 0 & 0 & 2 & 0 & 0 & 0 & 0 & 0 & 0 & 0 & 0 & 0\\
        \hline
C4V & 0 & 1 & 0 & 2 & 0 & 0 & 0 & 0 & 0 & 0 & 0 & 0 & 0 & 0 & 0 & 0 & 0\\
        \hline
D2D & 0 & 0 & 0 & 3 & 0 & 0 & 0 & 0 & 0 & 0 & 0 & 0 & 0 & 0 & 0 & 0 & 0\\
        \hline
D4H & 3 & 30 & 0 & 26 & 0 & 1 & 0 & 1 & 0 & 0 & 0 & 0 & 0 & 0 & 0 & 1 & 0\\
        \hline
C3 & 1 & 0 & 0 & 0 & 0 & 0 & 0 & 0 & 0 & 0 & 0 & 0 & 0 & 0 & 0 & 0 & 0\\
        \hline
C3I & 0 & 0 & 2 & 0 & 0 & 0 & 0 & 0 & 0 & 0 & 0 & 0 & 0 & 0 & 0 & 0 & 0\\
        \hline
D3 & 0 & 0 & 1 & 0 & 0 & 0 & 0 & 0 & 0 & 0 & 0 & 0 & 0 & 0 & 0 & 0 & 0\\
        \hline
C3V & 2 & 0 & 1 & 0 & 0 & 0 & 0 & 0 & 0 & 0 & 0 & 0 & 0 & 0 & 0 & 0 & 0\\
        \hline
D3D & 10 & 3 & 7 & 0 & 0 & 1 & 0 & 0 & 0 & 0 & 0 & 0 & 0 & 0 & 0 & 0 & 0\\
        \hline
C6 & 0 & 0 & 0 & 0 & 0 & 0 & 0 & 0 & 0 & 0 & 0 & 0 & 0 & 0 & 0 & 0 & 0\\
        \hline
C3H & 2 & 0 & 0 & 0 & 0 & 0 & 0 & 0 & 0 & 0 & 0 & 0 & 0 & 0 & 0 & 0 & 0\\
        \hline
C6H & 0 & 0 & 0 & 0 & 0 & 0 & 0 & 0 & 0 & 0 & 0 & 0 & 0 & 0 & 0 & 0 & 0\\
        \hline
D6 & 1 & 0 & 7 & 0 & 0 & 1 & 0 & 0 & 0 & 0 & 0 & 0 & 0 & 0 & 0 & 0 & 0\\
        \hline
C6V & 0 & 0 & 0 & 0 & 0 & 0 & 0 & 0 & 0 & 0 & 0 & 0 & 0 & 0 & 0 & 0 & 0\\
        \hline
D3H & 2 & 0 & 4 & 1 & 0 & 1 & 0 & 0 & 0 & 0 & 0 & 1 & 0 & 0 & 0 & 0 & 0\\
        \hline
D6H & 49 & 12 & 0 & 6 & 0 & 0 & 0 & 1 & 0 & 0 & 0 & 1 & 0 & 0 & 0 & 0 & 0\\
        \hline
T & 0 & 0 & 0 & 0 & 0 & 1 & 0 & 0 & 0 & 0 & 0 & 0 & 0 & 0 & 0 & 0 & 0\\
        \hline
TH & 0 & 1 & 0 & 12 & 0 & 0 & 0 & 0 & 0 & 0 & 0 & 0 & 0 & 0 & 0 & 1 & 0\\
        \hline
O & 0 & 0 & 0 & 3 & 0 & 0 & 0 & 0 & 0 & 0 & 0 & 0 & 0 & 0 & 0 & 0 & 0\\
        \hline
TD & 1 & 0 & 0 & 0 & 0 & 0 & 0 & 0 & 0 & 0 & 0 & 0 & 0 & 0 & 0 & 0 & 0\\
        \hline
OH & 0 & 0 & 0 & 17 & 0 & 0 & 0 & 3 & 0 & 0 & 0 & 0 & 0 & 0 & 0 & 0 & 0\\
        \hline
TOTAL & 73 & 84 & 23 & 113 & 0 & 8 & 0 & 25 & 0 & 0 & 0 & 6 & 0 & 0 & 0 & 5 & 0\\
        \hline
{\footnotesize Original } & 85 & 118 & 26 & 172 & 0 & 9 & 0 & 34 & 0 & 0 & 0 & 6 & 0 & 0 & 0 & 7 & 0\\
        \hline
        \end{tabular}
}
\newpage
\setlength{\oddsidemargin}{-0.3cm}
{\small
        \begin{tabular}{*{17}{|r}|}
        \hline
NO2 & Z=18 & \hspace*{0.7em}19 & \hspace*{0.7em}20 & \hspace*{0.7em}21 & \hspace*{0.7em}22 & \hspace*{0.7em}23 & \hspace*{0.7em}24 & \hspace*{0.7em}25 & \hspace*{0.7em}26 & \hspace*{0.7em}27 & \hspace*{0.7em}28 & \hspace*{0.7em}29 & \hspace*{0.7em}30 & \hspace*{0.7em}31 & \hspace*{0.7em}32 & {\scriptsize TOTAL }\\
        \hline
        \hline
C1 & 0 & 0 & 0 & 0 & 0 & 0 & 0 & 0 & 0 & 0 & 0 & 0 & 0 & 0 & 0 & 1\\
        \hline
CI & 0 & 0 & 0 & 0 & 0 & 0 & 0 & 0 & 0 & 0 & 0 & 0 & 0 & 0 & 0 & 0\\
        \hline
C2 & 0 & 0 & 0 & 0 & 0 & 0 & 0 & 0 & 0 & 0 & 0 & 0 & 0 & 0 & 0 & 5\\
        \hline
CS & 0 & 0 & 0 & 0 & 0 & 0 & 0 & 0 & 0 & 0 & 0 & 0 & 0 & 0 & 0 & 2\\
        \hline
C2H & 0 & 0 & 0 & 0 & 0 & 0 & 0 & 0 & 0 & 0 & 0 & 0 & 0 & 0 & 0 & 22\\
        \hline
D2 & 0 & 0 & 0 & 0 & 0 & 0 & 0 & 0 & 0 & 0 & 0 & 0 & 0 & 0 & 0 & 3\\
        \hline
C2V & 0 & 0 & 0 & 0 & 0 & 0 & 0 & 0 & 0 & 0 & 0 & 0 & 0 & 0 & 0 & 8\\
        \hline
D2H & 1 & 0 & 0 & 0 & 0 & 0 & 0 & 0 & 0 & 0 & 0 & 0 & 0 & 0 & 0 & 59\\
        \hline
C4 & 0 & 0 & 0 & 0 & 0 & 0 & 0 & 0 & 0 & 0 & 0 & 0 & 0 & 0 & 0 & 2\\
        \hline
S4 & 0 & 0 & 0 & 0 & 0 & 0 & 0 & 0 & 0 & 0 & 0 & 0 & 0 & 0 & 0 & 0\\
        \hline
C4H & 0 & 0 & 0 & 0 & 0 & 0 & 0 & 0 & 0 & 0 & 0 & 0 & 0 & 0 & 0 & 9\\
        \hline
D4 & 0 & 0 & 0 & 0 & 0 & 0 & 0 & 0 & 0 & 0 & 0 & 0 & 0 & 0 & 0 & 3\\
        \hline
C4V & 0 & 0 & 0 & 0 & 0 & 0 & 0 & 0 & 0 & 0 & 0 & 0 & 0 & 0 & 0 & 3\\
        \hline
D2D & 0 & 0 & 0 & 0 & 0 & 0 & 0 & 0 & 0 & 0 & 0 & 0 & 0 & 0 & 0 & 3\\
        \hline
D4H & 0 & 0 & 0 & 0 & 0 & 0 & 0 & 0 & 0 & 0 & 0 & 0 & 0 & 0 & 0 & 62\\
        \hline
C3 & 0 & 0 & 0 & 0 & 0 & 0 & 0 & 0 & 0 & 0 & 0 & 0 & 0 & 0 & 0 & 1\\
        \hline
C3I & 0 & 0 & 0 & 0 & 0 & 0 & 0 & 0 & 0 & 0 & 0 & 0 & 0 & 0 & 0 & 2\\
        \hline
D3 & 0 & 0 & 0 & 0 & 0 & 0 & 0 & 0 & 0 & 0 & 0 & 0 & 0 & 0 & 0 & 1\\
        \hline
C3V & 0 & 0 & 0 & 0 & 0 & 0 & 0 & 0 & 0 & 0 & 0 & 0 & 0 & 0 & 0 & 3\\
        \hline
D3D & 0 & 0 & 0 & 0 & 0 & 0 & 0 & 0 & 0 & 0 & 0 & 0 & 0 & 0 & 0 & 21\\
        \hline
C6 & 0 & 0 & 0 & 0 & 0 & 0 & 0 & 0 & 0 & 0 & 0 & 0 & 0 & 0 & 0 & 0\\
        \hline
C3H & 0 & 0 & 0 & 0 & 0 & 0 & 0 & 0 & 0 & 0 & 0 & 0 & 0 & 0 & 0 & 2\\
        \hline
C6H & 0 & 0 & 0 & 0 & 0 & 0 & 0 & 0 & 0 & 0 & 0 & 0 & 0 & 0 & 0 & 0\\
        \hline
D6 & 0 & 0 & 0 & 0 & 0 & 0 & 0 & 0 & 0 & 0 & 0 & 0 & 0 & 0 & 0 & 9\\
        \hline
C6V & 0 & 0 & 0 & 0 & 0 & 0 & 0 & 0 & 0 & 0 & 0 & 0 & 0 & 0 & 0 & 0\\
        \hline
D3H & 0 & 0 & 0 & 0 & 0 & 0 & 0 & 0 & 0 & 0 & 0 & 0 & 0 & 0 & 0 & 9\\
        \hline
D6H & 0 & 0 & 1 & 0 & 0 & 0 & 0 & 0 & 0 & 0 & 0 & 0 & 0 & 0 & 0 & 70\\
        \hline
T & 0 & 0 & 0 & 0 & 0 & 0 & 0 & 0 & 0 & 0 & 0 & 0 & 0 & 0 & 0 & 1\\
        \hline
TH & 0 & 0 & 0 & 0 & 0 & 0 & 0 & 0 & 0 & 0 & 0 & 0 & 0 & 0 & 0 & 14\\
        \hline
O & 0 & 0 & 0 & 0 & 0 & 0 & 0 & 0 & 0 & 0 & 0 & 0 & 0 & 0 & 0 & 3\\
        \hline
TD & 0 & 0 & 0 & 0 & 0 & 0 & 0 & 0 & 0 & 0 & 0 & 0 & 0 & 0 & 0 & 1\\
        \hline
OH & 0 & 0 & 0 & 0 & 0 & 0 & 0 & 0 & 0 & 0 & 0 & 0 & 0 & 0 & 1 & 21\\
        \hline
TOTAL & 1 & 0 & 1 & 0 & 0 & 0 & 0 & 0 & 0 & 0 & 0 & 0 & 0 & 0 & 1 & 340\\
        \hline
{\footnotesize Original } & 1 & 0 & 1 & 0 & 0 & 0 & 0 & 0 & 0 & 0 & 0 & 0 & 0 & 0 & 1 & 460\\
        \hline
        \end{tabular}
}
\newpage
\setlength{\oddsidemargin}{2.54cm}

\newpage

\newpage

\newpage

\newpage

\newpage

\newpage

\newpage

\newpage

\newpage

\newpage

\newpage
\setlength{\oddsidemargin}{-0.1cm}
ANX=A2X3\\
{\small
        \begin{tabular}{*{18}{|r}|}
        \hline
A2X3 & Z=1 & 2 & \hspace*{0.9em}3 & 4 & \hspace*{0.9em}5 & \hspace*{0.9em}6 & \hspace*{0.9em}7 & \hspace*{0.9em}8 & \hspace*{0.9em}9 & \hspace*{0.7em}10 & \hspace*{0.7em}11 & \hspace*{0.7em}12 & \hspace*{0.7em}13 & \hspace*{0.7em}14 & \hspace*{0.7em}15 & \hspace*{0.7em}16 & \hspace*{0.7em}17\\
        \hline
        \hline
C1 & 0 & 0 & 0 & 0 & 0 & 0 & 0 & 0 & 0 & 0 & 0 & 1 & 0 & 0 & 0 & 0 & 0\\
        \hline
CI & 0 & 0 & 0 & 1 & 0 & 0 & 0 & 0 & 0 & 0 & 0 & 0 & 0 & 0 & 0 & 0 & 0\\
        \hline
C2 & 0 & 1 & 0 & 0 & 0 & 0 & 0 & 0 & 0 & 0 & 0 & 0 & 0 & 0 & 0 & 0 & 0\\
        \hline
CS & 1 & 0 & 1 & 5 & 0 & 0 & 0 & 0 & 0 & 0 & 0 & 0 & 0 & 0 & 0 & 0 & 0\\
        \hline
C2H & 0 & 5 & 1 & 10 & 0 & 9 & 0 & 2 & 0 & 0 & 0 & 0 & 0 & 0 & 0 & 0 & 0\\
        \hline
D2 & 0 & 0 & 0 & 1 & 0 & 0 & 0 & 0 & 0 & 0 & 0 & 0 & 0 & 0 & 0 & 0 & 0\\
        \hline
C2V & 0 & 0 & 0 & 5 & 0 & 0 & 0 & 2 & 0 & 0 & 0 & 0 & 0 & 0 & 0 & 0 & 0\\
        \hline
D2H & 1 & 0 & 0 & 21 & 0 & 0 & 0 & 1 & 0 & 0 & 0 & 0 & 0 & 0 & 0 & 2 & 0\\
        \hline
C4 & 0 & 0 & 0 & 0 & 0 & 0 & 0 & 0 & 0 & 0 & 0 & 0 & 0 & 0 & 0 & 0 & 0\\
        \hline
S4 & 0 & 0 & 0 & 0 & 0 & 0 & 0 & 0 & 0 & 0 & 0 & 0 & 0 & 0 & 0 & 0 & 0\\
        \hline
C4H & 0 & 0 & 0 & 0 & 0 & 0 & 0 & 0 & 0 & 0 & 0 & 0 & 0 & 0 & 0 & 0 & 0\\
        \hline
D4 & 0 & 0 & 0 & 0 & 0 & 0 & 0 & 0 & 0 & 0 & 0 & 0 & 0 & 0 & 0 & 0 & 0\\
        \hline
C4V & 0 & 0 & 0 & 0 & 0 & 0 & 0 & 0 & 0 & 0 & 0 & 0 & 0 & 0 & 0 & 0 & 0\\
        \hline
D2D & 0 & 0 & 0 & 1 & 0 & 0 & 0 & 2 & 0 & 0 & 0 & 0 & 0 & 0 & 0 & 0 & 0\\
        \hline
D4H & 1 & 0 & 0 & 0 & 0 & 0 & 0 & 1 & 0 & 0 & 0 & 0 & 0 & 0 & 0 & 1 & 0\\
        \hline
C3 & 0 & 0 & 1 & 0 & 0 & 1 & 0 & 0 & 0 & 0 & 0 & 0 & 0 & 0 & 0 & 0 & 0\\
        \hline
C3I & 0 & 0 & 1 & 0 & 0 & 1 & 0 & 0 & 0 & 0 & 0 & 0 & 0 & 0 & 0 & 0 & 0\\
        \hline
D3 & 6 & 0 & 0 & 0 & 0 & 0 & 0 & 0 & 0 & 0 & 0 & 0 & 0 & 0 & 0 & 0 & 0\\
        \hline
C3V & 0 & 0 & 2 & 0 & 0 & 0 & 0 & 0 & 0 & 0 & 0 & 0 & 0 & 0 & 0 & 0 & 0\\
        \hline
D3D & 9 & 11 & 4 & 3 & 0 & 12 & 0 & 0 & 0 & 0 & 0 & 1 & 0 & 0 & 0 & 0 & 0\\
        \hline
C6 & 0 & 0 & 0 & 0 & 0 & 1 & 0 & 0 & 0 & 0 & 0 & 0 & 0 & 0 & 0 & 0 & 0\\
        \hline
C3H & 0 & 0 & 0 & 0 & 0 & 0 & 0 & 0 & 0 & 0 & 0 & 0 & 0 & 0 & 0 & 0 & 0\\
        \hline
C6H & 0 & 1 & 0 & 0 & 0 & 0 & 0 & 0 & 0 & 0 & 0 & 0 & 0 & 0 & 0 & 0 & 0\\
        \hline
D6 & 0 & 0 & 0 & 0 & 0 & 0 & 0 & 0 & 0 & 0 & 0 & 0 & 0 & 0 & 0 & 0 & 0\\
        \hline
C6V & 0 & 1 & 0 & 0 & 0 & 0 & 0 & 0 & 0 & 0 & 0 & 0 & 0 & 0 & 0 & 0 & 0\\
        \hline
D3H & 0 & 0 & 0 & 0 & 0 & 0 & 0 & 0 & 0 & 0 & 0 & 0 & 0 & 0 & 0 & 0 & 0\\
        \hline
D6H & 1 & 3 & 0 & 0 & 0 & 0 & 0 & 0 & 0 & 0 & 0 & 0 & 0 & 0 & 0 & 0 & 0\\
        \hline
T & 0 & 0 & 0 & 0 & 0 & 0 & 0 & 0 & 0 & 0 & 0 & 1 & 0 & 0 & 0 & 15 & 0\\
        \hline
TH & 0 & 0 & 0 & 0 & 0 & 0 & 0 & 0 & 0 & 0 & 0 & 0 & 0 & 0 & 0 & 17 & 0\\
        \hline
O & 0 & 0 & 0 & 0 & 0 & 0 & 0 & 0 & 0 & 0 & 0 & 0 & 0 & 0 & 0 & 0 & 0\\
        \hline
TD & 1 & 0 & 0 & 2 & 0 & 0 & 0 & 2 & 0 & 0 & 0 & 0 & 0 & 0 & 0 & 0 & 0\\
        \hline
OH & 1 & 4 & 0 & 3 & 0 & 0 & 0 & 2 & 0 & 0 & 0 & 0 & 0 & 0 & 0 & 2 & 0\\
        \hline
TOTAL & 21 & 26 & 10 & 52 & 0 & 24 & 0 & 12 & 0 & 0 & 0 & 3 & 0 & 0 & 0 & 37 & 0\\
        \hline
{\footnotesize Original } & 26 & 54 & 16 & 78 & 0 & 128 & 0 & 14 & 0 & 0 & 0 & 4 & 0 & 0 & 0 & 101 & 0\\
        \hline
        \end{tabular}
}
\newpage
\setlength{\oddsidemargin}{-0.3cm}
{\small
        \begin{tabular}{*{17}{|r}|}
        \hline
A2X3 & Z=18 & \hspace*{0.7em}19 & \hspace*{0.7em}20 & \hspace*{0.7em}21 & \hspace*{0.7em}22 & \hspace*{0.7em}23 & \hspace*{0.7em}24 & \hspace*{0.7em}25 & \hspace*{0.7em}26 & \hspace*{0.7em}27 & \hspace*{0.7em}28 & \hspace*{0.7em}29 & \hspace*{0.7em}30 & \hspace*{0.7em}31 & \hspace*{0.7em}32 & {\scriptsize TOTAL }\\
        \hline
        \hline
C1 & 0 & 0 & 0 & 0 & 0 & 0 & 0 & 0 & 0 & 0 & 0 & 0 & 0 & 0 & 0 & 1\\
        \hline
CI & 0 & 0 & 0 & 0 & 0 & 0 & 0 & 0 & 0 & 0 & 0 & 0 & 0 & 0 & 0 & 1\\
        \hline
C2 & 0 & 0 & 0 & 0 & 0 & 0 & 0 & 0 & 0 & 0 & 0 & 0 & 0 & 0 & 0 & 1\\
        \hline
CS & 0 & 0 & 0 & 0 & 0 & 0 & 0 & 0 & 0 & 0 & 0 & 0 & 0 & 0 & 0 & 7\\
        \hline
C2H & 0 & 0 & 0 & 0 & 0 & 0 & 0 & 0 & 0 & 0 & 0 & 0 & 0 & 0 & 0 & 27\\
        \hline
D2 & 0 & 0 & 0 & 0 & 0 & 0 & 0 & 0 & 0 & 0 & 0 & 0 & 0 & 0 & 0 & 1\\
        \hline
C2V & 0 & 0 & 0 & 0 & 0 & 0 & 0 & 0 & 0 & 0 & 0 & 0 & 0 & 0 & 0 & 7\\
        \hline
D2H & 0 & 0 & 0 & 0 & 0 & 0 & 0 & 0 & 0 & 0 & 0 & 0 & 0 & 0 & 0 & 25\\
        \hline
C4 & 0 & 0 & 0 & 0 & 0 & 0 & 0 & 0 & 0 & 0 & 0 & 0 & 0 & 0 & 0 & 0\\
        \hline
S4 & 0 & 0 & 0 & 0 & 0 & 0 & 0 & 0 & 0 & 0 & 0 & 0 & 0 & 0 & 0 & 0\\
        \hline
C4H & 0 & 0 & 0 & 0 & 0 & 0 & 0 & 0 & 0 & 0 & 0 & 0 & 0 & 0 & 0 & 0\\
        \hline
D4 & 0 & 0 & 0 & 0 & 0 & 0 & 0 & 0 & 0 & 0 & 0 & 0 & 0 & 0 & 0 & 0\\
        \hline
C4V & 0 & 0 & 0 & 0 & 0 & 0 & 0 & 0 & 0 & 0 & 0 & 0 & 0 & 0 & 0 & 0\\
        \hline
D2D & 0 & 0 & 0 & 0 & 0 & 0 & 0 & 0 & 0 & 0 & 0 & 0 & 0 & 0 & 0 & 3\\
        \hline
D4H & 0 & 0 & 0 & 0 & 0 & 0 & 0 & 0 & 0 & 0 & 0 & 0 & 0 & 0 & 0 & 3\\
        \hline
C3 & 0 & 0 & 0 & 0 & 0 & 0 & 0 & 0 & 0 & 0 & 0 & 0 & 0 & 0 & 0 & 2\\
        \hline
C3I & 0 & 0 & 0 & 0 & 0 & 0 & 0 & 0 & 0 & 0 & 0 & 0 & 0 & 0 & 0 & 2\\
        \hline
D3 & 0 & 0 & 0 & 0 & 0 & 0 & 0 & 0 & 0 & 0 & 0 & 0 & 0 & 0 & 0 & 6\\
        \hline
C3V & 0 & 0 & 0 & 0 & 0 & 0 & 0 & 0 & 0 & 0 & 0 & 0 & 0 & 0 & 0 & 2\\
        \hline
D3D & 0 & 0 & 0 & 0 & 0 & 0 & 0 & 0 & 0 & 0 & 0 & 0 & 0 & 0 & 0 & 40\\
        \hline
C6 & 0 & 0 & 0 & 0 & 0 & 0 & 0 & 0 & 0 & 0 & 0 & 0 & 0 & 0 & 0 & 1\\
        \hline
C3H & 0 & 0 & 0 & 0 & 0 & 0 & 0 & 0 & 0 & 0 & 0 & 0 & 0 & 0 & 0 & 0\\
        \hline
C6H & 0 & 0 & 0 & 0 & 0 & 0 & 0 & 0 & 0 & 0 & 0 & 0 & 0 & 0 & 0 & 1\\
        \hline
D6 & 0 & 0 & 0 & 0 & 0 & 0 & 0 & 0 & 0 & 0 & 0 & 0 & 0 & 0 & 0 & 0\\
        \hline
C6V & 0 & 0 & 0 & 0 & 0 & 0 & 0 & 0 & 0 & 0 & 0 & 0 & 0 & 0 & 0 & 1\\
        \hline
D3H & 0 & 0 & 0 & 0 & 0 & 0 & 0 & 0 & 0 & 0 & 0 & 0 & 0 & 0 & 0 & 0\\
        \hline
D6H & 0 & 0 & 0 & 0 & 0 & 0 & 0 & 0 & 0 & 0 & 0 & 0 & 0 & 0 & 0 & 4\\
        \hline
T & 0 & 0 & 0 & 0 & 0 & 0 & 0 & 0 & 0 & 0 & 0 & 0 & 0 & 0 & 0 & 16\\
        \hline
TH & 0 & 0 & 0 & 0 & 0 & 0 & 0 & 0 & 0 & 0 & 0 & 0 & 0 & 0 & 0 & 17\\
        \hline
O & 0 & 0 & 0 & 0 & 0 & 0 & 0 & 0 & 0 & 0 & 0 & 0 & 0 & 0 & 0 & 0\\
        \hline
TD & 0 & 0 & 0 & 0 & 0 & 0 & 0 & 0 & 0 & 0 & 0 & 0 & 0 & 0 & 0 & 5\\
        \hline
OH & 0 & 0 & 0 & 0 & 0 & 0 & 0 & 0 & 0 & 0 & 0 & 0 & 0 & 0 & 0 & 12\\
        \hline
TOTAL & 0 & 0 & 0 & 0 & 0 & 0 & 0 & 0 & 0 & 0 & 0 & 0 & 0 & 0 & 0 & 185\\
        \hline
{\footnotesize Original } & 0 & 0 & 0 & 0 & 0 & 0 & 0 & 0 & 0 & 0 & 0 & 0 & 0 & 0 & 0 & 421\\
        \hline
        \end{tabular}
} 
\newpage
\setlength{\oddsidemargin}{2.54cm}

\newpage

\newpage

\newpage

\newpage

\newpage

\newpage

\newpage

\newpage
\setlength{\oddsidemargin}{-0.1cm}
ANX=NO\\
{\small
        \begin{tabular}{*{18}{|r}|}
        \hline
NO & Z=1 & 2 & \hspace*{0.9em}3 & 4 & \hspace*{0.9em}5 & \hspace*{0.9em}6 & \hspace*{0.9em}7 & \hspace*{0.9em}8 & \hspace*{0.9em}9 & \hspace*{0.7em}10 & \hspace*{0.7em}11 & \hspace*{0.7em}12 & \hspace*{0.7em}13 & \hspace*{0.7em}14 & \hspace*{0.7em}15 & \hspace*{0.7em}16 & \hspace*{0.7em}17\\
        \hline
        \hline
C1 & 0 & 0 & 0 & 0 & 0 & 0 & 0 & 1 & 0 & 0 & 0 & 0 & 0 & 0 & 0 & 0 & 0\\
        \hline
CI & 0 & 0 & 0 & 0 & 0 & 0 & 0 & 0 & 0 & 0 & 0 & 0 & 0 & 0 & 0 & 0 & 0\\
        \hline
C2 & 0 & 0 & 0 & 0 & 0 & 0 & 0 & 0 & 0 & 0 & 0 & 0 & 0 & 0 & 0 & 0 & 0\\
        \hline
CS & 0 & 0 & 0 & 0 & 0 & 0 & 0 & 0 & 0 & 0 & 0 & 0 & 0 & 0 & 0 & 0 & 0\\
        \hline
C2H & 1 & 0 & 0 & 8 & 0 & 0 & 0 & 2 & 0 & 1 & 0 & 1 & 0 & 0 & 0 & 0 & 0\\
        \hline
D2 & 0 & 0 & 0 & 0 & 0 & 1 & 0 & 2 & 0 & 0 & 0 & 0 & 0 & 0 & 0 & 0 & 0\\
        \hline
C2V & 1 & 0 & 0 & 0 & 0 & 0 & 0 & 1 & 0 & 0 & 0 & 1 & 0 & 0 & 0 & 0 & 0\\
        \hline
D2H & 0 & 1 & 0 & 48 & 0 & 0 & 0 & 3 & 0 & 0 & 0 & 2 & 0 & 0 & 0 & 1 & 0\\
        \hline
C4 & 1 & 0 & 0 & 0 & 0 & 0 & 0 & 0 & 0 & 0 & 0 & 0 & 0 & 0 & 0 & 0 & 0\\
        \hline
S4 & 0 & 0 & 0 & 0 & 0 & 0 & 0 & 0 & 0 & 0 & 0 & 0 & 0 & 0 & 0 & 0 & 0\\
        \hline
C4H & 0 & 0 & 0 & 0 & 0 & 0 & 0 & 0 & 0 & 0 & 0 & 0 & 0 & 0 & 0 & 0 & 0\\
        \hline
D4 & 0 & 0 & 0 & 0 & 0 & 0 & 0 & 0 & 0 & 0 & 0 & 0 & 0 & 0 & 0 & 0 & 0\\
        \hline
C4V & 0 & 3 & 0 & 0 & 0 & 0 & 0 & 0 & 0 & 0 & 0 & 0 & 0 & 0 & 0 & 0 & 0\\
        \hline
D2D & 0 & 0 & 0 & 0 & 0 & 0 & 0 & 1 & 0 & 0 & 0 & 0 & 0 & 0 & 0 & 0 & 0\\
        \hline
D4H & 0 & 5 & 0 & 2 & 0 & 0 & 0 & 3 & 0 & 0 & 0 & 1 & 0 & 0 & 0 & 1 & 03\\
        \hline
C3 & 1 & 0 & 1 & 0 & 0 & 0 & 0 & 0 & 0 & 0 & 0 & 0 & 0 & 0 & 0 & 0 & 0\\
        \hline
C3I & 0 & 0 & 0 & 0 & 0 & 0 & 0 & 0 & 0 & 0 & 0 & 0 & 0 & 0 & 0 & 0 & 0\\
        \hline
D3 & 1 & 0 & 1 & 0 & 0 & 0 & 0 & 0 & 0 & 0 & 0 & 0 & 0 & 0 & 0 & 0 & 0\\
        \hline
C3V & 0 & 0 & 2 & 0 & 0 & 2 & 0 & 0 & 0 & 0 & 0 & 0 & 0 & 0 & 0 & 0 & 0\\
        \hline
D3D & 3 & 1 & 7 & 0 & 0 & 1 & 0 & 0 & 0 & 0 & 0 & 0 & 0 & 0 & 0 & 0 & 0\\
        \hline
C6 & 0 & 0 & 0 & 0 & 0 & 0 & 0 & 0 & 0 & 0 & 0 & 0 & 0 & 0 & 0 & 0 & 0\\
        \hline
C3H & 0 & 0 & 0 & 0 & 0 & 0 & 0 & 0 & 0 & 0 & 0 & 0 & 0 & 0 & 0 & 0 & 0\\
        \hline
C6H & 0 & 0 & 0 & 0 & 0 & 0 & 0 & 0 & 0 & 0 & 0 & 0 & 0 & 0 & 0 & 0 & 0\\
        \hline
D6 & 0 & 0 & 1 & 0 & 0 & 0 & 0 & 0 & 0 & 0 & 0 & 0 & 0 & 0 & 0 & 0 & 0\\
        \hline
C6V & 2 & 0 & 0 & 0 & 0 & 1 & 0 & 1 & 0 & 0 & 0 & 0 & 0 & 0 & 0 & 0 & 0\\
        \hline
D3H & 1 & 0 & 2 & 0 & 0 & 3 & 0 & 0 & 0 & 0 & 0 & 0 & 0 & 0 & 0 & 0 & 0\\
        \hline
D6H & 3 & 12 & 1 & 7 & 0 & 0 & 0 & 0 & 0 & 0 & 0 & 0 & 0 & 0 & 0 & 0 & 0\\
        \hline
T & 2 & 0 & 0 & 8 & 0 & 0 & 0 & 0 & 0 & 0 & 0 & 0 & 0 & 0 & 0 & 0 & 0\\
        \hline
TH & 0 & 0 & 0 & 0 & 0 & 0 & 0 & 0 & 0 & 0 & 0 & 0 & 0 & 0 & 0 & 0 & 0\\
        \hline
O & 0 & 0 & 0 & 0 & 0 & 0 & 0 & 0 & 0 & 0 & 0 & 0 & 0 & 0 & 0 & 0 & 0\\
        \hline
TD & 1 & 1 & 0 & 1 & 0 & 0 & 0 & 0 & 0 & 0 & 0 & 0 & 0 & 0 & 0 & 0 & 0\\
        \hline
OH & 2 & 2 & 0 & 33 & 0 & 0 & 0 & 2 & 0 & 0 & 0 & 0 & 0 & 0 & 0 & 0 & 0\\
        \hline
TOTAL & 19 & 25 & 15 & 107 & 0 & 8 & 0 & 16 & 0 & 1 & 0 & 5 & 0 & 0 & 0 & 2 & 0\\
        \hline
{\footnotesize Original } & 21 & 35 & 16 & 214 & 0 & 10 & 0 & 16 & 0 & 1 & 0 & 6 & 0 & 0 & 0 & 2 & 0\\
        \hline
        \end{tabular}
}
\newpage
\setlength{\oddsidemargin}{-0.3cm}
{\small
        \begin{tabular}{*{17}{|r}|}
        \hline
NO & Z=18 & \hspace*{0.7em}19 & \hspace*{0.7em}20 & \hspace*{0.7em}21 & \hspace*{0.7em}22 & \hspace*{0.7em}23 & \hspace*{0.7em}24 & \hspace*{0.7em}25 & \hspace*{0.7em}26 & \hspace*{0.7em}27 & \hspace*{0.7em}28 & \hspace*{0.7em}29 & \hspace*{0.7em}30 & \hspace*{0.7em}31 & \hspace*{0.7em}32 & {\scriptsize TOTAL }\\
        \hline
        \hline
C1 & 0 & 0 & 0 & 0 & 0 & 0 & 0 & 0 & 0 & 0 & 0 & 0 & 0 & 0 & 0 & 1\\
        \hline
CI & 0 & 0 & 0 & 0 & 0 & 0 & 0 & 0 & 0 & 0 & 0 & 0 & 0 & 0 & 0 & 0\\
        \hline
C2 & 0 & 0 & 0 & 0 & 0 & 0 & 0 & 0 & 0 & 0 & 0 & 0 & 0 & 0 & 0 & 0\\
        \hline
CS & 0 & 0 & 0 & 0 & 0 & 0 & 0 & 0 & 0 & 0 & 0 & 0 & 0 & 0 & 0 & 0\\
        \hline
C2H & 0 & 0 & 0 & 0 & 0 & 0 & 0 & 0 & 0 & 0 & 0 & 0 & 0 & 0 & 0 & 13\\
        \hline
D2 & 0 & 0 & 0 & 0 & 0 & 0 & 0 & 0 & 0 & 0 & 0 & 0 & 0 & 0 & 0 & 3\\
        \hline
C2V & 0 & 0 & 0 & 0 & 0 & 0 & 1 & 0 & 0 & 0 & 0 & 0 & 0 & 0 & 0 & 4\\
        \hline
D2H & 0 & 0 & 1 & 0 & 0 & 0 & 0 & 0 & 0 & 0 & 0 & 0 & 0 & 0 & 0 & 56\\
        \hline
C4 & 0 & 0 & 0 & 0 & 0 & 0 & 0 & 0 & 0 & 0 & 0 & 0 & 0 & 0 & 0 & 1\\
        \hline
S4 & 0 & 0 & 0 & 0 & 0 & 0 & 0 & 0 & 0 & 0 & 0 & 0 & 0 & 0 & 0 & 0\\
        \hline
C4H & 0 & 0 & 0 & 0 & 0 & 0 & 0 & 0 & 0 & 0 & 0 & 0 & 0 & 0 & 0 & 0\\
        \hline
D4 & 0 & 0 & 0 & 0 & 0 & 0 & 0 & 0 & 0 & 0 & 0 & 0 & 0 & 0 & 0 & 0\\
        \hline
C4V & 0 & 0 & 0 & 0 & 0 & 0 & 0 & 0 & 0 & 0 & 0 & 0 & 0 & 0 & 0 & 3\\
        \hline
D2D & 0 & 0 & 0 & 0 & 0 & 0 & 0 & 0 & 0 & 0 & 0 & 0 & 0 & 0 & 0 & 1\\
        \hline
D4H & 0 & 0 & 0 & 0 & 0 & 0 & 0 & 0 & 0 & 0 & 0 & 0 & 0 & 0 & 1 & 13\\
        \hline
C3 & 0 & 0 & 0 & 0 & 0 & 0 & 0 & 0 & 0 & 0 & 0 & 0 & 0 & 0 & 0 & 2\\
        \hline
C3I & 0 & 0 & 0 & 0 & 0 & 0 & 0 & 0 & 0 & 0 & 0 & 0 & 0 & 0 & 0 & 0\\
        \hline
D3 & 0 & 0 & 0 & 0 & 0 & 0 & 0 & 0 & 0 & 0 & 0 & 0 & 0 & 0 & 0 & 2\\
        \hline
C3V & 0 & 0 & 0 & 0 & 0 & 0 & 0 & 0 & 0 & 0 & 0 & 0 & 0 & 0 & 0 & 4\\
        \hline
D3D & 0 & 0 & 0 & 0 & 0 & 0 & 0 & 0 & 0 & 0 & 0 & 0 & 0 & 0 & 0 & 12\\
        \hline
C6 & 0 & 0 & 0 & 0 & 0 & 0 & 0 & 0 & 0 & 0 & 0 & 0 & 0 & 0 & 0 & 0\\
        \hline
C3H & 0 & 0 & 0 & 0 & 0 & 0 & 0 & 0 & 0 & 0 & 0 & 0 & 0 & 0 & 0 & 0\\
        \hline
C6H & 0 & 0 & 0 & 0 & 0 & 0 & 0 & 0 & 0 & 0 & 0 & 0 & 0 & 0 & 0 & 0\\
        \hline
D6 & 0 & 0 & 0 & 0 & 0 & 0 & 0 & 0 & 0 & 0 & 0 & 0 & 0 & 0 & 0 & 1\\
        \hline
C6V & 0 & 0 & 0 & 0 & 0 & 0 & 0 & 0 & 0 & 0 & 0 & 0 & 0 & 0 & 0 & 4\\
        \hline
D3H & 0 & 0 & 0 & 0 & 0 & 0 & 0 & 0 & 0 & 0 & 0 & 0 & 0 & 0 & 0 & 6\\
        \hline
D6H & 0 & 0 & 0 & 0 & 0 & 0 & 0 & 0 & 0 & 0 & 0 & 0 & 0 & 0 & 0 & 23\\
        \hline
T & 0 & 0 & 0 & 0 & 0 & 0 & 0 & 0 & 0 & 0 & 0 & 0 & 0 & 0 & 1 & 11\\
        \hline
TH & 0 & 0 & 0 & 0 & 0 & 0 & 0 & 0 & 0 & 0 & 0 & 0 & 0 & 0 & 0 & 0\\
        \hline
O & 0 & 0 & 0 & 0 & 0 & 0 & 0 & 0 & 0 & 0 & 0 & 0 & 0 & 0 & 0 & 0\\
        \hline
TD & 0 & 0 & 0 & 0 & 0 & 0 & 0 & 0 & 0 & 0 & 0 & 0 & 0 & 0 & 0 & 3\\
        \hline
OH & 0 & 0 & 0 & 0 & 0 & 0 & 0 & 0 & 0 & 0 & 0 & 0 & 0 & 0 & 0 & 39\\
        \hline
TOTAL & 0 & 0 & 1 & 0 & 0 & 0 & 1 & 0 & 0 & 0 & 0 & 0 & 0 & 0 & 2 & 202\\
        \hline
{\footnotesize Original } & 0 & 0 & 1 & 0 & 0 & 0 & 1 & 0 & 0 & 0 & 0 & 0 & 0 & 0 & 2 & 325\\
        \hline
        \end{tabular}
}
\newpage
\setlength{\oddsidemargin}{2.54cm}

\newpage

\newpage

\newpage

\newpage

\newpage

\newpage

\newpage

\newpage

\newpage

\newpage

\newpage

\newpage
\setlength{\oddsidemargin}{-0.1cm}
ANX=AXY\\
{\small
        \begin{tabular}{*{18}{|r}|}
        \hline
AXY & Z=1 & 2 & \hspace*{0.9em}3 & 4 & \hspace*{0.9em}5 & \hspace*{0.9em}6 & \hspace*{0.9em}7 & \hspace*{0.9em}8 & \hspace*{0.9em}9 & \hspace*{0.7em}10 & \hspace*{0.7em}11 & \hspace*{0.7em}12 & \hspace*{0.7em}13 & \hspace*{0.7em}14 & \hspace*{0.7em}15 & \hspace*{0.7em}16 & \hspace*{0.7em}17\\
        \hline
        \hline
C1 & 0 & 0 & 0 & 0 & 0 & 0 & 0 & 0 & 0 & 0 & 0 & 0 & 0 & 0 & 0 & 0 & 0\\
        \hline
CI & 0 & 1 & 0 & 0 & 0 & 0 & 0 & 0 & 0 & 0 & 0 & 0 & 0 & 0 & 0 & 0 & 0\\
        \hline
C2 & 0 & 0 & 0 & 0 & 0 & 0 & 0 & 0 & 0 & 0 & 0 & 0 & 0 & 0 & 0 & 0 & 0\\
        \hline
CS & 0 & 0 & 0 & 0 & 0 & 0 & 0 & 0 & 0 & 0 & 0 & 0 & 0 & 0 & 0 & 0 & 0\\
        \hline
C2H & 0 & 6 & 0 & 13 & 0 & 0 & 0 & 2 & 0 & 0 & 0 & 1 & 0 & 0 & 0 & 0 & 0\\
        \hline
D2 & 0 & 0 & 0 & 2 & 0 & 0 & 0 & 0 & 0 & 0 & 0 & 0 & 0 & 0 & 0 & 0 & 0\\
        \hline
C2V & 0 & 2 & 0 & 2 & 0 & 0 & 0 & 0 & 0 & 0 & 0 & 0 & 0 & 0 & 0 & 1 & 0\\
        \hline
D2H & 0 & 22 & 0 & 29 & 0 & 0 & 0 & 4 & 0 & 0 & 0 & 1 & 0 & 0 & 0 & 2 & 0\\
        \hline
C4 & 0 & 0 & 0 & 0 & 0 & 0 & 0 & 0 & 0 & 0 & 0 & 0 & 0 & 0 & 0 & 0 & 0\\
        \hline
S4 & 0 & 0 & 0 & 1 & 0 & 0 & 0 & 0 & 0 & 0 & 0 & 0 & 0 & 0 & 0 & 0 & 0\\
        \hline
C4H & 1 & 0 & 0 & 0 & 0 & 0 & 0 & 0 & 0 & 0 & 0 & 0 & 0 & 0 & 0 & 0 & 0\\
        \hline
D4 & 0 & 0 & 0 & 0 & 0 & 0 & 0 & 0 & 0 & 0 & 0 & 0 & 0 & 0 & 0 & 0 & 0\\
        \hline
C4V & 0 & 0 & 0 & 0 & 0 & 0 & 0 & 0 & 0 & 0 & 0 & 0 & 0 & 0 & 0 & 0 & 0\\
        \hline
D2D & 0 & 2 & 0 & 0 & 0 & 0 & 0 & 0 & 0 & 0 & 0 & 0 & 0 & 0 & 0 & 0 & 0\\
        \hline
D4H & 1 & 69 & 0 & 0 & 0 & 0 & 0 & 1 & 0 & 0 & 0 & 0 & 0 & 0 & 0 & 0 & 0\\
        \hline
C3 & 1 & 0 & 0 & 0 & 0 & 0 & 0 & 0 & 0 & 0 & 0 & 0 & 0 & 0 & 0 & 0 & 0\\
        \hline
C3I & 0 & 0 & 0 & 0 & 0 & 2 & 0 & 0 & 0 & 0 & 0 & 0 & 0 & 0 & 0 & 0 & 0\\
        \hline
D3 & 0 & 0 & 0 & 0 & 0 & 0 & 0 & 0 & 0 & 0 & 0 & 0 & 0 & 0 & 0 & 0 & 0\\
        \hline
C3V & 1 & 0 & 0 & 0 & 0 & 0 & 0 & 0 & 0 & 0 & 0 & 0 & 0 & 0 & 0 & 0 & 0\\
        \hline
D3D & 2 & 7 & 0 & 0 & 0 & 5 & 0 & 0 & 0 & 0 & 0 & 0 & 0 & 0 & 0 & 0 & 04\\
        \hline
C6 & 0 & 0 & 0 & 0 & 0 & 0 & 0 & 0 & 0 & 0 & 0 & 0 & 0 & 0 & 0 & 0 & 0\\
        \hline
C3H & 0 & 0 & 1 & 0 & 0 & 0 & 0 & 0 & 0 & 0 & 0 & 0 & 0 & 0 & 0 & 0 & 0\\
        \hline
C6H & 0 & 0 & 0 & 0 & 0 & 0 & 0 & 0 & 0 & 0 & 0 & 0 & 0 & 0 & 0 & 0 & 0\\
        \hline
D6 & 0 & 0 & 0 & 0 & 0 & 0 & 0 & 0 & 0 & 0 & 0 & 0 & 0 & 0 & 0 & 0 & 0\\
        \hline
C6V & 0 & 2 & 0 & 0 & 0 & 0 & 0 & 0 & 0 & 0 & 0 & 0 & 0 & 0 & 0 & 0 & 0\\
        \hline
D3H & 1 & 0 & 0 & 0 & 0 & 0 & 0 & 0 & 0 & 0 & 0 & 0 & 0 & 0 & 0 & 0 & 0\\
        \hline
D6H & 0 & 1 & 1 & 1 & 0 & 0 & 0 & 0 & 0 & 0 & 0 & 0 & 0 & 0 & 0 & 0 & 0\\
        \hline
T & 2 & 0 & 0 & 2 & 0 & 0 & 0 & 0 & 0 & 0 & 0 & 0 & 0 & 0 & 0 & 0 & 0\\
        \hline
TH & 0 & 0 & 0 & 0 & 0 & 0 & 0 & 0 & 0 & 0 & 0 & 0 & 0 & 0 & 0 & 0 & 0\\
        \hline
O & 0 & 0 & 0 & 0 & 0 & 0 & 0 & 0 & 0 & 0 & 0 & 0 & 0 & 0 & 0 & 0 & 0\\
        \hline
TD & 0 & 0 & 0 & 2 & 0 & 0 & 0 & 0 & 0 & 0 & 0 & 0 & 0 & 0 & 0 & 1 & 0\\
        \hline
OH & 2 & 0 & 0 & 0 & 0 & 0 & 0 & 0 & 0 & 0 & 0 & 0 & 0 & 0 & 0 & 0 & 0\\
        \hline
TOTAL & 11 & 112 & 2 & 52 & 0 & 7 & 0 & 7 & 0 & 0 & 0 & 2 & 0 & 0 & 0 & 4 & 0\\
        \hline
{\footnotesize Original } & 12 & 173 & 2 & 90 & 0 & 8 & 0 & 11 & 0 & 0 & 0 & 2 & 0 & 0 & 0 & 4 & 0\\
        \hline
        \end{tabular}
}
\newpage
\setlength{\oddsidemargin}{-0.3cm}
{\small
        \begin{tabular}{*{17}{|r}|}
        \hline
AXY & Z=18 & \hspace*{0.7em}19 & \hspace*{0.7em}20 & \hspace*{0.7em}21 & \hspace*{0.7em}22 & \hspace*{0.7em}23 & \hspace*{0.7em}24 & \hspace*{0.7em}25 & \hspace*{0.7em}26 & \hspace*{0.7em}27 & \hspace*{0.7em}28 & \hspace*{0.7em}29 & \hspace*{0.7em}30 & \hspace*{0.7em}31 & \hspace*{0.7em}32 & {\scriptsize TOTAL }\\
        \hline
        \hline
C1 & 0 & 0 & 0 & 0 & 0 & 0 & 0 & 0 & 0 & 0 & 0 & 0 & 0 & 0 & 0 & 0\\
        \hline
CI & 0 & 0 & 0 & 0 & 0 & 0 & 0 & 0 & 0 & 0 & 0 & 0 & 0 & 0 & 0 & 1\\
        \hline
C2 & 0 & 0 & 0 & 0 & 0 & 0 & 0 & 0 & 0 & 0 & 0 & 0 & 0 & 0 & 0 & 0\\
        \hline
CS & 0 & 0 & 0 & 0 & 0 & 0 & 0 & 0 & 0 & 0 & 0 & 0 & 0 & 0 & 0 & 0\\
        \hline
C2H & 0 & 0 & 1 & 0 & 0 & 0 & 0 & 0 & 0 & 0 & 0 & 0 & 0 & 0 & 0 & 23\\
        \hline
D2 & 0 & 0 & 0 & 0 & 0 & 0 & 0 & 0 & 0 & 0 & 0 & 0 & 0 & 0 & 0 & 2\\
        \hline
C2V & 0 & 0 & 0 & 0 & 0 & 0 & 0 & 0 & 0 & 0 & 0 & 0 & 0 & 0 & 0 & 5\\
        \hline
D2H & 0 & 0 & 0 & 0 & 0 & 0 & 0 & 0 & 0 & 0 & 0 & 0 & 0 & 0 & 0 & 58\\
        \hline
C4 & 0 & 0 & 0 & 0 & 0 & 0 & 0 & 0 & 0 & 0 & 0 & 0 & 0 & 0 & 0 & 0\\
        \hline
S4 & 0 & 0 & 0 & 0 & 0 & 0 & 0 & 0 & 0 & 0 & 0 & 0 & 0 & 0 & 0 & 1\\
        \hline
C4H & 0 & 0 & 0 & 0 & 0 & 0 & 0 & 0 & 0 & 0 & 0 & 0 & 0 & 0 & 1 & 2\\
        \hline
D4 & 0 & 0 & 0 & 0 & 0 & 0 & 0 & 0 & 0 & 0 & 0 & 0 & 0 & 0 & 0 & 0\\
        \hline
C4V & 0 & 0 & 0 & 0 & 0 & 0 & 0 & 0 & 0 & 0 & 0 & 0 & 0 & 0 & 0 & 0\\
        \hline
D2D & 0 & 0 & 0 & 0 & 0 & 0 & 0 & 0 & 0 & 0 & 0 & 0 & 0 & 0 & 0 & 2\\
        \hline
D4H & 0 & 0 & 0 & 0 & 0 & 0 & 0 & 0 & 0 & 0 & 0 & 0 & 0 & 0 & 0 & 71\\
        \hline
C3 & 0 & 0 & 0 & 0 & 0 & 0 & 0 & 0 & 0 & 0 & 0 & 0 & 0 & 0 & 0 & 1\\
        \hline
C3I & 0 & 0 & 0 & 0 & 0 & 0 & 0 & 0 & 0 & 0 & 0 & 0 & 0 & 0 & 0 & 2\\
        \hline
D3 & 0 & 0 & 0 & 0 & 0 & 0 & 0 & 0 & 0 & 0 & 0 & 0 & 0 & 0 & 0 & 0\\
        \hline
C3V & 0 & 0 & 0 & 0 & 0 & 0 & 0 & 0 & 0 & 0 & 0 & 0 & 0 & 0 & 0 & 1\\
        \hline
D3D & 0 & 0 & 0 & 0 & 0 & 0 & 0 & 0 & 0 & 0 & 0 & 0 & 0 & 0 & 0 & 14\\
        \hline
C6 & 0 & 0 & 0 & 0 & 0 & 0 & 0 & 0 & 0 & 0 & 0 & 0 & 0 & 0 & 0 & 0\\
        \hline
C3H & 0 & 0 & 0 & 0 & 0 & 0 & 0 & 0 & 0 & 0 & 0 & 0 & 0 & 0 & 0 & 1\\
        \hline
C6H & 0 & 0 & 0 & 0 & 0 & 0 & 0 & 0 & 0 & 0 & 0 & 0 & 0 & 0 & 0 & 0\\
        \hline
D6 & 0 & 0 & 0 & 0 & 0 & 0 & 0 & 0 & 0 & 0 & 0 & 0 & 0 & 0 & 0 & 0\\
        \hline
C6V & 0 & 0 & 0 & 0 & 0 & 0 & 0 & 0 & 0 & 0 & 0 & 0 & 0 & 0 & 0 & 2\\
        \hline
D3H & 0 & 0 & 0 & 0 & 0 & 0 & 0 & 0 & 0 & 0 & 0 & 0 & 0 & 0 & 0 & 1\\
        \hline
D6H & 0 & 0 & 0 & 0 & 0 & 0 & 0 & 0 & 0 & 0 & 0 & 0 & 0 & 0 & 0 & 3\\
        \hline
T & 0 & 0 & 0 & 0 & 0 & 0 & 0 & 0 & 0 & 0 & 0 & 0 & 0 & 0 & 0 & 4\\
        \hline
TH & 0 & 0 & 0 & 0 & 0 & 0 & 0 & 0 & 0 & 0 & 0 & 0 & 0 & 0 & 0 & 0\\
        \hline
O & 0 & 0 & 0 & 0 & 0 & 0 & 0 & 0 & 0 & 0 & 0 & 0 & 0 & 0 & 0 & 0\\
        \hline
TD & 0 & 0 & 0 & 0 & 0 & 0 & 0 & 0 & 0 & 0 & 0 & 0 & 0 & 0 & 0 & 3\\
        \hline
OH & 0 & 0 & 0 & 0 & 0 & 0 & 0 & 0 & 0 & 0 & 0 & 0 & 0 & 0 & 0 & 2\\
        \hline
TOTAL & 0 & 0 & 1 & 0 & 0 & 0 & 0 & 0 & 0 & 0 & 0 & 0 & 0 & 0 & 1 & 199\\
        \hline
{\footnotesize Original } & 0 & 0 & 1 & 0 & 0 & 0 & 0 & 0 & 0 & 0 & 0 & 0 & 0 & 0 & 1 & 304\\
        \hline
        \end{tabular}
}
\newpage
\setlength{\oddsidemargin}{2.54cm}

\newpage

\newpage

\newpage

\newpage

\newpage

\newpage

\newpage

\newpage

\newpage

\newpage
\setlength{\oddsidemargin}{-0.1cm}
ANX=AB2X5\\
{\small
        \begin{tabular}{*{18}{|r}|}
        \hline
{\footnotesize AB2X5 } & Z=1 & 2 & \hspace*{0.9em}3 & 4 & \hspace*{0.9em}5 & \hspace*{0.9em}6 & \hspace*{0.9em}7 & \hspace*{0.9em}8 & \hspace*{0.9em}9 & \hspace*{0.7em}10 & \hspace*{0.7em}11 & \hspace*{0.7em}12 & \hspace*{0.7em}13 & \hspace*{0.7em}14 & \hspace*{0.7em}15 & \hspace*{0.7em}16 & \hspace*{0.7em}17\\
        \hline
        \hline
C1 & 1 & 2 & 0 & 0 & 0 & 0 & 0 & 0 & 0 & 0 & 0 & 0 & 0 & 0 & 0 & 0 & 0\\
        \hline
CI & 0 & 4 & 0 & 4 & 0 & 0 & 0 & 0 & 0 & 0 & 0 & 0 & 0 & 0 & 0 & 0 & 0\\
        \hline
C2 & 0 & 1 & 0 & 1 & 0 & 1 & 0 & 0 & 0 & 0 & 0 & 0 & 0 & 0 & 0 & 0 & 0\\
        \hline
CS & 0 & 0 & 0 & 1 & 0 & 0 & 0 & 0 & 0 & 0 & 0 & 0 & 0 & 0 & 0 & 0 & 0\\
        \hline
C2H & 2 & 4 & 0 & 39 & 0 & 3 & 0 & 7 & 0 & 0 & 0 & 2 & 0 & 0 & 0 & 3 & 0\\
        \hline
D2 & 0 & 0 & 0 & 5 & 0 & 0 & 0 & 1 & 0 & 0 & 0 & 0 & 0 & 0 & 0 & 0 & 0\\
        \hline
C2V & 0 & 2 & 0 & 7 & 0 & 0 & 0 & 3 & 0 & 0 & 0 & 0 & 0 & 0 & 0 & 1 & 0\\
        \hline
D2H & 2 & 5 & 0 & 54 & 0 & 0 & 0 & 7 & 0 & 0 & 0 & 0 & 0 & 0 & 0 & 1 & 0\\
        \hline
C4 & 0 & 0 & 0 & 0 & 0 & 0 & 0 & 0 & 0 & 0 & 0 & 0 & 0 & 0 & 0 & 0 & 0\\
        \hline
S4 & 0 & 0 & 0 & 0 & 0 & 0 & 0 & 0 & 0 & 0 & 0 & 0 & 0 & 0 & 0 & 0 & 0\\
        \hline
C4H & 0 & 0 & 0 & 0 & 0 & 0 & 0 & 0 & 0 & 0 & 0 & 0 & 0 & 0 & 0 & 0 & 0\\
        \hline
D4 & 0 & 0 & 0 & 1 & 0 & 0 & 0 & 0 & 0 & 0 & 0 & 0 & 0 & 0 & 0 & 0 & 0\\
        \hline
C4V & 0 & 0 & 0 & 1 & 0 & 0 & 0 & 0 & 0 & 0 & 0 & 0 & 0 & 0 & 0 & 0 & 0\\
        \hline
D2D & 0 & 0 & 0 & 0 & 0 & 0 & 0 & 1 & 0 & 0 & 0 & 0 & 0 & 0 & 0 & 0 & 0\\
        \hline
D4H & 1 & 2 & 0 & 8 & 0 & 0 & 0 & 2 & 0 & 0 & 0 & 0 & 0 & 0 & 0 & 0 & 0\\
        \hline
C3 & 1 & 0 & 0 & 0 & 0 & 0 & 0 & 0 & 0 & 0 & 0 & 0 & 0 & 0 & 0 & 0 & 0\\
        \hline
C3I & 0 & 0 & 0 & 0 & 0 & 0 & 0 & 0 & 0 & 0 & 0 & 0 & 0 & 0 & 0 & 0 & 0\\
        \hline
D3 & 1 & 0 & 0 & 0 & 0 & 0 & 0 & 0 & 0 & 0 & 0 & 0 & 0 & 0 & 0 & 0 & 0\\
        \hline
C3V & 0 & 1 & 0 & 0 & 0 & 0 & 0 & 0 & 0 & 0 & 0 & 0 & 0 & 0 & 0 & 0 & 0\\
        \hline
D3D & 0 & 0 & 0 & 0 & 0 & 0 & 0 & 0 & 0 & 0 & 0 & 0 & 0 & 0 & 0 & 0 & 0\\
        \hline
C6 & 0 & 1 & 0 & 0 & 0 & 2 & 0 & 0 & 0 & 0 & 0 & 0 & 0 & 0 & 0 & 0 & 0\\
        \hline
C3H & 0 & 0 & 0 & 0 & 0 & 0 & 0 & 0 & 0 & 0 & 0 & 0 & 0 & 0 & 0 & 0 & 0\\
        \hline
C6H & 0 & 0 & 0 & 0 & 0 & 0 & 0 & 0 & 0 & 0 & 0 & 0 & 0 & 0 & 0 & 0 & 0\\
        \hline
D6 & 0 & 0 & 0 & 0 & 0 & 0 & 0 & 0 & 0 & 0 & 0 & 0 & 0 & 0 & 0 & 0 & 0\\
        \hline
C6V & 0 & 0 & 0 & 0 & 0 & 0 & 0 & 0 & 0 & 0 & 0 & 0 & 0 & 0 & 0 & 0 & 0\\
        \hline
D3H & 0 & 0 & 0 & 0 & 0 & 0 & 0 & 0 & 0 & 0 & 0 & 0 & 0 & 0 & 0 & 0 & 0\\
        \hline
D6H & 0 & 2 & 0 & 0 & 0 & 0 & 0 & 0 & 0 & 0 & 0 & 0 & 0 & 0 & 0 & 0 & 0\\
        \hline
T & 0 & 0 & 0 & 0 & 0 & 0 & 0 & 0 & 0 & 0 & 0 & 0 & 0 & 0 & 0 & 0 & 0\\
        \hline
TH & 0 & 0 & 0 & 0 & 0 & 0 & 0 & 0 & 0 & 0 & 0 & 0 & 0 & 0 & 0 & 0 & 0\\
        \hline
O & 0 & 0 & 0 & 1 & 0 & 0 & 0 & 0 & 0 & 0 & 0 & 0 & 0 & 0 & 0 & 0 & 0\\
        \hline
TD & 0 & 0 & 0 & 0 & 0 & 0 & 0 & 0 & 0 & 0 & 0 & 0 & 0 & 0 & 0 & 0 & 0\\
        \hline
OH & 0 & 0 & 0 & 1 & 0 & 0 & 0 & 0 & 0 & 0 & 0 & 0 & 0 & 0 & 0 & 0 & 0\\
        \hline
TOTAL & 8 & 24 & 0 & 123 & 0 & 6 & 0 & 21 & 0 & 0 & 0 & 2 & 0 & 0 & 0 & 5 & 0\\
        \hline
{\footnotesize Original } & 12 & 43 & 0 & 211 & 0 & 15 & 0 & 27 & 0 & 0 & 0 & 2 & 0 & 0 & 0 & 6 & 0\\
        \hline
        \end{tabular}
}
\newpage
\setlength{\oddsidemargin}{-0.3cm}
{\small
        \begin{tabular}{*{17}{|r}|}
        \hline
{\footnotesize AB2X5 } & Z=18 & \hspace*{0.7em}19 & \hspace*{0.7em}20 & \hspace*{0.7em}21 & \hspace*{0.7em}22 & \hspace*{0.7em}23 & \hspace*{0.7em}24 & \hspace*{0.7em}25 & \hspace*{0.7em}26 & \hspace*{0.7em}27 & \hspace*{0.7em}28 & \hspace*{0.7em}29 & \hspace*{0.7em}30 & \hspace*{0.7em}31 & \hspace*{0.7em}32 & {\scriptsize TOTAL } \\
        \hline
        \hline
C1 & 0 & 0 & 0 & 0 & 0 & 0 & 0 & 0 & 0 & 0 & 0 & 0 & 0 & 0 & 0 & 3\\
        \hline
CI & 0 & 0 & 0 & 0 & 0 & 0 & 0 & 0 & 0 & 0 & 0 & 0 & 0 & 0 & 0 & 8\\
        \hline
C2 & 0 & 0 & 0 & 0 & 1 & 0 & 0 & 0 & 0 & 0 & 0 & 0 & 0 & 0 & 0 & 4\\
        \hline
CS & 0 & 0 & 0 & 0 & 0 & 0 & 1 & 0 & 0 & 0 & 0 & 0 & 0 & 0 & 0 & 2\\
        \hline
C2H & 0 & 0 & 0 & 0 & 0 & 0 & 0 & 0 & 0 & 0 & 0 & 0 & 0 & 0 & 0 & 60\\
        \hline
D2 & 0 & 0 & 0 & 0 & 0 & 0 & 0 & 0 & 0 & 0 & 0 & 0 & 0 & 0 & 0 & 6\\
        \hline
C2V & 0 & 0 & 0 & 0 & 0 & 0 & 0 & 0 & 0 & 0 & 0 & 0 & 0 & 0 & 0 & 13\\
        \hline
D2H & 0 & 0 & 0 & 0 & 0 & 0 & 0 & 0 & 0 & 0 & 0 & 0 & 0 & 0 & 0 & 69\\
        \hline
C4 & 0 & 0 & 0 & 0 & 0 & 0 & 0 & 0 & 0 & 0 & 0 & 0 & 0 & 0 & 0 & 0\\
        \hline
S4 & 0 & 0 & 0 & 0 & 0 & 0 & 0 & 0 & 0 & 0 & 0 & 0 & 0 & 0 & 0 & 0\\
        \hline
C4H & 0 & 0 & 0 & 0 & 0 & 0 & 0 & 0 & 0 & 0 & 0 & 0 & 0 & 0 & 0 & 0\\
        \hline
D4 & 0 & 0 & 0 & 0 & 0 & 0 & 0 & 0 & 0 & 0 & 0 & 0 & 0 & 0 & 0 & 1\\
        \hline
C4V & 0 & 0 & 0 & 0 & 0 & 0 & 0 & 0 & 0 & 0 & 0 & 0 & 0 & 0 & 0 & 1\\
        \hline
D2D & 0 & 0 & 0 & 0 & 0 & 0 & 0 & 0 & 0 & 0 & 0 & 0 & 0 & 0 & 0 & 1\\
        \hline
D4H & 0 & 0 & 0 & 0 & 0 & 0 & 0 & 0 & 0 & 0 & 0 & 0 & 0 & 0 & 0 & 13\\
        \hline
C3 & 0 & 0 & 0 & 0 & 0 & 0 & 0 & 0 & 0 & 0 & 0 & 0 & 0 & 0 & 0 & 1\\
        \hline
C3I & 0 & 0 & 0 & 0 & 0 & 0 & 0 & 0 & 0 & 0 & 0 & 0 & 0 & 0 & 0 & 0\\
        \hline
D3 & 0 & 0 & 0 & 0 & 0 & 0 & 0 & 0 & 0 & 0 & 0 & 0 & 0 & 0 & 0 & 1\\
        \hline
C3V & 0 & 0 & 0 & 0 & 0 & 0 & 0 & 0 & 0 & 0 & 0 & 0 & 0 & 0 & 0 & 1\\
        \hline
D3D & 0 & 0 & 0 & 0 & 0 & 0 & 0 & 0 & 0 & 0 & 0 & 0 & 0 & 0 & 0 & 0\\
        \hline
C6 & 0 & 0 & 0 & 0 & 0 & 0 & 0 & 0 & 0 & 0 & 0 & 0 & 0 & 0 & 0 & 3\\
        \hline
C3H & 0 & 0 & 0 & 0 & 0 & 0 & 0 & 0 & 0 & 0 & 0 & 0 & 0 & 0 & 0 & 0\\
        \hline
C6H & 0 & 0 & 0 & 0 & 0 & 0 & 0 & 0 & 0 & 0 & 0 & 0 & 0 & 0 & 0 & 0\\
        \hline
D6 & 0 & 0 & 0 & 0 & 0 & 0 & 0 & 0 & 0 & 0 & 0 & 0 & 0 & 0 & 0 & 0\\
        \hline
C6V & 0 & 0 & 0 & 0 & 0 & 0 & 0 & 0 & 0 & 0 & 0 & 0 & 0 & 0 & 0 & 0\\
        \hline
D3H & 0 & 0 & 0 & 0 & 0 & 0 & 0 & 0 & 0 & 0 & 0 & 0 & 0 & 0 & 0 & 0\\
        \hline
D6H & 0 & 0 & 0 & 0 & 0 & 0 & 0 & 0 & 0 & 0 & 0 & 0 & 0 & 0 & 0 & 2\\
        \hline
T & 0 & 0 & 0 & 0 & 0 & 0 & 0 & 0 & 0 & 0 & 0 & 0 & 0 & 0 & 0 & 0\\
        \hline
TH & 0 & 0 & 0 & 0 & 0 & 0 & 0 & 0 & 0 & 0 & 0 & 0 & 0 & 0 & 0 & 0\\
        \hline
O & 0 & 0 & 0 & 0 & 0 & 0 & 0 & 0 & 0 & 0 & 0 & 0 & 0 & 0 & 0 & 1\\
        \hline
TD & 0 & 0 & 0 & 0 & 0 & 0 & 0 & 0 & 0 & 0 & 0 & 0 & 0 & 0 & 0 & 0\\
        \hline
OH & 0 & 0 & 0 & 0 & 0 & 0 & 0 & 0 & 0 & 0 & 0 & 0 & 0 & 0 & 0 & 1\\
        \hline
TOTAL & 0 & 0 & 0 & 0 & 1 & 0 & 1 & 0 & 0 & 0 & 0 & 0 & 0 & 0 & 0 & 191\\
        \hline
{\footnotesize Original } & 0 & 0 & 0 & 0 & 1 & 0 & 1 & 0 & 0 & 0 & 0 & 0 & 0 & 0 & 0 & 319\\
        \hline
        \end{tabular}
}
\newpage
\setlength{\oddsidemargin}{2.54cm}

\newpage

\newpage

\newpage

\newpage

\newpage

\newpage

\newpage

\newpage

\newpage
\setlength{\oddsidemargin}{-0.1cm}
ANX=A3X4\\
{\small
        \begin{tabular}{*{18}{|r}|}
        \hline
A3X4 & Z=1 & 2 & \hspace*{0.9em}3 & 4 & \hspace*{0.9em}5 & \hspace*{0.9em}6 & \hspace*{0.9em}7 & \hspace*{0.9em}8 & \hspace*{0.9em}9 & \hspace*{0.7em}10 & \hspace*{0.7em}11 & \hspace*{0.7em}12 & \hspace*{0.7em}13 & \hspace*{0.7em}14 & \hspace*{0.7em}15 & \hspace*{0.7em}16 & \hspace*{0.7em}17\\
        \hline
        \hline
C1 & 0 & 0 & 0 & 0 & 0 & 0 & 0 & 0 & 0 & 0 & 0 & 0 & 0 & 0 & 0 & 0 & 0\\
        \hline
CI & 0 & 0 & 0 & 0 & 0 & 0 & 0 & 0 & 0 & 0 & 0 & 0 & 0 & 0 & 0 & 0 & 0\\
        \hline
C2 & 0 & 1 & 0 & 0 & 0 & 0 & 0 & 0 & 0 & 0 & 0 & 0 & 0 & 0 & 0 & 0 & 0\\
        \hline
CS & 0 & 0 & 0 & 0 & 0 & 0 & 0 & 0 & 0 & 0 & 0 & 0 & 0 & 0 & 0 & 0 & 0\\
        \hline
C2H & 1 & 3 & 0 & 1 & 0 & 0 & 0 & 0 & 0 & 0 & 0 & 0 & 0 & 0 & 0 & 0 & 0\\
        \hline
D2 & 0 & 0 & 0 & 0 & 0 & 0 & 0 & 0 & 0 & 0 & 0 & 0 & 0 & 0 & 0 & 0 & 0\\
        \hline
C2V & 0 & 0 & 0 & 3 & 0 & 0 & 0 & 4 & 0 & 0 & 0 & 0 & 0 & 0 & 0 & 0 & 0\\
        \hline
D2H & 0 & 1 & 1 & 1 & 0 & 0 & 0 & 0 & 0 & 0 & 0 & 1 & 0 & 0 & 0 & 0 & 0\\
        \hline
C4 & 0 & 0 & 0 & 0 & 0 & 0 & 0 & 0 & 0 & 0 & 0 & 0 & 0 & 0 & 0 & 0 & 0\\
        \hline
S4 & 0 & 0 & 0 & 0 & 0 & 0 & 0 & 0 & 0 & 0 & 0 & 0 & 0 & 0 & 0 & 0 & 0\\
        \hline
C4H & 0 & 0 & 0 & 0 & 0 & 0 & 0 & 0 & 0 & 0 & 0 & 0 & 0 & 0 & 0 & 0 & 0\\
        \hline
D4 & 1 & 0 & 0 & 0 & 0 & 0 & 0 & 0 & 0 & 0 & 0 & 0 & 0 & 0 & 0 & 0 & 0\\
        \hline
C4V & 0 & 0 & 0 & 0 & 0 & 0 & 0 & 0 & 0 & 0 & 0 & 0 & 0 & 0 & 0 & 0 & 0\\
        \hline
D2D & 0 & 3 & 0 & 1 & 0 & 0 & 0 & 0 & 0 & 0 & 0 & 0 & 0 & 0 & 0 & 0 & 0\\
        \hline
D4H & 1 & 0 & 0 & 2 & 0 & 0 & 0 & 0 & 0 & 0 & 0 & 0 & 0 & 0 & 0 & 0 & 0\\
        \hline
C3 & 0 & 0 & 0 & 0 & 0 & 0 & 0 & 0 & 0 & 0 & 0 & 0 & 0 & 0 & 0 & 0 & 0\\
        \hline
C3I & 1 & 1 & 2 & 0 & 0 & 0 & 0 & 0 & 0 & 0 & 0 & 0 & 0 & 0 & 0 & 0 & 0\\
        \hline
D3 & 0 & 0 & 0 & 0 & 0 & 0 & 0 & 0 & 0 & 0 & 0 & 0 & 0 & 0 & 0 & 0 & 0\\
        \hline
C3V & 0 & 0 & 0 & 2 & 0 & 0 & 0 & 0 & 0 & 0 & 0 & 0 & 0 & 0 & 0 & 0 & 0\\
        \hline
D3D & 2 & 0 & 1 & 0 & 0 & 1 & 0 & 0 & 0 & 0 & 0 & 0 & 0 & 0 & 0 & 0 & 0\\
        \hline
C6 & 1 & 1 & 0 & 0 & 0 & 1 & 0 & 0 & 0 & 0 & 0 & 0 & 0 & 0 & 0 & 0 & 0\\
        \hline
C3H & 0 & 0 & 0 & 0 & 0 & 0 & 0 & 0 & 0 & 0 & 0 & 0 & 0 & 0 & 0 & 0 & 0\\
        \hline
C6H & 1 & 5 & 0 & 0 & 0 & 0 & 0 & 0 & 0 & 0 & 0 & 0 & 0 & 0 & 0 & 0 & 0\\
        \hline
D6 & 0 & 0 & 0 & 0 & 0 & 0 & 0 & 0 & 0 & 0 & 0 & 0 & 0 & 0 & 0 & 0 & 0\\
        \hline
C6V & 0 & 0 & 0 & 0 & 0 & 0 & 0 & 0 & 0 & 0 & 0 & 0 & 0 & 0 & 0 & 0 & 0\\
        \hline
D3H & 0 & 0 & 0 & 0 & 0 & 0 & 0 & 0 & 0 & 0 & 0 & 0 & 0 & 0 & 0 & 0 & 0\\
        \hline
D6H & 1 & 0 & 0 & 0 & 0 & 0 & 0 & 0 & 0 & 0 & 0 & 0 & 0 & 0 & 0 & 0 & 0\\
        \hline
T & 0 & 0 & 0 & 0 & 0 & 0 & 0 & 0 & 0 & 0 & 0 & 0 & 0 & 0 & 0 & 0 & 0\\
        \hline
TH & 0 & 0 & 0 & 0 & 0 & 0 & 0 & 0 & 0 & 0 & 0 & 0 & 0 & 0 & 0 & 0 & 0\\
        \hline
O & 0 & 0 & 0 & 1 & 0 & 0 & 0 & 1 & 0 & 0 & 0 & 0 & 0 & 0 & 0 & 0 & 0\\
        \hline
TD & 13 & 0 & 0 & 5 & 0 & 1 & 0 & 2 & 0 & 0 & 0 & 0 & 0 & 0 & 0 & 0 & 0\\
        \hline
OH & 7 & 2 & 0 & 2 & 0 & 0 & 0 & 48 & 0 & 0 & 0 & 0 & 0 & 0 & 0 & 0 & 0\\
        \hline
TOTAL & 29 & 17 & 4 & 18 & 0 & 3 & 0 & 55 & 0 & 0 & 0 & 1 & 0 & 0 & 0 & 0 & 0\\
        \hline
{\footnotesize Original } & 47 & 52 & 4 & 37 & 0 & 3 & 0 & 185 & 0 & 0 & 0 & 1 & 0 & 0 & 0 & 0 & 0\\
        \hline
        \end{tabular}
}
\newpage
\setlength{\oddsidemargin}{-0.3cm}
{\small
        \begin{tabular}{*{17}{|r}|}
        \hline
A3X4 & Z=18 & \hspace*{0.7em}19 & \hspace*{0.7em}20 & \hspace*{0.7em}21 & \hspace*{0.7em}22 & \hspace*{0.7em}23 & \hspace*{0.7em}24 & \hspace*{0.7em}25 & \hspace*{0.7em}26 & \hspace*{0.7em}27 & \hspace*{0.7em}28 & \hspace*{0.7em}29 & \hspace*{0.7em}30 & \hspace*{0.7em}31 & \hspace*{0.7em}32 & {\scriptsize TOTAL }\\
        \hline
        \hline
C1 & 0 & 0 & 0 & 0 & 0 & 0 & 0 & 0 & 0 & 0 & 0 & 0 & 0 & 0 & 0 & 0\\
        \hline
CI & 0 & 0 & 0 & 0 & 0 & 0 & 0 & 0 & 0 & 0 & 0 & 0 & 0 & 0 & 0 & 0\\
        \hline
C2 & 0 & 0 & 0 & 0 & 0 & 0 & 0 & 0 & 0 & 0 & 0 & 0 & 0 & 0 & 0 & 1\\
        \hline
CS & 0 & 0 & 0 & 0 & 0 & 0 & 0 & 0 & 0 & 0 & 0 & 0 & 0 & 0 & 0 & 0\\
        \hline
C2H & 0 & 0 & 0 & 0 & 0 & 0 & 0 & 0 & 0 & 0 & 0 & 0 & 0 & 0 & 0 & 5\\
        \hline
D2 & 0 & 0 & 0 & 0 & 0 & 0 & 0 & 0 & 0 & 0 & 0 & 0 & 0 & 0 & 0 & 0\\
        \hline
C2V & 0 & 0 & 0 & 0 & 0 & 0 & 0 & 0 & 0 & 0 & 0 & 0 & 0 & 0 & 0 & 7\\
        \hline
D2H & 0 & 0 & 0 & 0 & 0 & 0 & 0 & 0 & 0 & 0 & 0 & 0 & 0 & 0 & 0 & 4\\
        \hline
C4 & 0 & 0 & 0 & 0 & 0 & 0 & 0 & 0 & 0 & 0 & 0 & 0 & 0 & 0 & 0 & 0\\
        \hline
S4 & 0 & 0 & 0 & 0 & 0 & 0 & 0 & 0 & 0 & 0 & 0 & 0 & 0 & 0 & 0 & 0\\
        \hline
C4H & 0 & 0 & 0 & 0 & 0 & 0 & 0 & 0 & 0 & 0 & 0 & 0 & 0 & 0 & 0 & 0\\
        \hline
D4 & 0 & 0 & 0 & 0 & 0 & 0 & 0 & 0 & 0 & 0 & 0 & 0 & 0 & 0 & 0 & 1\\
        \hline
C4V & 0 & 0 & 0 & 0 & 0 & 0 & 0 & 0 & 0 & 0 & 0 & 0 & 0 & 0 & 0 & 0\\
        \hline
D2D & 0 & 0 & 0 & 0 & 0 & 0 & 0 & 0 & 0 & 0 & 0 & 0 & 0 & 0 & 0 & 4\\
        \hline
D4H & 0 & 0 & 0 & 0 & 0 & 0 & 0 & 0 & 0 & 0 & 0 & 0 & 0 & 0 & 0 & 3\\
        \hline
C3 & 0 & 0 & 0 & 0 & 0 & 0 & 0 & 0 & 0 & 0 & 0 & 0 & 0 & 0 & 0 & 0\\
        \hline
C3I & 0 & 0 & 0 & 0 & 0 & 0 & 0 & 0 & 0 & 0 & 0 & 0 & 0 & 0 & 0 & 4\\
        \hline
D3 & 0 & 0 & 0 & 0 & 0 & 0 & 0 & 0 & 0 & 0 & 0 & 0 & 0 & 0 & 0 & 0\\
        \hline
C3V & 0 & 0 & 0 & 0 & 0 & 0 & 0 & 0 & 0 & 0 & 0 & 0 & 0 & 0 & 0 & 2\\
        \hline
D3D & 0 & 0 & 0 & 0 & 0 & 0 & 0 & 0 & 0 & 0 & 0 & 0 & 0 & 0 & 0 & 4\\
        \hline
C6 & 0 & 0 & 0 & 0 & 0 & 0 & 0 & 0 & 0 & 0 & 0 & 0 & 0 & 0 & 0 & 3\\
        \hline
C3H & 0 & 0 & 0 & 0 & 0 & 0 & 0 & 0 & 0 & 0 & 0 & 0 & 0 & 0 & 0 & 0\\
        \hline
C6H & 0 & 0 & 0 & 0 & 0 & 0 & 0 & 0 & 0 & 0 & 0 & 0 & 0 & 0 & 0 & 6\\
        \hline
D6 & 0 & 0 & 0 & 0 & 0 & 0 & 0 & 0 & 0 & 0 & 0 & 0 & 0 & 0 & 0 & 0\\
        \hline
C6V & 0 & 0 & 0 & 0 & 0 & 0 & 0 & 0 & 0 & 0 & 0 & 0 & 0 & 0 & 0 & 0\\
        \hline
D3H & 0 & 0 & 0 & 0 & 0 & 0 & 0 & 0 & 0 & 0 & 0 & 0 & 0 & 0 & 0 & 0\\
        \hline
D6H & 0 & 0 & 0 & 0 & 0 & 0 & 0 & 0 & 0 & 0 & 0 & 0 & 0 & 0 & 0 & 1\\
        \hline
T & 0 & 0 & 0 & 0 & 0 & 0 & 0 & 0 & 0 & 0 & 0 & 0 & 0 & 0 & 0 & 0\\
        \hline
TH & 0 & 0 & 0 & 0 & 0 & 0 & 0 & 0 & 0 & 0 & 0 & 0 & 0 & 0 & 0 & 0\\
        \hline
O & 0 & 0 & 0 & 0 & 0 & 0 & 0 & 0 & 0 & 0 & 0 & 0 & 0 & 0 & 0 & 2\\
        \hline
TD & 0 & 0 & 0 & 0 & 0 & 0 & 0 & 0 & 0 & 0 & 0 & 0 & 0 & 0 & 0 & 21\\
        \hline
OH & 0 & 0 & 0 & 0 & 0 & 0 & 0 & 0 & 0 & 0 & 0 & 0 & 0 & 0 & 0 & 59\\
        \hline
TOTAL & 0 & 0 & 0 & 0 & 0 & 0 & 0 & 0 & 0 & 0 & 0 & 0 & 0 & 0 & 0 & 127\\
        \hline
{\footnotesize Original } & 0 & 0 & 0 & 0 & 0 & 0 & 0 & 0 & 0 & 0 & 0 & 0 & 0 & 0 & 0 & 329\\
        \hline
        \end{tabular}
}
\newpage
\setlength{\oddsidemargin}{2.54cm}

\newpage

\newpage

\newpage

\newpage

\newpage

\newpage

\newpage
\setlength{\oddsidemargin}{-0.1cm}
ANX=A2B2X7\\
{\small
        \begin{tabular}{*{18}{|r}|}
        \hline
{\footnotesize A2B2X7 } & Z=1 & 2 & \hspace*{0.9em}3 & 4 & \hspace*{0.9em}5 & \hspace*{0.9em}6 & \hspace*{0.9em}7 & \hspace*{0.9em}8 & \hspace*{0.9em}9 & \hspace*{0.7em}10 & \hspace*{0.7em}11 & \hspace*{0.7em}12 & \hspace*{0.7em}13 & \hspace*{0.7em}14 & \hspace*{0.7em}15 & \hspace*{0.7em}16 & \hspace*{0.7em}17\\
        \hline
        \hline
C1 & 1 & 1 & 0 & 0 & 0 & 1 & 0 & 0 & 0 & 0 & 0 & 0 & 0 & 0 & 0 & 0 & 0\\
        \hline
CI & 0 & 14 & 0 & 8 & 0 & 0 & 0 & 0 & 0 & 1 & 0 & 1 & 0 & 0 & 0 & 0 & 0\\
        \hline
C2 & 0 & 4 & 0 & 2 & 0 & 0 & 0 & 1 & 0 & 0 & 0 & 0 & 0 & 0 & 0 & 0 & 0\\
        \hline
CS & 0 & 1 & 0 & 0 & 0 & 0 & 0 & 0 & 0 & 0 & 0 & 0 & 0 & 0 & 0 & 0 & 0\\
        \hline
C2H & 1 & 30 & 0 & 20 & 0 & 0 & 0 & 1 & 0 & 0 & 0 & 1 & 0 & 0 & 0 & 0 & 0\\
        \hline
D2 & 0 & 0 & 0 & 1 & 0 & 0 & 0 & 0 & 0 & 0 & 0 & 0 & 0 & 0 & 0 & 0 & 0\\
        \hline
C2V & 0 & 1 & 0 & 12 & 0 & 0 & 0 & 3 & 0 & 0 & 0 & 0 & 0 & 0 & 0 & 0 & 0\\
        \hline
D2H & 0 & 1 & 0 & 5 & 0 & 0 & 0 & 4 & 0 & 0 & 0 & 0 & 0 & 0 & 0 & 2 & 0\\
        \hline
C4 & 0 & 0 & 0 & 0 & 0 & 0 & 0 & 5 & 0 & 0 & 0 & 0 & 0 & 0 & 0 & 0 & 0\\
        \hline
S4 & 0 & 0 & 0 & 0 & 0 & 0 & 0 & 0 & 0 & 0 & 0 & 0 & 0 & 0 & 0 & 0 & 0\\
        \hline
C4H & 0 & 0 & 0 & 0 & 0 & 0 & 0 & 0 & 0 & 0 & 0 & 0 & 0 & 0 & 0 & 0 & 0\\
        \hline
D4 & 0 & 0 & 0 & 4 & 0 & 0 & 0 & 0 & 0 & 0 & 0 & 0 & 0 & 0 & 0 & 0 & 0\\
        \hline
C4V & 0 & 0 & 0 & 0 & 0 & 0 & 0 & 0 & 0 & 0 & 0 & 0 & 0 & 0 & 0 & 0 & 0\\
        \hline
D2D & 1 & 0 & 0 & 0 & 0 & 0 & 0 & 0 & 0 & 0 & 0 & 0 & 0 & 0 & 0 & 0 & 0\\
        \hline
D4H & 0 & 0 & 0 & 0 & 0 & 0 & 0 & 0 & 0 & 0 & 0 & 0 & 0 & 0 & 0 & 0 & 0\\
        \hline
C3 & 0 & 0 & 0 & 0 & 0 & 0 & 0 & 0 & 0 & 0 & 0 & 0 & 0 & 0 & 0 & 0 & 0\\
        \hline
C3I & 0 & 0 & 0 & 0 & 0 & 0 & 0 & 0 & 0 & 0 & 0 & 0 & 0 & 0 & 0 & 0 & 0\\
        \hline
D3 & 0 & 0 & 0 & 0 & 0 & 1 & 0 & 0 & 0 & 0 & 0 & 0 & 0 & 0 & 0 & 0 & 0\\
        \hline
C3V & 0 & 0 & 0 & 0 & 0 & 0 & 0 & 0 & 0 & 0 & 0 & 0 & 0 & 0 & 0 & 0 & 0\\
        \hline
D3D & 0 & 0 & 0 & 0 & 0 & 0 & 0 & 0 & 0 & 0 & 0 & 0 & 0 & 0 & 0 & 0 & 0\\
        \hline
C6 & 0 & 0 & 0 & 0 & 0 & 0 & 0 & 0 & 0 & 0 & 0 & 0 & 0 & 0 & 0 & 0 & 0\\
        \hline
C3H & 0 & 0 & 0 & 0 & 0 & 0 & 0 & 0 & 0 & 0 & 0 & 0 & 0 & 0 & 0 & 0 & 0\\
        \hline
C6H & 0 & 0 & 0 & 0 & 0 & 0 & 0 & 0 & 0 & 0 & 0 & 0 & 0 & 0 & 0 & 0 & 0\\
        \hline
D6 & 0 & 0 & 0 & 0 & 0 & 0 & 0 & 0 & 0 & 0 & 0 & 0 & 0 & 0 & 0 & 0 & 0\\
        \hline
C6V & 0 & 0 & 0 & 0 & 0 & 0 & 0 & 0 & 0 & 0 & 0 & 0 & 0 & 0 & 0 & 0 & 0\\
        \hline
D3H & 0 & 0 & 0 & 0 & 0 & 0 & 0 & 0 & 0 & 0 & 0 & 0 & 0 & 0 & 0 & 0 & 0\\
        \hline
D6H & 0 & 0 & 0 & 0 & 0 & 0 & 0 & 0 & 0 & 0 & 0 & 0 & 0 & 0 & 0 & 0 & 0\\
        \hline
T & 0 & 0 & 0 & 0 & 0 & 0 & 0 & 0 & 0 & 0 & 0 & 0 & 0 & 0 & 0 & 0 & 0\\
        \hline
TH & 0 & 0 & 0 & 0 & 0 & 0 & 0 & 0 & 0 & 0 & 0 & 0 & 0 & 0 & 0 & 0 & 0\\
        \hline
O & 0 & 0 & 0 & 0 & 0 & 0 & 0 & 0 & 0 & 0 & 0 & 0 & 0 & 0 & 0 & 0 & 0\\
        \hline
TD & 0 & 0 & 0 & 0 & 0 & 0 & 0 & 0 & 0 & 1 & 0 & 0 & 0 & 0 & 0 & 0 & 0\\
        \hline
OH & 1 & 0 & 0 & 0 & 0 & 0 & 0 & 54 & 0 & 0 & 0 & 0 & 0 & 0 & 0 & 3 & 0\\
        \hline
TOTAL & 4 & 52 & 0 & 52 & 0 & 2 & 0 & 68 & 0 & 2 & 0 & 2 & 0 & 0 & 0 & 5 & 0\\
        \hline
{\footnotesize Original } & 4 & 77 & 0 & 75 & 0 & 3 & 0 & 176 & 0 & 2 & 0 & 2 & 0 & 0 & 0 & 5 & 0\\
        \hline
        \end{tabular}
}
\newpage
\setlength{\oddsidemargin}{-0.3cm}
{\small
        \begin{tabular}{*{17}{|r}|}
        \hline
{\footnotesize A2B2X7 } & Z=18 & \hspace*{0.7em}19 & \hspace*{0.7em}20 & \hspace*{0.7em}21 & \hspace*{0.7em}22 & \hspace*{0.7em}23 & \hspace*{0.7em}24 & \hspace*{0.7em}25 & \hspace*{0.7em}26 & \hspace*{0.7em}27 & \hspace*{0.7em}28 & \hspace*{0.7em}29 & \hspace*{0.7em}30 & \hspace*{0.7em}31 & \hspace*{0.7em}32 & {\scriptsize TOTAL }\\
        \hline
        \hline
C1 & 0 & 0 & 0 & 0 & 0 & 0 & 0 & 0 & 0 & 0 & 0 & 0 & 0 & 0 & 0 & 3\\
        \hline
CI & 0 & 0 & 0 & 0 & 0 & 0 & 0 & 0 & 0 & 0 & 0 & 0 & 0 & 0 & 0 & 24\\
        \hline
C2 & 0 & 0 & 0 & 0 & 0 & 0 & 0 & 0 & 0 & 0 & 0 & 0 & 0 & 0 & 0 & 7\\
        \hline
CS & 0 & 0 & 0 & 0 & 0 & 0 & 0 & 0 & 0 & 0 & 0 & 0 & 0 & 0 & 0 & 1\\
        \hline
C2H & 0 & 0 & 0 & 0 & 0 & 0 & 0 & 0 & 0 & 0 & 0 & 0 & 0 & 0 & 0 & 53\\
        \hline
D2 & 0 & 0 & 0 & 0 & 0 & 0 & 0 & 0 & 0 & 0 & 0 & 0 & 0 & 0 & 0 & 1\\
        \hline
C2V & 0 & 0 & 0 & 0 & 0 & 0 & 0 & 0 & 0 & 0 & 0 & 0 & 0 & 0 & 0 & 16\\
        \hline
D2H & 0 & 0 & 0 & 0 & 0 & 0 & 0 & 0 & 0 & 0 & 0 & 0 & 0 & 0 & 0 & 12\\
        \hline
C4 & 0 & 0 & 0 & 0 & 0 & 0 & 0 & 0 & 0 & 0 & 0 & 0 & 0 & 0 & 0 & 5\\
        \hline
S4 & 0 & 0 & 0 & 0 & 0 & 0 & 0 & 0 & 0 & 0 & 0 & 0 & 0 & 0 & 0 & 0\\
        \hline
C4H & 0 & 0 & 0 & 0 & 0 & 0 & 0 & 0 & 0 & 0 & 0 & 0 & 0 & 0 & 0 & 0\\
        \hline
D4 & 0 & 0 & 0 & 0 & 0 & 0 & 0 & 0 & 0 & 0 & 0 & 0 & 0 & 0 & 0 & 4\\
        \hline
C4V & 0 & 0 & 0 & 0 & 0 & 0 & 0 & 0 & 0 & 0 & 0 & 0 & 0 & 0 & 0 & 0\\
        \hline
D2D & 0 & 0 & 0 & 0 & 0 & 0 & 0 & 0 & 0 & 0 & 0 & 0 & 0 & 0 & 0 & 1\\
        \hline
D4H & 0 & 0 & 0 & 0 & 0 & 0 & 0 & 0 & 0 & 0 & 0 & 0 & 0 & 0 & 0 & 0\\
        \hline
C3 & 0 & 0 & 0 & 0 & 0 & 0 & 0 & 0 & 0 & 0 & 0 & 0 & 0 & 0 & 0 & 0\\
        \hline
C3I & 0 & 0 & 0 & 0 & 0 & 0 & 0 & 0 & 0 & 0 & 0 & 0 & 0 & 0 & 0 & 0\\
        \hline
D3 & 0 & 0 & 0 & 0 & 0 & 0 & 0 & 0 & 0 & 0 & 0 & 0 & 0 & 0 & 0 & 1\\
        \hline
C3V & 0 & 0 & 0 & 0 & 0 & 0 & 0 & 0 & 0 & 0 & 0 & 0 & 0 & 0 & 0 & 0\\
        \hline
D3D & 0 & 0 & 0 & 0 & 0 & 0 & 0 & 0 & 0 & 0 & 0 & 0 & 0 & 0 & 0 & 0\\
        \hline
C6 & 0 & 0 & 0 & 0 & 0 & 0 & 0 & 0 & 0 & 0 & 0 & 0 & 0 & 0 & 0 & 0\\
        \hline
C3H & 0 & 0 & 0 & 0 & 0 & 0 & 0 & 0 & 0 & 0 & 0 & 0 & 0 & 0 & 0 & 0\\
        \hline
C6H & 0 & 0 & 0 & 0 & 0 & 0 & 0 & 0 & 0 & 0 & 0 & 0 & 0 & 0 & 0 & 0\\
        \hline
D6 & 0 & 0 & 0 & 0 & 0 & 0 & 0 & 0 & 0 & 0 & 0 & 0 & 0 & 0 & 0 & 0\\
        \hline
C6V & 0 & 0 & 0 & 0 & 0 & 0 & 0 & 0 & 0 & 0 & 0 & 0 & 0 & 0 & 0 & 0\\
        \hline
D3H & 0 & 0 & 0 & 0 & 0 & 0 & 0 & 0 & 0 & 0 & 0 & 0 & 0 & 0 & 0 & 0\\
        \hline
D6H & 0 & 0 & 0 & 0 & 0 & 0 & 0 & 0 & 0 & 0 & 0 & 0 & 0 & 0 & 0 & 0\\
        \hline
T & 0 & 0 & 0 & 0 & 0 & 0 & 0 & 0 & 0 & 0 & 0 & 0 & 0 & 0 & 0 & 0\\
        \hline
TH & 0 & 0 & 0 & 0 & 0 & 0 & 0 & 0 & 0 & 0 & 0 & 0 & 0 & 0 & 0 & 0\\
        \hline
O & 0 & 0 & 0 & 0 & 0 & 0 & 0 & 0 & 0 & 0 & 0 & 0 & 0 & 0 & 0 & 0\\
        \hline
TD & 0 & 0 & 0 & 0 & 0 & 0 & 0 & 0 & 0 & 0 & 0 & 0 & 0 & 0 & 0 & 1\\
        \hline
OH & 0 & 0 & 0 & 0 & 0 & 0 & 0 & 0 & 0 & 0 & 0 & 0 & 0 & 0 & 0 & 58\\
        \hline
TOTAL & 0 & 0 & 0 & 0 & 0 & 0 & 0 & 0 & 0 & 0 & 0 & 0 & 0 & 0 & 0 & 187\\
        \hline
{\footnotesize Original } & 0 & 0 & 0 & 0 & 0 & 0 & 0 & 0 & 0 & 0 & 0 & 0 & 0 & 0 & 0 & 344\\
        \hline
        \end{tabular}
}
\newpage
\setlength{\oddsidemargin}{2.54cm}

\newpage

\newpage

\newpage

\newpage

\newpage

\newpage

\newpage


}
\end{document}